\newcommand*\diff{\mathop{}\!\mathrm{d}}

\documentclass[a4paper,11pt]{article}
\pdfoutput=1 

\usepackage{jheppub} 

\usepackage[utf8]{inputenc}
\usepackage{tensor}
\usepackage{graphicx}
\usepackage{dcolumn}
\usepackage{bm}
\usepackage{comment}
\usepackage{xcolor}
\usepackage{booktabs}
\usepackage{hyperref}
\hypersetup{
    colorlinks,
    linkcolor=black,
    citecolor=black,
    urlcolor=black,
}

\usepackage{braket}
\newcommand{\virgolette}[1]{``#1''}

\begin{document}


\title{Effective relational cosmological dynamics from quantum gravity} 

\author[a,b,c]{Luca Marchetti}
\author[b]{Daniele Oriti}


\affiliation[a]{Università di Pisa,\\Lungarno Antonio Pacinotti 43, 56126 Pisa, Italy, EU}
\affiliation[b]{Arnold Sommerfeld Center for Theoretical Physics,\\ Ludwig-Maximilians-Universit\"at München \\ Theresienstrasse 37, 80333 M\"unchen, Germany, EU}
\affiliation[c]{Istituto Nazionale di Fisica Nucleare sez. Pisa,\\Largo Bruno Pontecorvo 3, 56127 Pisa, Italy, EU}


\emailAdd{luca.marchetti@phd.unipi.it}
\emailAdd{daniele.oriti@physik.lmu.de}

\date{\today}

\abstract{
We discuss the relational
strategy to solve the problem of time in quantum gravity and different ways in which it could be implemented, pointing out in particular the fundamentally new dimension that the problem takes in a quantum gravity context in which spacetime and geometry are understood as emergent.
We realize concretely the  relational strategy we have advocated in the context of the tensorial group field theory formalism for quantum gravity, leading to the extraction of an effective relational cosmological dynamics from quantum geometric models.
We analyze in detail the emergent cosmological dynamics, highlighting  the improvements over previous work, the contribution of the quantum properties of the relational clock to it, and the interplay between the conditions ensuring a bona fide relational dynamics throughout the cosmological evolution and the existence of a quantum bounce resolving the classical big bang singularity. }

\maketitle
\section{Introduction}
The background independence of classical General Relativity, which we expect to be carried over to the quantum domain, implies that only dynamical entities should determine the content of our physical description of the world. Any remaining background structure should be diffeomorphism invariant and thus devoid of any physical and local spacetime characterization itself \cite{Giulini:2006yg}. This means that, if present, background structures should play at most an auxiliary role. For instance, this applies to the differentiable manifold on which dynamical fields are defined, which should only enter the physics in its global (topological) characterization. In particular, any notion of local region or point (which may be referred to in terms of coordinate systems) is not physical because it is not diffeomorphism invariant.

This has many conceptual, mathematical and physical consequences, the most notable being that no external, fixed or preferred notion of time (nor space) can be assumed.
It is therefore a difficult task, in general, to extract from the theory a diffeomorphism-invariant, yet dynamical picture of the world in terms of observable quantities evolving in time.
In the classical setting, we often manage to avoid dealing with this troublesome feature by directly working 
with specific solutions characterized by special isometries, to which preferred temporal and spatial directions can be associated.  

In a quantum context, this way out is precluded. Thus, approaches to the quantization of gravity have to deal directly with the absence of preferred temporal (and spatial) directions.  
In the canonical description \cite{Rovelli:2004tv, Thiemann:2007pyv}, for example, this background independence manifests itself in the absence of a true Hamiltonian (in absence of boundaries) \cite{Rovelli:2004tv, Thiemann:2007pyv, Arnowitt:1962hi}. At the quantum level, the resulting picture is that of a \virgolette{frozen-time}, where states can not evolve in time. This fact, often referred to as the \lq\emph{problem of time}\rq$\,$ in quantum gravity \cite{Kuchar2011,Isham1992}, is actually just the statement that physical states should not evolve with respect to an external time, and it is inherited straight from the classical theory. 

Still, no evolution with respect to external parameters does not mean no evolution (or \virgolette{no change}) at all; it only means that physical systems, including the gravitational field, evolve with respect to other dynamical degrees of freedom of the theory. 
This, at least, is the {\it relational} point of view on the problem of time (and space, and observables more generally) in classical and quantum gravity. This is also the point of view we adopt in this work.
From this perspective, the reference frames (i.e. chosen clock and rods) that we are used to in the pre-relativistic context should now be recognised as internal objects of the theory and, in a quantum theory of gravity, these \virgolette{clocks} and \virgolette{rods} should therefore be themselves chosen among quantum degrees of freedom described by the theory. 

There are three main approaches to describe a (quantum) relational evolution in a generally covariant theory (see \cite{Hoehn:2019owq} and references therein). The first one is based on an appropriate definition of gauge invariant relational (Dirac) observables in the full Hilbert space of the theory (without a priori any rewriting of the same), expressing the evolution of all but one of the quantum degrees of freedom of the full system as a function of the values taken by a selected one used as a clock. The second one, known as Page-Wooters formalism, is based on the explicit separation of the Hilbert space into \virgolette{clock} and \virgolette{system} spaces, and on the introduction of system states which are \virgolette{conditioned} on the value of the clock. In this way, it realizes a relational Schr\"odinger picture. Lastly, the third one, often named \lq quantum symmetry reduction\rq, classically selects a time observable, which is then used to construct the quantum theory. This is close to a reduced phase space quantization, and gives a relational Heisenberg picture. These three frameworks, born from the same physical requirement of describing evolution of some degrees of freedom with respect to another one, can in fact be shown to be equivalent, when the \virgolette{clock} and the \virgolette{system} satisfy a certain number of conditions \cite{Hoehn:2019owq}.

However, none of these procedures can be straightforwardly applied to quantum gravity formalisms where spacetime and geometry are emergent. In these approaches, the fundamental degrees of freedom of the theory do not correspond directly to (quantized) fields (which are what ends up defining our physical rods and clock) and, as a consequence, the connection to standard continuum spacetime notions and to any classical gravitational theory is in this case more indirect. 
Therefore, in such quantum gravity formalisms, one further step is needed to link the fundamental objects of the theory to any spacetime notion and to reproduce continuum structures, like fields, to be then used as relational clock and rods. This is typically obtained via some form of coarse graining, based on collective states or some averaged observables (or possibly both of them). This complicates the extraction of an effective relational dynamics, whose definition is however of great importance for emergent quantum gravity theories. Indeed, since they lack any geometric structure, it provides a rather straightforward way to compare their resulting continuum physics with classical generally relativistic theories. In the following, we are going to explain these additional difficulties in some generality, as well as giving some more detail on the various strategies for extracting a relational dynamics in quantum gravity, before tackling the issue in a specific quantum gravity context.

The concrete context of our choice is the tensorial group field theory formalism (see \cite{Oriti:2011jm, Krajewski:2012aw, Carrozza:2016vsq, Gielen:2016dss} for general introductions)\footnote{In the following, TGFT or GFT, the latter usually labelling the specific class of models directly constructed by quantizing simplicial geometric structures, is used as a general label.}. This formalism is based in fact on this \virgolette{emergent spacetime} perspective and it is not, in itself, the result of a straighforward quantization of a classical gravitational theory.
TGFTs aim to describe the structure and dynamics of \virgolette{quanta of space}, identified with elementary discrete objects (usually, quantized tetrahedra), to each of which one can associate a classical phase space, but whose relation with continuum spacetime, the Hilbert space of canonical quantum gravity, and the classical phase space of General Relativity, is only indirect. This makes relational constructions based on a classical continuum spacetime phase space unavailable. The Page-Wooters approach, on the other hand, starts with the separation of the Hilbert space into one for the clock and another for the remaining degrees of freedom, and the (Fock) Hilbert space of TGFT does not admit such a decomposition. 

A first attempt to define a relational dynamics in the full TGFT framework has been made in \cite{Oriti:2016qtz}, in the context of so-called GFT condensate cosmology, by minimally coupling the degrees of freedom corresponding, in a continuum approximation, to a free massless scalar to the quantum pre-geometric ones. Then, \virgolette{relational} operators (close to complete observables \cite{Tambornino:2011vg, Dittrich:2004cb, Dittrich:2005kc, Rovelli:2001bq}) were constructed within the full quantum setting. The fundamental quantum dynamics has then been shown to imply, for such relational observables, an effective relational dynamics with a very interesting cosmological interpretation. However, as we will discuss in Subsection \ref{subsec:pregeometric}, the definition of such relational observables is plagued by some ambiguities, following from some conceptual shortcomings of the construction, and the ensuing \virgolette{relational dynamics} presents some problematic aspects. 

For instance, variances of \virgolette{relational} quantum operators defined as in \cite{Oriti:2016qtz} are plagued by divergences. 
A proper evaluation of variances of relational observables, in turn, is crucial to assess the liability of the mean field approximation used in \cite{Oriti:2016qtz} to extract the same effective (relational) cosmological evolution. Moreover, one of the most intriguing features of the TGFT condensate cosmology approach, i.e.\ the resolution of the initial singularity into a bounce, is obtained within such mean field approximation, which should then be tested for robustness. The same is true for the semi-classical limit itself, which can be trusted as long as quantum fluctuations are negligible. 

This kind of technical issues has been already discussed in the literature (see for example \cite{Adjei:2017bfm,Gielen:2018xph,Assanioussi:2020hwf}) and tackled with different approaches. In \cite{Adjei:2017bfm,Gielen:2018xph}, new \virgolette{equal-time} commutation relations (with respect to a scalar field clock)
have been postulated. Similar commutation relations have been instead derived from a canonical quantization of a (class of) TGFT model(s) in which the theory is \virgolette{ deparametrised} with respect to the scalar field clock at the coarse-grained continuum level. In this way the aforementioned divergences disappear because of the distributional nature of the more fundamental commutators of the TGFT formalism is suppressed. However, this is a rather non-trivial modification of the kinematic structure of the theory, which one should expect to be valid only at some effective level, with respect to the fundamental TGFT theory (see Section \ref{sec:relationalgft}).  
 
Another approach is the one used in \cite{Assanioussi:2020hwf}, where, once acknowledged the distributional nature of the TGFT field, a smearing of the operators with appropriate test functions has been performed. Again, this solves the issues with divergences, but it leads to a functional dynamics which is difficult to interpret at a physical level.

We will discuss further these difficulties in the following, clarifying also how they can be seen as ultimately due to the ambiguities in defining a relational dynamics at a full quantum level in a theory characterized by \virgolette{clock-neutral} or \virgolette{timeless} commutation relations, such as those between the fundamental TGFT fields (see Subsection \ref{subsec:secondquantization}).
We will show that if relational dynamics is defined in a different, more physical way, improving on the procedure adopted in \cite{Oriti:2016qtz}, these issues do not arise and also the conceptual setup becomes clearer. The resulting relational dynamics has again a good semi-classical limit, while maintaining an interesting signature of the underlying quantum geometry. 



\section{Relational dynamics in quantum gravity}\label{sec:reldynamicsqg}
As we have mentioned, in a background independent theory, where by definition a preferred notion of time is lacking, any meaningful notion of evolution must be relational. Extracting such a relational dynamics from (any given candidate to) the fundamental quantum theory is a hard challenge. We sketch in Subsection \ref{subsec:generalpicture} some of the main elements of this issue, while, in Subsection \ref{subsec:definingeffectivedynamics}, we will describe which conditions are needed in order to implement an effective relational dynamics framework for theories in which gravity is expected to appear as an emergent phenomenon.
\subsection{The general picture}\label{subsec:generalpicture}
There is a vast literature discussing the issue of relational dynamics in quantum gravity. The problem is however mostly studied in a canonical setting (in particular see \cite{Hoehn:2019owq} for a more careful treatment of the general scheme, and, for example, \cite{Bojowald:2010xp,Bojowald:2010qw,Bojowald:2009zzc} for canonical systems, with related applications to cosmological systems in \cite{Bojowald:2012xy}). While we also refer to the canonical case in the following discussion, the aim of this subsection is more general (and thus necessarily less formal), including also the case of theories which are not a direct quantization of a classical theory of geometry and gravity.
\subsubsection{\virgolette{Quantum General Relativity} theories}
In the context of theories obtained from a direct quantization of a classical theory of geometry and gravity (for instance, Quantum General Relativity), there are basically two different routes that can be followed. 

One could select a clock variable $X_0\equiv T$ at the classical level, among the dynamical fields of the theory, singling it out as an \virgolette{external structure} (this may require solving some of the dynamical, or, perhaps constraint equations). Schematically: 
\begin{equation}
\left\{ X_0,P_0; X_1, P_1; ...; X_N, P_N\right\}\quad\longrightarrow \quad \left\{ X_1, P_1; ...; X_N, P_N\right\}_{X_0\equiv  T; P_0=P_0(X_i,P_i)} \,,
\end{equation} 
where $X_i,P_i$ are the dynamical variables of the theory in a phase space formulation, i.e. all fields (including the metric) and their conjugate momenta. 
Then one can quantize the resulting theory in terms of the chosen classical relational time (\virgolette{tempus ante quantum} \cite{Isham1992}). 

Alternatively, one can look for a notion of relational time after a clock-neutral quantization of the full background independent theory (\virgolette{tempus post quantum} \cite{Isham1992}). This implies identifying a relevant quantum observable $\hat{X}_0$, constructed out of the classical variable $X_0\equiv T$ as the one \virgolette{ measuring time}, thus defining a \virgolette{quantum clock} , with its eigenstates corresponding to its readings
. The fact that one can not simply work with the quantum operator corresponding to the classical phase space variable chosen as a clock is a consequence of the need to have a well-defined (Hamiltonian) evolution and a well-defined evolution operator generating it \cite{Hoehn:2019owq}. It is however possible, as one may expect, to relate the more rigorously defined \virgolette{quantum clock} observable to the classical clock variable, at an effective level. This can be done in terms of observables or of quantum states on which such observables are evaluated. To do so requires additional conditions on the relevant class of quantum states to focus on, for example enforcing appropriate semi-classicality properties, basically restricting oneself to the regime in which the chosen clock subsystem behaves nicely enough to be traded for a good time label. 
Schematically, one would look for $\Psi_T\left(X_0,X_1,...,X_N\right) \in \mathcal{H}$ such that
\begin{equation}
   \langle\hat{X}_0\rangle_{\Psi_T} \simeq T\,, \quad \delta_{\Psi_T} \hat{X}_0 \ll 1\,\qquad \text{and}\qquad
   \langle \Psi_T \mid \hat{O} \mid \Psi_T \rangle \simeq O(T)\,,
\end{equation}
where with $\delta_{\Psi_T}\hat{X}_0$ we mean generic quantum fluctuations of the clock operator $\hat{X}_0$ on the state $\Psi_T$. Notice that there could be in general several possible choices of dynamical variables that could be promoted to a (relational) clock. Different choices may produce a different dynamics, all equally valid in principle. This \virgolette{clock covariance} is an important feature that completely relational frameworks are expected to possess (see \cite{Bojowald:2010xp, Bojowald:2010qw, Hoehn2011} for pioneering works about the issue of changing clocks in generally covariant quantum systems in a quantum phase space langauge and at an effective level, and \cite{Hoehn2018} for a more general and systematic approach to the problem). We will come back to this point later on. This feature remains true in the quantum theory and in all approaches to relational dynamics, and the relative merits of one clock over another have to be judged case by case.

Notice also that the simple form of the phase space, which nicely separates into the variables corresponding to the would-be clock and the rest, is not always available. In fact, this is not the case in the presence of gauge symmetries like diffeomorphism invariance after imposition of constraints. In the quantum theory this is reflected in the fact that the physical Hilbert space of quantum states, i.e., those solving the dynamics, does not generally factorize into a direct product of quantum states for the clock and those for the rest of the physical system. Such factorization may be at best an approximate one. This is a crucial technical (as well as conceptual) complication that has to be dealt with when constructing clock/time observables and the corresponding relational evolution in quantum gravity. In our present context we will not need to deal directly with this issue, due to the peculiarities of the TGFT formalism we work with, but we refer to \cite{Hoehn:2019owq} for a in-depth discussion of these and other issues.
\

While a \virgolette{tempus ante quantum} approach turns out to be technically easier for deparametrizable systems (i.e. in presence of some dynamical variables whose dynamics and coupling is simple enough to be attributed the role of external clock), it is an approach where a specific clock is somehow preferred, in order to canonically quantize the reduced theory. Since, however, different choices of the clock may in general produce different quantum theories (this is the so-called \virgolette{multiple choice problem} \cite{Kuchar2011, Isham1992}), a truly clock-covariant approach, as a tempus post quantum approach, where the clock variables are all treated on the same footing, should be preferred \cite{Hoehn2018}. In both approaches, however, for the chosen subsystem to behave nicely enough to be used as a clock, several restrictions should be in place, at least approximately: weak interactions between clock subsystem and the remaining degrees of freedom, weak self-interactions of the clock itself, semi-classical behaviour in the clock values, etc.

\subsubsection{\virgolette{Emergent quantum gravity} theories}

Further complications arise in quantum gravity theories based on different types of degrees of freedom than straightforwardly quantized continuum fields. In these theories, the notions of spacetime, geometry and gravity should \emph{emerge} from the collective behavior of some \emph{pre-geometric}, not directly spatiotemporal \virgolette{atoms of space}, to be only indirectly related to the continuum fields we define space and time with respect to\footnote{It should be noted, however, that the distinction between these two categories is not sharp: some structures that arise from the quantization of fields can also be understood more radically, and of course the notion of emergence can play an important role also in more traditional canonical or covariant quantizations of classical field theories like GR.}. Examples of structures admitting such more radical interpretation include the spin networks
of loop quantum gravity \cite{Ashtekar:2004eh, Perez:2012wv} (though they were introduced first within a straightforward canonical quantization of the gravitational field), the simplicial (piecewise-flat) geometries of lattice quantum gravity approaches \cite{Hamber:2009mt, Ambjorn:2012jv}, the quanta
of group field theories \cite{Oriti:2011jm, Krajewski:2012aw, Oriti:2017ave}, which as we will discuss in the following can be understood both as spin networks and as
simplicial building blocks of piecewise-flat geometries, causal sets
\cite{Dowker:aza}, and possibly the underlying fundamental degrees of freedom of String Theory \cite{Blau:1900zza}. 

In such approaches, one expects the existence of a \lq proto-geometric\rq$\,$ phase in which the pre-geometric degrees of freedom behave in a collective way, ultimately conspiring to the re-appearence of continuum spacetime notions (among which, there is of course any notion of relational dynamics) at least at some effective, approximate level. As we have just discussed, this presupposes some internal degree of freedom well-behaved enough, so to speak, to be trusted as a good clock. Now, in an emergent spacetime context, all (classical or quantum) dynamical variables of usual spacetime-based field theories are understood as the result of suitable averaging/coarse-graining procedures applied to the fundamental pre-geometric entities, and may well correspond to only a sub-set of the relevant collective quantities one may define from them. The same applies to the would-be (classical or quantum) clock subsystem: we need an additional coarse-graining/averaging step to arrive at something approximately continuous and regular enough to label the evolution of other degrees of freedom in the theory. Again, this additional difficulty is present independently of whether we are dealing with quantum or classical non-spatiotemporal pre-geometric entities.  Thus, we are dealing with a genuinely new dimension of the \lq problem of time\rq$\,$ in quantum gravity.

Schematically, in the classical case, we can intuitively understand the needed extra step as:
\begin{equation}
  \left\{ \mathfrak{x}_1,\mathfrak{p}_1; ...; \mathfrak{x}_n,\mathfrak{p}_n\right\}  \longrightarrow \left\{ X_0,P_0; X_1, P_1; ...; X_N, P_N\right\} ,
\end{equation}
where we have indicated the number of fundamental degrees of freedom (each corresponding to a subset of phase space variables) by $n$, with $N$ expected to be much smaller than $n$. Notice that the coarse-graining step is best understood at the level of observables or of their associated phase space. Still, it is usually accompanied by a switch to a formulation of the theory in terms of coarse-grained distributions over the fundamental phase space, which become the new relevant dynamical variables, and which we then use to compute expectation values of the effective quantities $(X_i,P_i)$.
 
At the quantum level, the analogous step is, intuitively:
\begin{equation}
    \mathcal{H}_{\text{fund}} \ni \Phi\left( \mathfrak{x}_1; ...; \mathfrak{x}_n\right) \longrightarrow \Psi_T\left(X_0,X_1,...,X_N\right),
\end{equation}
where now the resulting function to be used to compute expectation values of the effective quantities  $(X_i,P_i)$ is an effective probability distribution. Depending on the specific formalism, this distribution can be understood as a quantum state (element of some Hilbert space) for an effective quantum system described only in terms of the coarse-grained observables, or as a classical, hydrodynamic type distribution accounting at the effective level for the quantum properties of the fundamental degrees of freedom (which in turn remain the only ones to which a Hilbert space of quantum states is associated). This second possibility is, in fact, the one we will see realized in the case of TGFT condensate cosmology.

Beyond technicalities and particular realizations, the general point is the following: what was the classical description of the system in a formulation in terms of continuum fields, that we want to manipulate to recast it in the form of a relational dynamics, is now itself the result of some previous treatment of more fundamental entities, which, in general, would not allow any identification of a relational clock.

\begin{figure}
    \centering
    \includegraphics[width=0.5\linewidth]{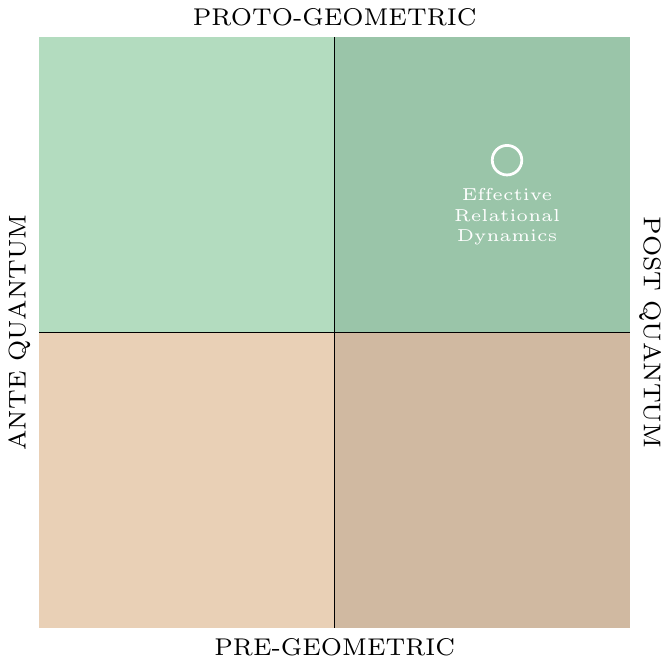}
    \caption{The four possible contexts for relational dynamics in emergent theories of quantum gravity. Besides the classical choice of quantizing before or after having chosen a quantum clock, the emergent perspective adds another layer: defining relational dynamics in a pre-geometric or in a proto-geometric phase of the theory. The preferable choice (both for practical and conceptual reasons) is to define an effective and emergent notion of relational dynamics in the upper right side of the diagram, but as an effective description of states and observables defined in the lower right corner.}
    \label{fig:diagram}
\end{figure}

The situation, therefore, can be represented as in Figure \ref{fig:diagram}.
For a discussion of some conceptual issues raised in these emergent quantum gravity scenarios, see \cite{Oriti:2018dsg}.
\

One way to appreciate these additional difficulties is to realize that a proper extraction of an effective relational dynamics in quantum gravity formalisms based on fundamental non-spatiotemporal entities requires two distinct limits/approximations: continuum and semi-classical.
These two limits/approximations have to be considered conceptually different in emergent theories of quantum gravity, and they are not expected, in general, to commute with each other \cite{Oriti:2017ave}, implying that the final approximate description of the system may well depend on the order in which the two approximations have been implemented. 
In particular, it might be the case that the quantum properties of the fundamental degrees of freedom are actually necessary in order to obtain the correct continuum general relativistic description of the quantum gravity system. 

This suggest that the most appropriate and general path toward the extraction of a well-defined relational dynamics from a fundamental theory of quantum gravity would start from the bottom-right sector of the diagram in Figure \ref{fig:diagram} and move to the top-right quadrant by means of some coarse-graining or other continuum approximation scheme, while staying in the quantum half of the diagram. Once a potentially good clock has been found at this level, and thus a good definition of (quantum) relational dynamics, one can move towards the top-left quadrant via some semi-classicality restriction, where the quantum properties of the clock can be neglected and the usual time evolution is re-obtained.



\subsection{Defining an emergent effective relational dynamics}\label{subsec:definingeffectivedynamics}

Having outlined the general problem and different approaches one can take for solving it, we now clarify further what we mean by \emph{emergent effective relational dynamics} in what we called a proto-geometric phase of the theory, in the context of an underlying pre-geometric formalism. 
The conditions that we are going to give below should be understood, of course, in addition to the fundamental requirements that an internal time variable, corresponding to one of the dynamical degrees of freedom of the theory, can be identified and that a well-defined (e.g. Hamiltonian) evolution can be specified for quantum states and other observables with respect to it. 

The relational framework that we are interested in defining should be characterized by the following features:
\begin{description}
\item[\emph{Emergence}] The effective dynamics should emerge as a collective phenomenon: therefore, it should be formulated in terms of operators corresponding to \emph{collective observables} and states encoding \emph{collective behavior} of the underlying degrees of freedom. 
\item[\emph{Effectiveness}] The relational evolution should be intended to hold \emph{on average}. Operators used to define the internal clock should have small quantum (and thermal, when relevant) fluctuations (semi-classicality condition on the internal clock). Whenever these are large, the effective relational dynamics could not be trusted.
\end{description}
The requirement of effectiveness implies that the emergent relational dynamics we are trying to define is approximate only. Considering just an averaged relational evolution is one of such approximations, due, as we have already argued, to the fact that a notion of relational dynamics is only supposed to make sense in a proto-geometric regime and when the chosen clock is \virgolette{ideal enough}. However, in this proto-geometric regime it is also likely for other approximations to be justified. For instance, this would be the case for the imposition of only the averaged quantum dynamics of the microscopic degrees of freedom (mean field approximation), in light of the fact that we are interested only in an averaged relational evolution of geometric observables. Other types of approximations (like the aforementioned \virgolette{good-behavior} of the internal clock) are instead not expected to hold in an arbitrary proto-geometric regime. As a consequence, requiring their validity will likely specify an even more peculiar regime of theory (in practice, it will impose further constraints on the quantities concretely realizing the effective relational dynamics). The importance of approximations in this framework will become manifest when we will show its concrete implementation in the GFT condensate cosmology scenario.

Now, let us spell out other general ingredients of an effective relational dynamics, before going to realize it concretely in the TGFT context. 
In order to fix the ideas, suppose that we are interested in defining the dynamics of geometric degrees of freedom with respect to some matter degree of freedom, for example the simplest possible type of matter, i.e., a minimally coupled massless free scalar field (classically, gravity plus a minimally coupled massless scalar field is a deparametrizable system \cite{Tambornino:2011vg}, with the massless scalar field representing a \virgolette{good clock}).   

Let us assume that we are able to identify a class of states, in the fundamental theory, which encode collective behavior and can be given a continuum proto-geometric interpretation. Call these states $\ket{\Psi}$. Let us further assume that we have at our disposal a set of collective observables, say $\{\hat{O}_{a}\}_{a\in S}$ and $\hat{\chi}$, whose expectation values on the proto-geometric states $\ket{\Psi}$ have a continuum interpretation as geometric observables (e.g. volumes, curvature invariants, etc) and massless scalar field, respectively. Given the emergent nature of the theory, another relevant quantity for the description of the system is the \virgolette{number operator} $\hat{N}$ counting the number of fundamental \virgolette{atoms of space}, and useful to characterize a continuum approximation, that we could expect to require some form of thermodynamic limit.  Notice that both the states $\ket{\Psi}$ and the observables $\hat{O}_a$ are constructed at the \virgolette{kinematic} level, in the sense of not having imposed on them the quantum dynamics of the theory, yet\footnote{\label{footnote:statesobs}Here, we are using the word \virgolette{kinematical} in a slightly different sense from what is usually done in a classical or canonical setting. Indeed, in that case, the meaning of \virgolette{kinematical} and \virgolette{physical} is strictly related to diffeomorphism invariance. In our emergent quantum gravity setting, however, the theory is formulated without referring to any differentiable manifold, nor fields living on it. Also, in this sense, it is defined in absence of any notion of spacetime. As such, diffeomorphisms are simply not defined at this level. Kinematical states and observables, therefore, are quantities defined with respect to the abstract Hilbert space of the quantum theory, while physical quantities are intended as quantities restricted to solutions of the quantum dynamics. Whether and how these quantities are related to diffeomorphism invariance can only be understood once the appropriate emergent limit is taken. In order to verify this, one should check that relational observables built in the pre-geometric theory and their dynamics correspond (at least approximately) with relational (thus diffeo-invariant) observables constructed in continuum GR; this correspondence would imply that they can also be understood as diffeo-invariant observables in a theory formulated in terms of continuum fields where diffeomorphisms are defined, and satisfying a diffeo-invariant dynamics. We should not necessarily expect, however, to find a diffeomorphism invariance symmetry in the fundamental theory, e.g.\ in its quantum dynamics, although it would be indeed interesting if such symmetry can be identified and put in correspondence with any symmetry property of specific (classes of) GFT models.}. For instance, this means that the states $\ket{\Psi}$ are not required to solve the full quantum dynamics of the theory, but they are certainly required to solve it in an averaged sense, as it will become clearer when we will discuss the concrete example of GFT condensate cosmology.

The states $\ket{\Psi}$ can be said to implement a notion of effective and emergent relational dynamics if they also satisfy the following conditions \emph{at least on-shell}, i.e. after imposing, approximately, the quantum equations of motion of the fundamental theory:
\begin{description}
\item[\emph{Averaged relational evolution}] It exists an Hermitean operator $\hat{H}$ such that, for each geometric collective observable $\hat{O}_{a}$, 
\begin{subequations}\label{eqn:averagedrelationaldynamics}
\begin{equation}\label{eqn:averageddynamics}
i\frac{\diff}{\diff\braket{\hat{\chi}}_{\Psi}}\braket{\hat{O}_{a}}_{\Psi}=\braket{[\hat{H},\hat{O}_{a}]}_{\Psi}\,.
\end{equation}
Moreover, at effective semi-classical level, the operator $\hat{H}$ should be equal to the momentum operator $\hat{\Pi}$ canonically conjugated to $\hat{\chi}$, which means that \emph{all the moments of $\hat{H}$ and $\hat{\Pi}$ on $\ket{\Psi}$ should be equal}. In particular, this implies that the averages of these two operators on $\ket{\Psi}$ should be equal,
\begin{equation}\label{eqn:piequalh}
\braket{\hat{H}}_{\Psi}=\braket{\hat{\Pi}}_{\Psi}\,.
\end{equation}
\end{subequations}
This \virgolette{effective equality}  approximately implements the idea of $\hat{\Pi}$ generating the relational evolution. 
\item[\emph{Semi-classicality condition}] Assuming that the expectation value of $\hat{\chi}$ is non-zero, we require its variance on $\ket{\Psi}$ to be much smaller than one, and to have the characteristic $\braket{\hat{N}}^{-1}$ behavior, i.e., 
\begin{equation}\label{eqn:smallfluctuationsclock}
    \sigma^2_\chi\ll 1\,,\qquad \sigma^2_\chi\sim \braket{\hat{N}}^{-1}\,,
\end{equation}
where the relative variance on $\ket{\Psi}$ is defined as
\begin{equation*}
\sigma^2_{O}=\frac{\braket{\hat{O}^2}_{\Psi}-\braket{\hat{O}}^2_{\Psi}}{\braket{\hat{O}}^2_{\Psi}}\,.
\end{equation*}
\end{description}
Equation \eqref{eqn:averageddynamics} is of course the embodiment of the averaged effective relational dynamics, describing the evolution of the expectation value of a given geometric collective operator in terms of the expectation value of the massless scalar field. 

Conditions \eqref{eqn:smallfluctuationsclock} instead, are a formalization of the requirement that the averaged relational dynamics is not obscured by quantum fluctuations (in which case our relational clock would be a bad choice because \virgolette{too quantum} to label evolution). In particular, notice that while the first condition in \eqref{eqn:smallfluctuationsclock} is usually enough to guarantee a semi-classical behavior of quantities in the standard frameworks (e.g., the simple harmonic oscillator) where coherent states are employed (because of their Gaussian form in the phase space), in this case this might not be enough. Still, if the above relative variance has the characteristic behavior of $N^{-1}$ expected for collective observables, as required by the second condition in \eqref{eqn:smallfluctuationsclock}, this can be taken as a strong indication that indeed even higher moments will be somehow negligible when the number of fundamental degrees of freedom in the state is large enough, which is expected to be the case in the relevant proto-geometric regime of the theory. Thus, it is the very large number of fundamental degrees of freedom accommodated in the states $\ket{\Psi}$ that can make fluctuations arbitrarily small. 

However, the \virgolette{clock} resulting from the above conditions might be very far from an \virgolette{ideal} one, for different reasons. First, the clock could feature turning points, in which case it would not be possible to separate positive and negative frequency (or forward and backward evolving) modes relative to it (see, for instance, \cite{Hoehn2011}). For such clocks, equation \eqref{eqn:averagedrelationaldynamics} should not be inteded to hold globally, but only on a local, transient level, far enough from such a singular point. In the case of a minimally coupled massless scalar field used as a clock one does not expect this issue to appear, as we will see explicitly in the concrete example of GFT condensate cosmology below. Second, its momentum may suffer from large quantum fluctuations. In this sense, if one wants a relational description in terms of a \virgolette{good, classical clock}, one has to require also that quantum fluctuations on the momentum (and thus on the Hamiltonian operator, according to the averaged relational evolution conditions) are small. Assuming that the expectation values of $\hat{\Pi}$ and $\hat{H}$ on $\ket{\Psi}$ are small, this can be obtained by requiring that 
\begin{equation*}
\{\sigma^2_{\chi},\sigma^2_{\Pi},\sigma^2_{H}\}\ll 1\,,\qquad\{\sigma^2_{\chi},\sigma^2_{\Pi},\sigma^2_{H}\}\sim \braket{\hat{N}}^{-1}\,.
\end{equation*}
In such a case, condition \eqref{eqn:piequalh} is enough to define an \virgolette{approximate, effective equality} between $\hat{\Pi}$ and $\hat{H}$.

Of course, one may also require that quantum fluctuations of the geometric operators $\hat{O}_{a}$, as well as fluctuations of $\hat{N}$, are also negligible. Assuming that the expectation value of each $\hat{O}_a$ and of $\hat{N}$ on $\ket{\Psi}$ are non-zero, we can formulate this condition as a condition on the relative variances of the relevant operators:
\begin{subequations}\label{eqn:semi-classicality}
\begin{align}\label{eqn:smallvariances}
\{\sigma^2_{O_{a}},\sigma^2_{N},\sigma^2_{\chi},\sigma^2_{\Pi},\sigma^2_{H}\}&\ll 1&\quad&\forall a\in S\,,\\
\label{eqn:nbehavior}
\{\sigma^2_{O_{a}},\sigma^2_{N},\sigma^2_{\chi},\sigma^2_{\Pi},\sigma^2_{H}\}&\sim \braket{\hat{N}}^{-1}&\quad&\forall a\in S,
\end{align}
\end{subequations}
i.e. a fully semi-classical behavior of the system.

Let us conclude this section with three somewhat minor comments:
\begin{itemize}
    \item First, let us remark that the above conditions on relative variances are of no use in the case in which the expectation values are identically zero. In that case, as argued in \cite{Ashtekar:2005dm}, one has to define some thresholds $\delta_{i}^2$ and require that, for each operator $\hat{O}_{i}$ in the set $\{\hat{O}_{a},\hat{N},\hat{\chi},\hat{\Pi},\hat{H}\}$, $\Delta^2\hat{O}_{i}\equiv \braket{\hat{O}_{i}^2}_{\Psi}-\braket{\hat{O}_{i}}^2_{\Psi}<\delta^2_{i}$. However, notice that, contrarily to what is done in \cite{Ashtekar:2005dm}, we will not require that the expectation values of the desired operators peak on some precise value.
\item Second, we want to stress how non-trivial the above requirements are. In particular, imposing semi-classicality on different operators is a very strong one. A state can be semi-classical with respect to some operators and not semi-classical at all for others. For instance, coherent states of the harmonic oscillator are not semi-classical for its Hamiltonian operator \cite{Ashtekar:2005dm}. Another example is the quantum theory of the Einstein-Rosen waves in $4$-dimensional General Relativity \cite{Ashtekar:1996yk}. See \cite{Ashtekar:2005dm} for a detailed discussion of the issue of semi-classicality. For our purposes, i.e. defining an effective relational dynamics, it is important to focus on ensuring semi-classicality at least for the operators encoding properties of the chosen relational clock subsystem. 
\item Lastly, let us mention that effective approaches to the problem of relational dynamics in background-independent canonical systems have been already proposed in \cite{Bojowald:2010xp,Bojowald:2010qw,Bojowald:2009zzc}, and they were already been applied to cosmological settings with interesting results \cite{Bojowald:2012xy}. Some aspects of the construction highlighted above are indeed shared with the framework developed in \cite{Bojowald:2010xp,Bojowald:2010qw,Bojowald:2009zzc}. Besides the most evident one, i.e., the approximate nature of the approach (which in \cite{Bojowald:2010xp,Bojowald:2010qw,Bojowald:2009zzc} comes with a truncation in powers of $\hslash$), there are two more important similarities:
 \begin{enumerate}
 \item The use of expectation values (and moments) as basic quantities of interest. In \cite{Bojowald:2010xp,Bojowald:2010qw,Bojowald:2009zzc} this is because the quantum theory defined on a Hilbert space is reformulated in terms of a quantum phase space where quantum states, instead of being described by density matrices, are characterized in terms of all the expectation values that states assign to a basis set of observables. In particular, the Poisson structure of this quantum phase space is defined in terms of expectation values of the commutator of the corresponding operators (which here appear, for instance, in equation \ref{eqn:averagedrelationaldynamics}).
 \item The evolution of some expectation values relative to the expectation value of a clock. In \cite{Bojowald:2010xp,Bojowald:2010qw,Bojowald:2009zzc}, this comes about because the quantum phase space inherits a quantum constraint surface on which one can then formulate relational dynamics in the usual classical manner (except that one has to deal with more degrees of freedom to encode fluctuations).
 \end{enumerate}
Despite these important technical and conceptual similarities, it should however be stressed that the effective approach described here aims to go beyond the works \cite{Bojowald:2010xp,Bojowald:2010qw,Bojowald:2009zzc}, by addressing a field-theoretical and, most importantly, an emergent scenario\footnote{We are however currently working on an attempt to generalize the work in \cite{Bojowald:2010xp,Bojowald:2010qw,Bojowald:2009zzc} to a similar field-theoretic and emergent scenario, in order to highlight possible similarities and differences between the two approaches.}.
\end{itemize}
\section{GFT and effective cosmology}\label{sec:review}
In this section we review the basics of the GFT approach to quantum gravity (focusing on the quantum simplicial geometric aspects, but also highlighting the connection with the LQG kinematical space) and the framework of GFT condensate cosmology. In the latter context, we describe in which sense condensate states represent cosmological geometries and, importantly, how \virgolette{relational operators} are defined and their dynamics is obtained (see for example \cite{Gielen:2013naa,Oriti:2016qtz} and \cite{Pithis:2019tvp, Gielen:2014uga} for reviews) via the introduction of a \virgolette{massless scalar field clock}. 
\subsection{The GFT Fock space}\label{subsec:secondquantization}
GFTs are field theories of a (in general complex) field $\varphi: G^d\to\mathbb{C}$ defined on $d$ copies of a group manifold, $\varphi(g_{I})\equiv \varphi(g_1,\dots, g_d)$. With a careful choice of the dimension $d$, of the group manifold $G$, and of the (combinatorial) action, which may include additional restrictions on the fields, these theories can be understood as \virgolette{quantum field theories of spacetime} \cite{Reisenberger:2000zc}. On the one hand, the fundamental quanta of the theory can be seen as 3-simplices, i.e. building blocks of three-dimensional simplicial geometries representing the (boundary) states of the theory, with their quantum simplicial geometric properties encoded in the group-theoretic data. On the other hand, the perturbative expansion of the n-point functions produces a sum over Feynman diagrams associated to 4-dimensional cellular complexes, weighted by a discrete gravity path-integral with the same group-theoretic data as dynamical variables. It is then from this type of discrete structures that one should reconstruct continuum four-dimensional spacetimes and geometries, in a suitable approximation. In this sense, therefore, GFTs indeed are theories in which spacetime has dissolved into pre-geometric \virgolette{atoms of space}. Typical choices of $d$ and $G$ that allows for this interpretation are $d=4$ (i.e., the spacetime dimension), and $G=\text{SL}(2,\mathbb{C})$ (local gauge group of gravity) or its Euclidean version, $\text{Spin}(4)$. For most specific GFT models, the same group-theoretic data can also be mapped into data taken from $G=\text{SU}(2)$, corresponding to the rotation subgroup of the above groups. As we discuss below, this allows for an explicit connection of the GFT quantum states with those appearing in LQG, which gives additional guidelines for extracting continuum physics \cite{Gielen:2016dss}. From now on, therefore, we will specialize to $d=4$ and $G=\text{SU}(2)$.  
\paragraph*{Field operators.}
These field theories can be formulated in the language of second quantization. One defines the field operators satisfying the commutation relations:
\begin{subequations}
\begin{align}\label{eqn:basiccommutator}
[\hat{\varphi}(g_{I}),\hat{\varphi}^\dagger(g_{I}')]&=\mathbb{I}_{G}(g_{I},g_{I}')\,,\\ [\hat{\varphi}(g_{I}),\hat{\varphi}(g_{I}')]&=[\hat{\varphi}^\dagger(g_{I}),\hat{\varphi}^\dagger(g_{I}')]=0\,,
\end{align}
\end{subequations}
where $\mathbb{I}_G(g_I,g_I')$ is a Dirac delta distribution on the space $G^4/G$. 

Let us spend some words about the geometric interpretation of the quantities appearing in the two equations above. The field operator $\hat{\varphi}^\dagger(g_I)$, acting on the vacuum $\ket{0}$, creates a \virgolette{quantum of space} with data $\{g_I\}$. When such field satisfies the closure condition $\varphi(g_I)=\varphi(g_Ih)$ for each $h\in G$, and the GFT action encodes appropriate geometricity (\lq simplicity\rq) conditions (which also allow to map these $\text{SU}(2)$ data to $\text{SL}(2,\mathbb{C})$ ones), this \virgolette{quantum of space} can be interpreted as a $3$-simplex (tetrahedron) whose $4$ faces are decorated with an \emph{equivalence class} of geometrical data $[\{g_I\}]=\{\{g_Ih\},h\in G\}$. The group elements can be associated to the parallel transport of a gravitational connection associated to the group $G$ along the links dual to such faces, representing thus a discretization of the same.  In the dual picture, such a \virgolette{decorated tetrahedron} corresponds to an \emph{open spin-network}, i.e., a node from which four links emanate, each of which is assigned group-theoretical data. The closure condition encodes the invariance under local gauge transformations acting on the vertex of the spin-network. Such a local gauge invariance requires that the right-hand-side of equation \eqref{eqn:basiccommutator} is the identity in the space of gauge-invariant fields. For example, for compact groups, we can write this as $\mathbb{I}_{G}(g_{I},g_{I}')=\int\diff h\prod_{I=1}^4\delta(g_{I}hg_{I}^{-1})$, since this is essentially the projector onto that space. For non-compact groups, additional care with divergences associated to group integrations is needed, e.g. via gauge fixing \cite{Gielen:2013naa}.

The interpretation of the above \virgolette{quanta of space} as open spin-network states is made even clearer once one expands the field on a basis of functions on $L^2(G^4/G)$ labeled by group representations (which, here, for simplicity, we label with a single set of labels $\vec{x}$)
\begin{subequations}
\begin{align}
\hat{\varphi}(g_{I})&=\sum_{\vec{x}}\hat{c}_{\vec{x}}\psi_{\vec{x}}(g_{I})\,,\\ \hat{\varphi}^\dagger(g_{I})&=\sum_{\vec{x}}\hat{c}^\dagger_{\vec{x}}\psi^*_{\vec{x}}(g_{I})\,,
\end{align}
\end{subequations}
satisfying
\begin{equation}
[\hat{c}_{\vec{x}},\hat{c}^\dagger_{\vec{x}'}]=\delta_{\vec{x},\vec{x}'}\,,\qquad [\hat{c}_{\vec{x}},\hat{c}_{\vec{x}'}]=[\hat{c}^\dagger_{\vec{x}},\hat{c}^\dagger_{\vec{x}'}]=0\,.
\end{equation}
The quanta created by $\hat{c}^\dagger_{\vec{x}}$, can now be interpreted again as nodes from which $4$ links are emanating, but now they are explicitly decorated with spin-network vertex data,
\begin{equation}
\vec{x}=\{\vec{j},\vec{m},\iota\}\,,
\end{equation}
exactly because of gauge invariance and the choice of $G=\text{SU}(2)$. Here $\vec{j}$ and $\vec{m}$ are respectively spin and angular momentum projection associated to the open edges of a given vertex, while $\iota$ represents the intertwiner quantum number associated to the vertex itself. In this way we can write the \virgolette{spin-network wave function} $\psi_{\vec{x}}(g_I)$ as
\begin{equation}\label{eqn:spinnetworkwavefunction}
\psi_{\vec{x}}(g_I)\equiv \left\langle\, g_I\mid\vec{x}\,\right\rangle=\left[\prod_{i=1}^4\sqrt{d(j_i)}D^{j_i}_{m_in_i}(g_i)\right]\mathcal{I}^{\vec{j},\iota}_{\vec{n}}\,,
\end{equation}
where $\mathcal{I}$ is a normalized intertwiner and $\vec{j}\equiv \{j_1,\dots,j_4\}$, similarly for $\vec{n}$. The operators $\hat{c}_{\vec{x}}$ and $\hat{c}^\dagger_{\vec{x}}$ are \emph{creation} and \emph{annihilation} operators for open spin-network vertices. 

Starting from the above ladder operators, together with the vacuum state $\ket{0}$ annihilated by all $\hat{c}_{\vec{x}}$s (which represents a \virgolette{no-space state}), one can construct a Fock space, whose $n$-particle states satisfy
\begin{align*}
 \hat{c}_{\vec{x}}\ket{n_{\vec{x}}}&=\sqrt{n_{\vec{x}}}\ket{n_{\vec{x}}-1}\,,\\
 \hat{c}_{\vec{x}}\ket{n_{\vec{x}}}&=\sqrt{n_{\vec{x}}+1}\ket{n_{\vec{x}}+1}\,.
 \end{align*}
The Fock space introduced in this way is analogous to the kinematical Hilbert space of LQG \cite{Oriti:2013aqa}, in the sense that it encodes similar fundamental degrees of freedom. This connection is useful because, as we will see below, it offers further guidance (in addition to the one coming from simplicial geometry) to the geometric interpretation and definition of geometric operators.
\paragraph*{Second-quantized observables.}
Starting from the field operators, we can construct quantum observables of geometric interest. The simplest one is the \emph{number operator},
\begin{equation}\label{eqn:numberoperator}
\hat{N}\equiv \int\diff g_{I}\,\hat{\varphi}^\dagger(g_{I})\varphi(g_{I})\,,
\end{equation}
which counts the number of quanta present in a given state and whose eigenvalues distinguish between the $n$-body sectors of the GFT Fock space. More generally, one can consistently construct GFT \virgolette{$(m+n)$-body operators}  $\hat{O}_{n+m}$, as
\begin{equation}\label{eqn:secondquantizationoperator}
\hat{O}_{n+m}\equiv \int (\diff g_{I})^m(\diff h_{I})^n\,O_{m+n}(g_{I}^1,\dots, g_{I}^m,h_{I}^1,\dots, h_{I}^n)\prod_{i=1}^m\hat{\varphi}^\dagger(g_{I}^i)\prod_{j=1}^n\hat{\varphi}(h_{I}^j)\,,
\end{equation}
from the matrix elements $O_{m+n}$ defined either in a simplicial geometric context between states associated to quantized tetrahedra, or in the LQG context between spin-network vertex states. The same kind of construction can be performed of course in any representation of the relevant Hilbert space. For example, a generic two-body operator can be written as 
\begin{equation}\label{eqn:twobodyspin}
\hat{O}_2=\sum_{\vec{x}\vec{x}'}O_2(\vec{x},\vec{x}')c^\dagger_{\vec{x}}c_{\vec{x}'}\,,
\end{equation}
where again $O(\vec{x},\vec{x}')$ are matrix elements between, e.g., spin-network states. All operators we are interested in here (e.g., the volume operator) are two-body operators of this kind.
\paragraph*{Coupling to a scalar field.}
With the later goal of defining a notion of relational dynamics, it is useful to add to the pure quantum geometric data additional ones later to become a relational matter clock. The simplest choice \cite{Oriti:2016qtz} is a minimally coupled free massless scalar field\footnote{The choice of a \emph{minimally coupled free massless} scalar field remarkably simplifies the form of the dynamics, as discussed in more detail in Subsection \ref{subsec:dynamics}.} (see \cite{Li:2017uao} for more details). The inclusion of a minimally coupled free massless scalar field is performed by adding to the GFT field and action the degrees of freedom corresponding to a scalar field in such a way that the perturbative expansion of the GFT partition function can be identified with the (discrete) path-integral of a model of simplicial gravity minimally coupled with a free massless scalar field (or, equivalently, with the corresponding spin-foam model). Following this procedure, the definition of the field operator is modified as follows:
\begin{equation}
\hat{\varphi}(g_I)\quad\longrightarrow\quad\hat{\varphi}(g_I,\chi)\,.
\end{equation}
In this way, the one-particle Hilbert space is now $L^2(\text{SU}(2)^4/\text{SU}(2)\times \mathbb{R})$. So, each GFT atom carries a value of the scalar field, which is then \virgolette{discretized} on the simplicial structures associated to GFT states and (perturbative) amplitudes. The commutation relation in \eqref{eqn:basiccommutator} has to be modified consistently, obtaining 
\begin{equation}\label{eqn:frozencommutations}
\left[\hat{\varphi}(g_I,\chi),\hat{\varphi}^\dagger(h_I,\chi')\right]=\mathbb{I}_G(g_I,h_I)\delta(\chi-\chi')\,.
\end{equation} 
Starting from this structure of the Fock space, operators in the second quantization picture now involve integrals over the possible values of the massless scalar field. For instance, the number operator \eqref{eqn:numberoperator} takes the form 
\begin{subequations}
\begin{equation}
\hat{N}=\int\diff\chi\int \diff g_I\hat{\varphi}^\dagger(g_I,\chi)\hat{\varphi}(g_I,\chi)\,.
\end{equation}
Another one is the volume operator:
\begin{equation}
\hat{V}=\int\diff\chi\int \diff g_I\diff g'_I\hat{\varphi}^\dagger(g_I,\chi)V(g_I,g'_I)\hat{\varphi}(g'_I,\chi)\,,
\end{equation}
defined in terms of matrix elements of the first quantized volume operator in the group representation (the first quantized volume operator is instead diagonal in the spin representation), and which adds up the volume contributions (individual 3-volumes) of all the tetrahedra in a given GFT state (themselves not dependent on the value of the discretized scalar field). 

Having introduced new \virgolette{pre-matter} degrees of freedom, one can find a new whole set of observables related to them, which are the second-quantized
GFT counterpart of the standard observables of a scalar field, namely polynomials in the scalar field and its derivatives. The two fundamental ones that can be constructed in this way are the scalar
field operator and the momentum operator \cite{Oriti:2016qtz}:
\begin{align}
\label{eqn:scalarfieldoperator}
\hat{X}&\equiv \int\diff g_I\int\diff \chi\,\chi\hat{\varphi}^\dagger(g_I,\chi)\hat{\varphi}(g_I,\chi)\,,\\
\label{eqn:momentumoperator}
\hat{\Pi}&=\frac{1}{i}\int\diff g_I\int\diff\chi \left[\hat{\varphi}^\dagger(g_I,\chi)\left(\frac{\partial}{\partial\chi}\hat{\varphi}(g_I,\chi)\right)\right].
\end{align}
\end{subequations}
From the scalar field momentum operator and the volume operator one can in principle define an operator corresponding to the energy density of the scalar field, of obvious relevance for cosmological dynamics. For technical reasons, however, it is more convenient to define a quantity with this interpretation in terms of expectation values, as done for instance in \cite{Oriti:2016qtz}.
Notice that all the above operators are self-adjoint, as it should be.

Starting from them, in \cite{Oriti:2016qtz} new \virgolette{relational operators} $\hat{O}(\chi)$ have been defined 
essentially as the integrand in the general expression for observables $\hat{O}\equiv \int\diff\chi\hat{O}(\chi)$. 
For instance, the relational number operator at \virgolette{a time $\chi$} was defined as
\begin{equation}\label{eqn:fakerelaitionalobservable}
\hat{N}(\chi)=\int \diff g_I\hat{\varphi}^\dagger(g_I,\chi)\hat{\varphi}(g_I,\chi)\,;
\end{equation}
similarly for volume or scalar field momentum operators.

This is therefore a definition of relational quantities, thus indirectly of an internal time variable, that applies at the level of the fundamental presentation of the theory. It is not preceded by any sort of coarse-graining procedure or continuum approximation. 

This definition allows to derive a number of interesting results, producing a promising effective cosmological dynamics from the fundamental quantum gravity formalism. We will review some of these results in the next subsection. At the same time, however, it is problematic, as we are also going to discuss in the following. The main difficulty is that these operators have a distributional nature, leading to divergences in the computation of several physically relevant quantities. These divergences, we argue, indicate a fundamental problem with such definition, rather than simply the need for some regularization, and therefore call for the more refined procedure we develop in this work.
A number of other, somewhat minor issues with the above definition arise, motivating further the search for an alternative route toward the extraction of a relational dynamics from the theory. For example, the operator corresponding to the scalar field momentum \virgolette{at given time $\chi$} it is not self-adjoint, and it has to be made so by adding to it its hermitian conjugate operator. 
\subsection{Homogeneous and isotropic geometries}
In order to obtain a quantum cosmological dynamics from a GFT, the first necessary step is to identify a class of states in the quantum theory which can be consistently interpreted as continuum cosmological spaces. Two criteria are fundamental for the construction of such states:
\begin{enumerate}
    \item First, since they are supposed to represent continuum geometries, they should be composed by a very large (possibly infinite) number of GFT quanta.
    \item Second, they should encode some notion of homogeneity (required in the coarse-grained cosmological setting), in some probabilistic sense.
\end{enumerate}
The second condition can be satisfied if the chosen quantum state is collectively described by a single function over the space of geometries associated to a single tetrahedron, since the latter is isomorphic (modulo an additional symmetry requirement that has to be imposed on the collective function) to the minisuperpsace of homogeneous geometries \cite{Gielen:2014ila}. In turn, one way to achieve this simplified collective description is by endowing each fundamental spin-network vertex/tetrahedron with the same information. This matches the intuitive idea of a condensate state, and it is often labeled \lq wavefunction homogeneity\rq$\,$ in the literature. However, many different states can be constructed with this same prescription, basically because GFT quanta, even if they are in the same configuration, can still be \virgolette{glued} one to another in different ways. 
\paragraph*{Coherent states.}
In \cite{Oriti:2016qtz}, the simplest choice satisfying the two criteria above has been studied: states which completely neglect all the connectivity information\footnote{Obviously, this could be at best an approximation to more realistic quantum states corresponding to continuum homogeneous quantum geometries.}. These are coherent states of the GFT field operator,
\begin{equation}\label{eqn:coherentstates}
    \ket{\sigma}=N_\sigma\exp\left[\int \diff \chi\int\diff g_I\,\sigma(g_I,\chi)\hat{\varphi}^\dagger(g_I,\chi)\right]\ket{0},
\end{equation}
where
\begin{subequations}
\begin{align}
N_\sigma&\equiv e^{-\Vert \sigma\Vert^2/2},\\\
\Vert\sigma\Vert^2&=\int \diff g_I\diff\chi\vert\sigma(g_I,\chi)\vert^2\equiv \braket{\,\sigma\mid \hat{N}\mid \,\sigma\,}\,.
\end{align}
\end{subequations}
By definition, such coherent states satisfy the important property
\begin{equation}\label{eqn:eigenstateannihilation}
    \hat{\varphi}(g_I,\chi)\ket{\sigma}=\sigma(g_I,\chi)\ket{\sigma}\,,
\end{equation}
i.e., they are eigenstates of the annihilation operator. Equations \eqref{eqn:coherentstates} and \eqref{eqn:eigenstateannihilation} can also be rewritten in the spin representation:
\begin{equation}\label{eqn:coherentspin}
    \ket{\sigma}=e^{-\Vert \sigma\Vert^2/2}\exp\left[\int \diff\phi\sum_{\vec{x}}\sigma_{\vec{x}}(\phi)\hat{c}_{\vec{x}}^\dagger(\phi)\right]\ket{0}\,,
\end{equation}
and
\begin{equation}\label{eqn:ccoherent}
    \hat{c}_{\vec{x}}(\phi)\ket{\sigma}=\sigma_{\vec{x}}(\phi)\ket{\sigma}\,.
\end{equation}
\paragraph*{Isotropy.}
Besides homogeneity, cosmological geometries are assumed to be (approximately) isotropic. In \cite{Oriti:2016qtz}, isotropy has been imposed as an additional restriction on the condensate wave function, drastically simplifying the effective continuum dynamics. Notice that imposing a particular symmetry on the condensate wave function is in general very different from the symmetry reduction of the microscopic deegrees of freedom, basically because the condensate wave function is a macroscopic variable (in the simple case of coherent condensate states this point is somewhat obscured by the fact that the colllective wavefunction is also, at the same time, the individual wavefunction of each tetrahedron in the system). In \cite{Oriti:2016qtz}, isotropy of the wave function has been imposed by requiring the associated tetrahedra to be equilateral, resulting in the following condensate wavefunction:
\begin{equation}
    \sigma(g_I,\chi)=\sum_{j=0}^\infty\sigma_j(\chi)\overline{\mathcal{I}}^{jjjj,\iota_+}_{m_1m_2m_3m_4}\mathcal{I}^{jjjj,\iota_+}_{n_1n_2n_3n_4}\sqrt{d^4(j)}\prod_{i=1}^4D^j_{m_in_i}(g_i)\,,
\end{equation}
where $d(j)=2j+1$, $j$ are spin labels, $D^j_{mn}$ are Wigner representation matrices, $\mathcal{I}$ are intertwiners, and $\iota_+$ is the largest eigenvalue of the volume operator compatible with $j$. 
For the condensate wavefunction in spin representation we then have 
\begin{equation}\label{eqn:sigmachi}
    \sigma_{\vec{x}}(\chi)\equiv \sigma_{\{j,\vec{m},\iota_+\}}(\chi)=\sigma_j(\chi)\overline{\mathcal{I}}^{jjjj,\iota_+}_{m_1m_2m_3m_4}\,.
\end{equation}
\subsection{Dynamics}\label{sec:previousdynamics}
In \cite{Oriti:2016qtz}, the effective dynamics of the condensate has been obtained using the connection between the path-integral and the operator formulation provided by the Schwinger-Dyson (SD) equations, i.e.,
\begin{equation}\label{eqn:schwingerdyson}
  0=\int\mathcal{D}\varphi\mathcal{D}\bar{\varphi}\frac{\delta}{\delta\bar{\varphi}(g_{I})}\left(O[\varphi,\bar{\varphi}]e^{-S[\varphi,\bar{\varphi}]}\right)=\left\langle\frac{\delta O[\varphi,\bar{\varphi}]}{\delta\bar{\varphi}(g_{I})}-O[\varphi,\bar{\varphi}]\frac{\delta S[\varphi,\bar{\varphi}]}{\delta\bar{\varphi}(g_{I})}\right\rangle\,,
  \end{equation}  
for any functional $O[\varphi,\bar{\varphi}]$ of the field and its complex conjugate. In the above equation, $S[\varphi,\bar{\varphi}]$ is the GFT action, typically including a quadratic kinetic term and some higher order (in powers of the field operator) interaction term, chosen so that the perturbative expansion of the GFT partition function around the Fock vacuum matches the spin-foam model one is attempting to reproduce (see also the discussion about the coupling to a scalar field in Subsection \ref{subsec:secondquantization}). The expectation value, here, is to be interpreted as taken in the \virgolette{ground state} of the full dynamics. If the ground state is assumed to be given, approximately, by the above isotropic condensate states, the resulting dynamics, truncated at the level of the simplest SD equation, i.e., considering only $O=1$ among the infinitely many possibilities, corresponds to the classical equation of motion of the underlying GFT action, with the field replaced by the condensate wavefunction. In fact, the same result could simply be understood as the mean field approximation of the full GFT quantum effective action, evaluated in the isotropic restriction. Such mean-field dynamics can also can be described by the following effective action \cite{Oriti:2016qtz}:
\begin{equation}\label{eqn:action}
    S_\text{eff}=\sum_{j=0}^\infty \int\diff\chi\biggl(A_j\vert\partial_\chi\sigma_j(\chi)\vert^2+B_j\vert\sigma_j(\chi)\vert^2-\frac{1}{5}w_j\sigma_j^5(\chi)-\frac{1}{5}\overline{w}_j\overline{\sigma}_j^5(\chi)\biggr),
\end{equation}
where the dependence on the details of the GFT model are encoded in the coefficient functions $A_j$, $B_j$ and $w_j$. 

It should also be pointed out that this dynamics, with a differential operator of second order with respect to the scalar field variable, results from a further approximation, obtained in the limit in which the GFT field varies slowly with respect to the scalar field variable. Whether or not this assumption is satisfied for the solutions of the equations of motion derived from the above action is not obvious. 
In particular, let us notice that this truncation is not guaranteed to be reasonable for exponential solutions, which indeed is the form of the classical solutions for the volume relational evolution first obtained in \cite{Oriti:2016qtz}. We will return to this issue below.

The interaction term corresponds, at the level of the discrete structures asssociated to GFT states, to the gluing of five tetrahedra to obtain a $4$-simplex. This interactions is typical in quantum geometric GFT models. 
\paragraph*{Symmetries.}
From the symmetries of the above action, one can deduce the following conserved quantities:
\begin{itemize}
    \item The first quantity which is conserved for every $j$, 
\begin{equation}\label{eqn:condensateenergy}
 \mathcal{E}_{j}=A_{j}\vert\partial_{\chi}\sigma_{j}(\chi)\vert^2-B_{j}\vert\sigma_{j}(\chi)\vert^2+\frac{2}{5}\Re\left(w_{j}\sigma_{j}^5(\chi)\right),
 \end{equation}
 can be interpreted as a \virgolette{condensate energy} for the wave function $\sigma_{j}$.
 \item In the limit where the interaction term is small, there is another conserved quantity, which is related to the $U(1)$ symmetry $\sigma_{j}(\chi)\to e^{i\alpha}\sigma_{j}(\chi)$, 
\begin{equation}\label{eqn:definitionqj}
 Q_{j}=-\frac{i}{2}\left[\bar{\sigma}_{j}(\chi)\partial_{\chi}\sigma_{j}(\chi)-\sigma_{j}(\chi)\partial_{\chi}\bar{\sigma}_{j}(\chi)\right].
 \end{equation}
  This quantity can be related to the expectation value of the momentum of the scalar field at given $\chi$ 
 in the condensate state $\sigma$:
  \begin{equation}\label{eqn:momentumq}
      \braket{\hat{\Pi}(\chi)}_\sigma\equiv \braket{\,\sigma\mid \hat{\Pi}(\chi)\mid \sigma\,}=\sum_{j}Q_{j}\,.
  \end{equation}
In the small-interaction limit, therefore, the quantity $\braket{\hat{\Pi}(\chi)}_\sigma$ is a constant. Modulo the mentioned issues with this definition of the relational scalar field momentum observable, this could be seen as the quantum geometric analogue of the continuity equation for the massless scalar field. 
\end{itemize}
\paragraph*{Negligible interactions.}
In the mesoscopic regime where interactions are negligible, characterized by a relatively \virgolette{small} $\vert\sigma_j(\chi)\vert^2$ (but not so small as to endanger the hydrodynamic approximation, since $\vert\sigma_j(\chi)\vert^2$ controls the average number of condensate quanta), the equations of motion from the action \eqref{eqn:action} can be written as
\begin{align*}
    0&=\partial_\chi^2\rho_j-[m_j^2+(\partial_\chi\theta_j)^2]\rho_j\\
    0&=2\partial_\chi\rho_j\partial_\chi\theta+\rho_j\partial^2_\chi\theta\,,
\end{align*}
where $\rho_j$ and $\theta_j$ are determined from $\sigma_j=\rho_j\exp[i\theta_j]$ and $m_j^2=B_j/A_j$. The second equation is nothing but the conservation of $Q_j=\rho_j^2\partial_\chi\theta$ (recall that we are neglecting interactions), while the first one, via the introduction of $Q_j$, can be rewritten as
\begin{equation}\label{eqn:effectivedynamicsrhoold}
    \partial^2_\chi\rho_j-\frac{Q_j^2}{\rho_j^3}-m_j^2\rho_j=0\,.
\end{equation}
Now, if one interprets $\hat{\Pi}(\chi)$ as the momentum operator of the massless scalar field at a given value of it, in order for it to have a non-zero expectation value, at least one of the $Q_j$ has to be non-zero. This, in turns, implies that $\rho_j$ stays finite at all times. Since, as we will see below, the expectation value of the volume operator is controlled by $\rho_j$, this in turn will imply that the average of the volume never reaches zero, thus solving (on average) the cosmological singularity.
\paragraph*{Volume dynamics.}
Given the above dynamics of the condensate wave function one can study the dynamics of the expectation value of the \virgolette{relational volume operator}. According to the rule of Subsection \ref{subsec:secondquantization}, such an operator is defined by
\begin{equation}
    \hat{V}(\chi)=\int\diff g_I\diff g_I'\hat{\varphi}^\dagger(g_I,\chi)V(g_I,g_I')\hat{\varphi}(g_I',\chi)\,,
\end{equation}
in the group representation. The action of the volume operator on spin-network states depends only on the intertwiner label $\iota$:
\begin{equation}\label{eqn:volumematrixelements}
    V(\vec{x},\vec{x}')=V(\iota,\iota')\delta_{\vec{x}-\{\iota\},\vec{x}'-\{\iota'\}}\,,
\end{equation}
where $\vec{x}$ and $\vec{x}'$ are spin-network labels. By using equations \eqref{eqn:twobodyspin} and \eqref{eqn:ccoherent}, together with the orthonormality of the intertwiners $\mathcal{I}$, we see immediately that, in the spin representation, we can write
\begin{align}\label{eqn:expectationvaluevold}
    V(\chi)&\equiv \braket{\,\sigma\mid \hat{V}\mid \sigma\,}=\sum_{\vec{x}\vec{x}'}\braket{\,\sigma\mid V(\iota,\iota')\delta_{\vec{x}-\{\iota\},\vec{x}'-\{\iota'\}}\hat{c}^\dagger_{\vec{x}}\hat{c}_{\vec{x}'}\mid \sigma\,}\nonumber\\
    &=\sum_{j,\vec{m}}V_j\vert\sigma_{\{j,\vec{m}\}}\vert^2=\sum_j V_j\vert\sigma_j\vert^2=\sum_j V_j\rho_j^2\,,
\end{align}
where we have used that when $\hat{c}_{\vec{x}}$ acts on $\ket{\sigma}$ the resulting wave function has support only on the intertwiner which is an eigenvalue of the volume operator with the highest possible eigenstate compatible with the spin quantum number $j$, which we call $V_j$, and where we have suppressed for notational simplicity the explicit dependence on $\chi$. Since the expectation value of the number of equilateral tetrahedrons with spin quantum number $j$ associated to each face is
\begin{equation*}
    \hat{n}_j\equiv \sum_{\vec{m}}\hat{c}^\dagger_{j,\vec{m}}c_{j,\vec{m}}\,,
\end{equation*}
so that $\braket{\,\sigma\mid \hat{n}_j\mid \sigma\,}=\vert\sigma_j\vert^2$, we see that equation \eqref{eqn:expectationvaluevold} means that the expectation value of the volume operator is given by the sum over all the possibles spins $j$ of the average number of \virgolette{isotropic atoms} with spin $j$ multiplied by their volume, $V_j$. 

The volume operator then satisfies the equations:
\begin{subequations}\label{eqn:oldgeneralizedfriedmann}
\begin{align}
    \left[\frac{\partial_\chi V}{3V}\right]^2&=\left[\frac{2\sum_jV_j\rho_j\text{sgn}(\partial_\chi\rho_j)\sqrt{\mathcal{E}_j-Q_j^2/\rho_j^2+m_j^2\rho_j^2}}{3\sum_jV_j\rho_j^2}\right]^2,\\
    \frac{\partial^2_\chi V}{V}&=\frac{2\sum_jV_j\left[\mathcal{E}_j+2m_j^2\rho_j^2\right]}{\sum_jV_j\rho_j^2}\,.
\end{align}
\end{subequations}
We mention three interesting features of this volume dynamics, already stressed in \cite{Oriti:2016qtz}:
\begin{description}
\item[\emph{Bounce}]
In the mesoscopic regime considered in \cite{Oriti:2016qtz}, where equation \eqref{eqn:effectivedynamicsrhoold} holds, the expectation value of the volume operator \emph{never reaches zero}, as long as at least one of the $Q_j$s is non-zero. In \cite{Oriti:2016qtz}, it was argued that in order to get both a meaningful relational dynamics and a proper FRW spacetime (rather than a Minkowski spacetime), the energy density of the massless scalar field has to be non-zero, in turns implying that the expectation value of the massless scalar field momentum has to be non-zero as well. Because of equation \eqref{eqn:momentumq}, \cite{Oriti:2016qtz} concluded that at least one of the $Q_j$s has to be non-zero. In this context, therefore, a bouncing scenario with an always non-vanishing volume seems very natural. 
\item[\emph{Classical limit}] Further, \cite{Oriti:2016qtz} observed that in the limit in which $\rho_j^2\gg\vert \mathcal{E}_j\vert/m_j^2$ and $\rho_j^4\gg Q_j^2/m_j^2$, the above equations become 
\begin{subequations}
\begin{align}
    \left[\frac{\partial_\chi V}{3V}\right]^2&=\left[\frac{2\sum_jV_jm_j\rho_j^2}{3\sum_jV_j\rho_j^2}\right]^2,\\
    \frac{\partial^2_\chi V}{V}&=\frac{4\sum_jV_jm_j^2\rho_j^2}{\sum_jV_j\rho_j^2}\,,
\end{align}
\end{subequations}
leading to the classical  flat space ($k=0$) Friedmann equations
\begin{equation*}
    \left(\frac{\partial_\chi V}{V}\right)^2=\frac{V''}{V}=12\pi \tilde{G}\,,
\end{equation*}
as long as all the $m_j^2$s satisfy $m_j^2=3\pi \tilde{G}$, where $\tilde{G}\equiv GM^2$ is the dimensionless gravitational constant. Also, the classical Friedmann equations are obtained in the limit in which one of the $j$s dominates the above sums, say $j_o$, satisfying $m_{j_o}^2=3\pi \tilde{G}$. 
\item[\emph{Single spin}]
Lastly, \cite{Oriti:2016qtz} considered the case of a single-spin scenario, i.e., with $\rho_j=0$ for each $j\neq j_o$. This situation, mirroring the Loop Quantum Cosmology (LQC) context, leads to a dynamics of the form
\begin{subequations}\label{eqn:oldsinglespin}
\begin{align}\label{eqn:oldisinglespin1}
    \left[\frac{\partial_\chi V}{3V}\right]^2&=\frac{4\pi G}{3}\left(1-\frac{\rho}{\rho_c}\right)+\frac{4 V_{j_o}\mathcal{E}_{j_o}}{9V}\,,\\
    \frac{\partial^2_\chi V}{V}&=12\pi \tilde{G}+\frac{2V_{j_o}\mathcal{E}_{j_o}}{V}\,,
\end{align}
\end{subequations}
where it was assumed that $m_{j_o}^2=3\pi \tilde{G}$ and $\rho=\braket{\hat{\Pi}(\chi)}_\sigma^2/(2V^2)$, with $\braket{\hat{\Pi}(\chi)}_\sigma=Q_{j_o}$ and $V=V_{j_o}\rho_{j_o}^2$. The quantity $\rho_c$ \cite{Oriti:2016qtz}, instead, is defined as in equation \eqref{eqn:rhoc}. Interestingly enough, this dynamics resembles the effective LQC dynamics, with additional terms due to the $\mathcal{E}_{j_o}$ contributions. 
\end{description}
As we will see in Subsection \ref{subsec:volumedynamics}, the volume dynamics obtained in the  \virgolette{improved} relational framework that we will construct below will be remarkably similar to the one described here. However, some of the parameters will differ (essentially because of the use of different states), and some of the interpretations proposed in \cite{Oriti:2016qtz} will not be justified anymore.
\section{Relational dynamics in GFT}\label{sec:relationalgft}
As we have seen, GFTs describe the universe as a quantum many-body system, from which General Relativity is expected to emerge as some kind of collective phenomenon \cite{Oriti:2016acw}. For GFTs therefore, the general arguments concerning the extraction of an effective relational dynamics from pre-geometric theories, discussed in Subsection \ref{sec:reldynamicsqg}, are very fitting. Some technical and conceptual difficulties in the definition of relational dynamics in GFTs, both in \virgolette{tempus ante quantum} and in \virgolette{tempus post quantum} approaches were discussed in \cite{Kotecha:2018gof, Oriti:2016qtz}, and we will review them in Section \ref{subsec:pregeometric}. 

At a technical level, the GFT Fock space and more generally their close-to-standard QFT formulation allow to deal with the continuum limit using powerful QFT methods, and to identify and manipulate more easily states with proto-geometric features. How these features can be exploited to define a relational evolution on such proto-geometric states will be discussed in Subsection \ref{subsec:protogeometrictime}.

We will consider GFT models which include among their degrees of freedom those corresponding to a discretized scalar field, as introduced above, and focus on how one could proceed to extract a notion of time, i.e. an internal clock, and relational dynamics using it. 

\subsection{On a \virgolette{pre-geometric  relational time}}\label{subsec:pregeometric}
In GFT models for discrete gravity/geometry coupled to a discretized scalar field, each GFT quantum has an internal variable that could be used in principle as its own \virgolette{relational clock}. As we argue in the following, this \virgolette{single-quantum time}, however, fails to provide a notion of relational dynamics for generic (many-body) quantum states (which can only have a pre-geometric interpretation), both from a \virgolette{tempus ante quantum} and a \virgolette{tempus post quantum} perspective. 
The main difficulty is that, at this pre-geometric level, the many \virgolette{single-quantum times} fail in general to give rise to a notion of relational time that is \virgolette{organized enough} to label the evolution in the whole Fock space. In fact, the same difficulty arises in a classical description of the same pre-geometric degrees of freedom, making it clear that the core difficulty does not lie in the quantum properties of the degrees of freedom, but in their pre-geometric nature. 

\subsubsection{Pre-geometric GFT \virgolette{tempus ante quantum}}
Let us give an example of a pre-geometric \virgolette{tempus ante quantum} approach based on the identification of the internal scalar field degree of freedom as a relational clock. Following \cite{Kotecha:2018gof}, let us consider a system of $N$ GFT atom, i.e.\ tetrahedra, each characterized at the classical level by its own extended phase space\footnote{Since GFT lacks a preferred time evolution, the study of its classical formulation is best done in the framework of extended phase space and presymplectic mechanics \cite{Rovelli:2004tv, Rovelli:2001bq}, which are manifestly independent of any notion of absolute time. One could also imagine to deal with an external time parameter for each tetrahedron, and work with usual symplectic mechanics, but since this external parameter would a priori be different in each tetrahedron, on top of not appearing in the dynamics of the theory, it is not useful to refer to it at all.} $\Gamma_{\text{ex}}^{(i)}$, possibly subject to some (e.g.\ dynamical) constraint $C_{\text{full}}^{(i)}:\Gamma_{\text{ex}}^{(i)}\to\mathbb{R}$. In the case of a GFT coupled with  $M$ massless scalar degrees of freedom, the extended configuration space of each GFT \virgolette{atom} is $\mathcal{C}^{(i)}_{\text{ex}}=G^d\times\mathbb{R}^M\ni(g^{(i)I},\chi^{(i)a})$, with $I=1,\dots, D$ $a=1,\dots M$, and the corresponding phase space is\footnote{Notice that, differently from what we do in the rest of the paper, here we are denoting the group elements $g^I$ instead then $g_I$ in order to maintain a clear phase space notation.} $\Gamma^{(i)}_{\text{ex}}=T^*(\mathcal{C}^{(i)}_{\text{ex}})\simeq G^d\times\mathbb{R}^M\times(\mathfrak{g}^d)\times\mathbb{R}^M\ni (g^{(i)I},\chi^{(i)a},x^{(i)}_{I},p_{\chi^{(i)a}})$.

Further, we assume that each single-atom subsystem is deparametrizable, meaning that each single-atom constraint can be rewritten as
\begin{equation}\label{eqn:deparametrizedconstraint}
 C^{(i)}=p_{\chi^{(i)c}}+H^{(i)}(g^{(i)I},\chi^{(i)\alpha},x^{(i)}_I,p_{\chi^{(i)\alpha}})\,,
\end{equation} 
where $\alpha=1,\dots,M-1$ labels all the scalar matter values except for $\chi^{(i)c}$. This form of the constraint matches that of a non-relativistic system: the constraint surface $\Sigma^{(i)}_{\text{full}}$ defined by $C_{\text{full}}^{(i)}=0$ in this case admits a foliation in clock time, $\Sigma_{\text{dep}}=\mathbb{R}\times\Gamma^{(i)}_{\text{can}}$ \cite{Rovelli:2004tv}, which in turns allows to identify a reduced canonical phase space $\Gamma^{(i)}_{\text{can}}=T^*(\mathcal{C}^{(i)}_{\text{can}})\ni (g^{(i)I},\chi^{(i)\alpha},x^{(i)}_I,p_{\chi^{(i)\alpha}})$ constructed out of the initial partial observables (and their momenta) but without the variable playing the role of time. The function $H^{(i)}:\Gamma^{(i)}_{\text{can}}\to\mathbb{R}$ is then the Hamiltonian defining the evolution with respect to the relational time\footnote{Notice, however, that in order for $C^{(i)}$ to be equivalent to $C_{\text{full}}^{(i)}$, one has to rely on two approximations: first that $C_{\text{full}}^{(i)}$ can be linearized in terms of $p_{\chi^{(i)c}}$, and second that the remaining part of the constraint is actually independent of $\chi^{(i)c}$, which thus behaves as a good \emph{global} clock (see also the discussion in Subsection \ref{subsec:definingeffectivedynamics}).}.
 
Once any time variable $\chi$ for each individual particle is chosen, the idea \cite{Rovelli:2001bq,Chirco:2013zwa, Chirco:2016wcs, Kotecha:2018gof} is to select a clock $t$ among them and to \virgolette{synchronize} the others by imposing $\chi^{(2)c_2}=F_{2}(t),\dots, \chi^{(N)c_N}=F_{N}(t)$, satisfying $F_{i}'(t)=k_{i}$, with $k_{i}$ non-zero real constants. The deparametrized system is now defined on $\mathcal{C}_{\text{ex}}=\mathbb{R}\times \Gamma_{\text{can}}\ni (t,g^{(1)I},\chi^{(1)\alpha},\dots g^{(N)I},\chi^{(N)\gamma})$, with $\Gamma_{\text{ex}}=T^*(\mathcal{C}_{\text{ex}})$, together with a single combined constraint function $C_{\text{dep,}N}=p_{t}+H_N$ on $\Gamma_{\text{ex}}$. The physical Hamiltonian $H_N=\sum_{i=1}^Nk_{i}H^{(i)}$ describes the relational evolution in terms of the single particle Hamiltonians $H^{(i)}$ acting on the single particle reduced phase space $\Gamma_{\text{can}}^{(i)}$.
\paragraph*{Quantization of the deparametrized system.}
After the system is deparametrized, one can perform a canonical quantization. The details of such quantization procedure will not be important\footnote{In particular, we will not describe the mathematical details of a quantization map of the geometric part of the phase space, which can be found instead in \cite{Guedes:2013vi}.}. Rather, we will focus on the main conceptual steps \cite{Kotecha:2018gof}. 
 
Quantizing means choosing a quantization map between the classical algebra of observables, which are real smooth functions on the phase space, and the quantum algebra of observables, then represented as self-adjoint operators acting on a Hilbert space. Correspondingly, classical Poisson brackets are mapped into commutators. After the deparametrization approximation, this procedure can be straightforwardly applied to map the $N$-atoms canonical phase space $\Gamma_{\text{can},N}$ to the Hilbert space $\mathcal{H}_{\text{can},N}\equiv \mathcal{H}_{\text{can}}^{\otimes N}$, where $\mathcal{H}_{\text{can}}=L^2(G^d\times \mathbb{R}^{M-1})$. Correspondingly, Poisson brackets on $\Gamma_{\text{can},N}$ $\{\chi^{(i)\alpha},p_{\chi^{(j)\beta}}\}=\delta_{ij}\delta_{\alpha\beta}$ are mapped into commutators $[\widehat{\chi^{(i)\alpha}},\widehat{p_{\chi^{(j)\beta}}}]=i\delta_{ij}\delta_{\alpha\beta}$ on $\mathcal{H}_{\text{can},N}$. Notice that, by construction, the single clock variable $t$ chosen to deparametrize the system, is now treated as a parameter and not quantized. The Hamiltonian operator defining the evolution along $t$ is $\hat{H}_{N}=\sum_{i=1}^Nk_{i}\hat{H}^{(i)}$. 
 
The resulting (bosonic) Fock space can be written as
 \begin{equation}
  \mathcal{F}_{\text{can}}=\bigoplus_{N\ge 0}\text{sym}\mathcal{H}_{\text{can}}^{\otimes N}\,,
  \end{equation} 
and it is generated by the action of operators $\hat{\varphi}$, $\hat{\varphi}^\dagger$ on the Fock vacuum $\ket{0}$ and satisfying the \emph{equal Fock-time} commutation relations 
\begin{equation}\label{eqn:equaltimefockcommutations}
[\hat{\varphi}(t_{F},\mathbf{g}_{1},\vec{\chi}_{1}),\hat{\varphi}^\dagger(t_{F},\mathbf{g}_{2},\vec{\chi}_{2})]=\mathbb{I}(\mathbf{g}_{1},\mathbf{g}_{2})\delta(\vec{\chi}_{1}-\vec{\chi}_{2})\,,
\end{equation}
all the other commutators being zero. Here we have defined $\vec{\chi}\equiv (\chi^1,\dots,\chi^{M-1})$, and $\mathbf{g}\equiv (g_{1},\dots,g_{d})$ in order to make the notation simpler. Operators in this Fock space are then defined following the usual procedure. For instance, the operator
\begin{equation}
\hat{N}=\int\diff g_I\diff\chi^\alpha\hat{\varphi}^\dagger(g_I,\chi^\alpha)\hat{\varphi}(g_I,\chi^\alpha)\,,
\end{equation}
is the occupation number operator at a given value of the relational time $t_{F}$. 

However, as emphasized in \cite{Kotecha:2018gof}, the nature of the Fock time $t_{F}$ is not entirely clear. We would like a time parameter to be common to all the multi-atom sectors of the Fock space. This is not the case for the relational time that we have constructed above because that time is related only to one specific sector. On the other hand, the notion of time in equation \eqref{eqn:equaltimefockcommutations} must be flexible enough to be compatible with a variable $N$. So, despite the deparametrization approximations and the explicit use of a \virgolette{tempus ante quantum} approach, the fact that \virgolette{each fundamental GFT atom has its own clock}, together with the desired Fock space structure of the resulting quantum theory, conspires to a lack of a clear notion of relational time in the resulting reduced Fock space.

\subsubsection{\virgolette{Tempus post quantum} in GFT}
Similar issues are expected to appear also in a \virgolette{tempus post quantum} approach based on the use of the internal single-particle $\chi$-variable as a relational clock, for instance as the one developed in \cite{Oriti:2016qtz} and briefly reviewed in Section \ref{sec:review}. The reason is that they are due to the difficulty in organizing (\lq synchronizing\rq) the individual clocks associated to each microscopic constituent of the system, and not to their classical or quantum nature. 

One should expect that the relational observables defined in Subsection \ref{subsec:secondquantization}, can only be true relational quantities in some sector of the theory. Indeed, let us notice first that for each one-particle state $\ket{g_I,\chi}\equiv \hat{\varphi}^\dagger(g_I,\chi)\ket{0}$, the massless scalar field operator acts as
\begin{equation}
\hat{\chi}\ket{g_I,\chi}=\chi\ket{g_I,\chi}\,.
\end{equation} 
Thus eigenvectors of the massless scalar field operator span the one-particle Hilbert space $\mathcal{H}_{1}$. Moreover, when interpreting $\chi$ as a relational time, as suggested from the definition of \virgolette{relational operators} given for instance in \eqref{eqn:fakerelaitionalobservable}, these eigenstates satisfy the desired Schr\"odinger equation,
\begin{equation}\label{eqn:schrodinger}
 -i\frac{\diff}{\diff\chi}\ket{g_I,\chi}=\hat{\Pi}\ket{g_I,\chi}\,,
 \end{equation} 
where the Hamiltonian generating relational evolution is indeed given by the momentum of the massless scalar field $\hat{\Pi}$. Therefore, if we were to consider just the one-particle Hilbert space $\mathcal{H}_{1}$, \virgolette{$\chi$-\emph{diagonal}} geometric (which we assume having $\chi$-independent matrix elements) two-body operators defined following the prescription in \eqref{eqn:fakerelaitionalobservable}, which we denote by $\hat{O}_{2}(\chi)$ and by construction satisfying
\begin{equation*}
[\hat{\Pi},\hat{O}_{2}(\chi)]=i\partial_{\chi}\hat{O}_{2}(\chi)\,,
\end{equation*}
would indeed have a good relational meaning\footnote{Notice that these operators are not the Heisenberg version of the Schr\"odinger operators defined in the full second quantization framework.}. 

By the same token, (geometric and $\chi$-diagonal) operators defined according to the prescription \eqref{eqn:fakerelaitionalobservable} would be proper relational quantities if we were restricting our attention to the space $\tilde{\mathcal{F}}\subset\mathcal{F}$, generated by the algebra of the GFT operator evaluated at the same eigenvalue of the massless scalar field operator. Indeed, for such states $\ket{\psi(g_I^i,\chi)}$, a relational Schr\"odinger equation
\begin{equation}
-i\frac{\diff}{\diff\chi}\ket{\psi(g_I^i,\chi)}=\hat{\Pi}\ket{\psi(g_I^i,\chi)}\,,
\end{equation}
holds. This is not surprising: the prescription of \eqref{eqn:fakerelaitionalobservable} defines relational operators according to an internal \virgolette{one-atom time}, so by considering only \virgolette{synchronized} atoms the construction is still satisfactory.

This, however, suggests that such prescription is not well-defined and fails to provide a meaningful relational dynamics for structures outside $\tilde{\mathcal{F}}$ (which, instead, have a \virgolette{multi-fingered} time), and for general (i.e., non-diagonal) $(n+m)$-body operator of the form
\begin{align}
\hat{O}_{n+m}&=\int(\diff\chi)^{m}\int(\diff\tilde{\chi})^n \int (\diff g_{I})^m(\diff h_{I})^n\prod_{i=1}^m\hat{\varphi}^\dagger(g_{I}^i,\chi_{i})\prod_{j=1}^n\hat{\varphi}(h_{I}^j,\tilde{\chi}_{j})\nonumber\\
&\qquad\times O_{m+n}(g_{I}^1,\dots, g_{I}^m,\chi_{1},\dots,\chi_{m},h_{I}^1,\dots, h_{I}^n,\tilde{\chi}_{1},\dots,\tilde{\chi}_{n})\,.\label{eqn:generaloperator}
\end{align} 
Indeed, consider any such geometric operator, for which
\begin{equation*}
O_{m+n}(g_{I}^1,\dots, g_{I}^m,\chi_{1},\dots,\chi_{m},h_{I}^1,\dots, h_{I}^n,\tilde{\chi}_{1},\dots,\tilde{\chi}_{n})=O_{m+n}(g_{I}^1,\dots, g_{I}^m,h_{I}^1,\dots, h_{I}^n)\,.
\end{equation*}
If we suppress (as suggested by \eqref{eqn:fakerelaitionalobservable}) all the integrals and leave the dependence on the various $\{\chi_{i}\}_{i=1,\dots,m}$, $\{\tilde{\chi_{j}}\}_{j=1,\dots,n}$, the commutator with $\hat{\Pi}$ gives
\begin{align*}
[\hat{O}_{m+n}\left(\{\chi_{i}\}_{i=1,\dots,m},\{\tilde{\chi_{j}}\}_{j=1,\dots,n}\right),\hat{\Pi}]&=i\sum_{i=1}^n\partial_{\chi_{i}}\hat{O}_{m+n}\left(\{\chi_{i}\}_{i=1,\dots,m},\{\tilde{\chi_{j}}\}_{j=1,\dots,n}\right)\\
&\quad+i\sum_{j=1}^n\partial_{\tilde{\chi}_{j}}\hat{O}_{m+n}\left(\{\chi_{i}\}_{i=1,\dots,m},\{\tilde{\chi_{j}}\}_{j=1,\dots,n}\right),
\end{align*}
generating indeed evolution, but now along all the possible \virgolette{time directions}. On the other hand, considering only the diagonal part (in terms of $\chi$-eigenvalues) of the operator $\hat{O}_{m+n}$, so that we obtain an operator $\hat{O}_{m+n}(\chi)$, amounts to rejecting a large amount of potentially relevant information without any good physical justification, in particular when such operators are applied to states outside $\tilde{\mathcal{F}}$. Of course, this observation applies also to \virgolette{non-diagonal} two-body operators of the form
\begin{equation*}
\hat{\tilde{O}}\equiv \int\diff\chi\diff\tilde{\chi}\diff g_I\diff h_I\,\tilde{O}(g_I;h_{I})\hat{\varphi}^\dagger(g_I,\chi)\hat{\varphi}(h_I,\tilde{\chi}),
\end{equation*}
making the construction not entirely clear even at the level of the one-atom space $\mathcal{H}_{1}$.

Moreover, states outside $\tilde{\mathcal{F}}$ played, in the construction of \cite{Oriti:2016qtz}, a crucial role, since the coherent states \eqref{eqn:coherentstates} do not live in $\tilde{\mathcal{F}}$. Therefore, not only they do not have a robust Schr\"odinger relational dynamics, but one could also be interested in computing expectation values of general ($m+n$)-body operators on these states, thus having to face all the aforementioned ambiguities.

The bottom line is no different from the one we have discussed in the classical case: generic \virgolette{pre-geometric} states in the Fock space have intrinsic \virgolette{multi-fingered} relational times. Defining a notion of relational dynamics for such states becomes therefore a remarkably complicated task.
\paragraph*{Divergences.}
Besides the mentioned conceptual difficulties, the prescription \eqref{eqn:fakerelaitionalobservable} leads also to some technical difficulties. One could be interested in the variances of quantum operators at a given value of the parameter $\chi$ on coherent states. But then, defining for instance $\hat{O}^2(\chi)=\hat{O}(\chi)\hat{O}(\chi)$ (thus using the diagonal prescription) one finds that the result is always divergent. 

A similar divergent behavior of the above \virgolette{relational} operators were reported in \cite{Assanioussi:2020hwf}, where it was noticed that, in presence of thermal fluctuations, even the one-atom diagonal operators are ill-defined. They observed, however, that those divergences can be kept under control by defining \emph{smeared} operators of the form
\begin{subequations}\label{eqn:smearedcreation}
\begin{align}
\hat{a}_{\vec{x}}(t)&\equiv \int\diff g_I\int\diff\chi\overline{D}_{\vec{x}}(g_I)\overline{t}(\chi)\hat{\varphi}(g_I,\chi)\,,\\
\hat{a}^\dagger_{\vec{x}}(t)&\equiv \int\diff g_I\int\diff\chi D_{\vec{x}}(g_I)t(\chi)\hat{\varphi}(g_I,\chi)\,,
\end{align}
\end{subequations}
for an arbitrary test function $t(\chi)$. Similarly, they defined (regularized) \virgolette{relational} operators of the form
\begin{equation*}
\hat{V}_{x}\equiv \sum_{\vec{x}}v_{\vec{x}}\hat{a}^\dagger_{\vec{x}}(t)\hat{a}_{\vec{x}}(t)\,.
\end{equation*}
The dynamics was then obtained as in \cite{Oriti:2016qtz}, i.e., by imposing the averaged quantum equations of motion (or, equivalently, only considering the SD equations for the operator $O=1$)
\begin{equation}\label{eqn:schwingerpast}
\left\langle\sigma\biggl\vert\frac{\delta \hat{S}[\varphi,\varphi^\dagger]}{\delta \hat{\varphi}^\dagger(g_I,\chi)}\biggr\vert\sigma\right\rangle=0\,,
\end{equation}
where $\ket{\sigma}$ was again a coherent (but thermal) state, neglecting interactions and by assuming a \emph{local kinetic term}. This leads to a functional dynamics for the condensate wavefunction of the form\footnote{Notice that for such a kinetic term which is also quadratic, as it was indeed assumed in \cite{Oriti:2016qtz}, the resulting equations of motion for $\sigma_{\vec{x}}$ are equivalent to the equations of motion for $\bar{\sigma}_{\vec{x}}$.}
\begin{equation}\label{eqn:functionaldynamics}
\nabla^2_{t}\sigma_{\vec{x}}(t)-M_{\vec{x}}\sigma_{\vec{x}}(t)=0\,,
\end{equation}
where $M_{\vec{x}}\equiv -B_{\vec{x}}/A_{\vec{x}}$, and
\begin{align*}
\sigma_{\vec{x}}(t)&\equiv\int\diff\chi \overline{t}(\chi)\sigma_{\vec{x}}(\chi)\,,\\
\nabla_{t}&\equiv-\int\diff\chi\left(\partial_{\chi}\overline{t}\frac{\delta}{\delta\overline{t}(\chi)}+\partial_{\chi}t\frac{\delta}{\delta t(\chi)}\right).
\end{align*}
As remarked in \cite{Assanioussi:2020hwf}, the use of a delta distribution peaked on $\chi$, i.e., $t(\chi)=\delta(\chi'-\chi)$ reproduces the framework defined in \cite{Oriti:2016qtz}. 

Let us mention that while the algebra \eqref{eqn:frozencommutations} is distributional, and needs to be smeared with test functions in order to produce meaningful results, expectation values of a general second quantized operator \eqref{eqn:generaloperator} (and of any product of them) on coherent states do not show a distributional behavior. The reason is that these operators are defined in such a way that there is (at least) one integration for any couple $\varphi$-$\varphi^\dagger$ on their domain of dependence which accounts for the distributional nature of their commutator. Thus, while divergences may appear because of redundant integrations on a non-compact region, $\delta$-like divergences are not expected. As a consequence, a smearing of the algebra \eqref{eqn:frozencommutations} is not needed as long as one is interested in expectation values of operators on coherent states. On the other hand, distributional behavior of expectation values on coherent states is expected to manifest itself if one modifies the definition \eqref{eqn:generaloperator} by suppressing some integrations. Therefore, the origin of this kind of divergences should be attributed to the chosen definition of relational many-body operators rather than on the distributional nature of the algebra \eqref{eqn:frozencommutations}.

Moreover, while the use of suitable smooth functions regularizes the algebra \eqref{eqn:frozencommutations}, it does not make any more clear in which sense this allows for a good relational dynamics, since all the ambiguities we discussed in defining (even regularized) relational operators are still present. Finally, one would like to have a clear physical meaning of the functional dynamics expressed by the function $t(\chi)$. This is instead missing at the moment, because we lack of a manifest physical interpretation for the clock it should represent.

These issues call for a different way of defining a relational dynamics in GFT. Given the aforementioned difficulties, we argue that this should be done in a \virgolette{proto-geometric regime}, thus tackling the problem of time from an effective point of view. 

\subsection{On a \virgolette{proto-geometric relational time}}\label{subsec:protogeometrictime}
In a sense, a \virgolette{proto-geometric} notion of relational dynamics has been already postulated in some previous works \cite{Wilson-Ewing:2018mrp, Adjei:2017bfm, Gielen:2019kae}. In particular, in \cite{Wilson-Ewing:2018mrp}, classical \virgolette{same-time} Poisson brackets of the form
\begin{equation}
    \left\{\varphi_{\vec{x}}(\chi),\pi_{\vec{x}'}(\chi)\right\}=\delta_{\vec{x}\vec{x}'}\,,
\end{equation}
(cfr. with equation \eqref{eqn:equaltimefockcommutations}) were assumed among the GFT field and its momentum (obtained from a Legendre transform of the GFT action, assuming the kinetic kernel can be expanded by retaining only second derivative contributions\footnote{This is not a quite general assumption: as explained in Subsection \ref{subsec:dynamics}, such truncation of the kinetic kernel can be seen as a approximation that might not be satisfied by solutions of the resulting dynamical equations. This is particularly relevant in the cosmological case.}). 
In other words, this followed from a canonical reformulation of the classical GFT action, which singled out one variable in the domain of the GFT field (the one corresponding to the scalar field degree of freedom) and treated it as a time parameter. This step has been done at the classical level. It has been performed also at the level of a collective, coarse-grained description of the microscopic GFT degrees of freedom, since the GFT field is in fact a collective variable.
Thus, the choice of a \virgolette{collective} relational time in the whole Fock space resulting from the above equation corresponds, in our classification, to a proto-geometric (collective) relational dynamics from a \virgolette{tempus ante quantum} perspective. It is subject to the general limitations (and therefore criticisms) that characterize \virgolette{tempus ante quantum} approaches, i.e., that they lack the important notion of clock-covariance.


On the other hand, as argued in Subsection \ref{subsec:definingeffectivedynamics}, it would be best to work at the post-quantum level. Indeed, we have at our disposal all the needed structures and ingredients: geometric observables with a collective character (e.g. the total volume operator, together with the number operator \eqref{eqn:numberoperator}), and candidate proto-geometric states encoding a notion of continuum, i.e. the condensate states used in GFT condensate cosmology framework. 

We also have operators related to the massless scalar degree of freedom, like $\hat{X}$ and $\hat{\Pi}$. However, it is important to notice that the operator $\hat{X}$ can not straightforwardly be interpreted as a massless scalar field, not even when the expectation value on a proto-geometric state is taken. Indeed, such an operator is extensive, while a scalar field is an intensive quantity (from the microscopic QG perspective). As already pointed out in \cite{Gielen:2014uga}, this is a standard feature even in non-relativistic quantum mechanics, where the canonically conjugate operators $\hat{x}$ and $\hat{p}$ become two extensive quantities, the \virgolette{total position} operator $\hat{X}$ and the total momentum operator $\hat{P}$, whose commutator would now be given by $[\hat{X},\hat{P}]=i\hat{N}$. The position variable, which should correspond to an intensive variable, has now become an extensive one. To obtain an intensive quantity, we can not just \virgolette{divide} the operator $\hat{X}$ by the number of particles, since the operator $\widehat{N^{-1}}$, containing zero in its spectrum, is not well defined in the Fock space\footnote{Moreover, the resulting operator would act non-linearly on a quantum state. Correspondingly, we could not identify it as an \virgolette{observable} in the sense of a self-adjoing (and therefore, linear) operator. Still, such non-linearities are expected to be negligible when the quantum properties of the number operator are negligible as well, as we will assume below.}. The best we can do is to define the \virgolette{center of mass variable} in terms of expectation values: $x_{\text{c.o.m}}\equiv \braket{\hat{X}}/\braket{\hat{N}}$. Similarly, we can define the intrinsic quantity corresponding to $\hat{\chi}$ as $\hat{\chi}\equiv \hat{X}/\braket{\hat{N}}_{\Psi}$. This definition is expected to be good as long as the quantum features of the number operator are not relevant, so that it can be approximated with its expectation value without losing physical information. This is one of the requirements already expressed by the small fluctuations conditions \eqref{eqn:semi-classicality}, as relevant for a good relational dynamics (in fact, it is needed, before that, for a sensible continuum proto-geometric interpretation of the theory).

We will therefore proceed to the definition of an effective relational dynamics in GFT condensate cosmology, implementing the general ideas defined in Section \ref{subsec:definingeffectivedynamics}. 

\section{Effective relational GFT cosmology}\label{sec:effectivevolumedynamics}
In the cosmological case, thanks to the assumptions of homogeneity and isotropy, we can explicitly construct states allowing for a meaningful notion of relational dynamics in an effective regime for the only relevant geometric observable: the volume operator. We do so in Subsection \ref{subsec:cps}, while their explicit dynamics is derived in Subsection \ref{subsec:dynamics}. Then, in Subsection \ref{subsec:validityaverageddynamics}, we discuss the validity of the conditions \eqref{eqn:averagedrelationaldynamics}, while a detailed study of the soundness of the conditions \eqref{eqn:semi-classicality} is presented in \cite{toappear}. Finally, in Subsection \ref{subsec:volumedynamics} we will study the relational dynamics of the volume operator, in particular checking whether it matches the classical expectation at least at a certain semi-classical effective level and whether the singularity is resolved at least in terms of expectation values. 

\subsection{States of a leaf: Coherent Peaked States}\label{subsec:cps}
In order to construct appropriate relational states, we can take inspiration from the classical spacetime intuition. The easiest way to define a relational dynamics of geometric quantities at the classical level is to fix a gauge and choose a foliation of spacetime adapted to the massless scalar field itself. Analogously, in our case, what we need is to construct states which can be interpreted as \virgolette{bona fide} leaves of a $\chi$-foliation\footnote{The connection between GFT coherent states and $3$-geometries was already suggested in \cite{Gielen:2013naa}.}. Starting from the class of coherent states \eqref{eqn:coherentstates}, we want to specialize to states which are sharply peaked on a given value of the massless scalar field variable $\chi$. In other words, we want states which represent an infinite superposition of atoms of space and which are associated in a precise way (i.e., with an a priori defined small margin of error) to a given value $\chi_{0}$ of the massless scalar field. In a sense, we are reconstructing collectively, coarse-grained synchronized states, similar to those living in $\tilde{\mathcal{F}}$ (but with important differences, see the comments below). Since, in the simple condensate states considered, all the tetrahedra share the same information, encoded in the condensate wavefunction $\sigma(g_I,\chi)$, the needed states can be constructed by assuming that the condensate wavefunction takes the following form:
\begin{equation}\label{eqn:wavefunctioncps}
 \sigma_{\epsilon}(g_I,\chi)\equiv \eta_{\epsilon}(g_I;\chi-\chi_{0},\pi_{0})\tilde{\sigma}(g_I,\chi)\,,
 \end{equation} 
 where $\eta_{\epsilon}$ is a \emph{peaking function} around $\chi_{0}$ with a typical width given by $\epsilon$. The simplest example of such peaking function is given by a Gaussian,
 \begin{equation}\label{eqn:peakingfunction}
 \eta_{\epsilon}(\chi-\chi_{0},\pi_{0})\equiv \mathcal{N}_{\epsilon}\exp\left[-\frac{(\chi-\chi_{0})^2}{2\epsilon}\right]\exp[i\pi_{0}(\chi-\chi_{0})]\,,
 \end{equation}
 where $\mathcal{N}_{\epsilon}$ is a normalization constant to be fixed later, and where we have assumed, for simplicity, that the peaking function does not depend on the group variables $g_{I}$. Of course, in order for the condensate wavefunction to be truly peaked around $\chi_0$ it is necessary for the \emph{reduced wavefunction} $\tilde{\sigma}$ not to spoil the peaking properties of $\eta_\epsilon$. 
 Since, as we will see below, $\tilde{\sigma}$ will be dynamically determined, only solutions which satisfy this requirement (if there are any) should be considered. 
\subsubsection{Comments on the properties of the CPSs}
Let us briefly comment on some features of these Coherent Peaked States (CPSs). 

In order to implement a notion of effective relational dynamics, these states should satisfy the conditions discussed in Subsection \ref{subsec:definingeffectivedynamics}, at least in some regimes and in some regions of the parameter space. A first condition that we impose on the above parameters in order for these CPSs to actually meet the requirements in Subsection \ref{subsec:definingeffectivedynamics} is
\begin{subequations}\label{eqn:goodclocksmallvariance}
\begin{equation}\label{eqn:goodclock}
    \epsilon\ll 1\,.
\end{equation}
This condition, as the computations below will clarify further, is what allows us to \virgolette{synchronize the internal clocks} of the fundamental GFT quanta\footnote{For generic condensate states, such synchronization realized at the level of the collective condensate wavefunction would be only performed at the coarse-grained level; for the simple coherent condensate states we use here, this is in fact also implemented at the level of the individual GFT quanta. It is important to distinguish the general rationale from the peculiarities of the specific implementation.}, and thus to consider the CPSs as some kind of discrete counterpart of proper leaves of a foliation based on the massless scalar field itself. As a consequence, one can interpret (at the effective level) the expectation value of an operator on a CPS characterized by $\chi_{0}$ as the same operator computed at a relational time $\chi_{0}$ (i.e., on a slice labelled by $\chi_{0}$). The resulting relational averages (see Subsection \ref{subsec:validityaverageddynamics}) are thus very similar in spirit to those that can be obtained from the relational operators constructed in \cite{Giddings:2005id, Bojowald:2010xp, Bojowald:2010qw, Hoehn2011}. Such a good peaking property on the clock value is expected to be necessary simply because only in such case one could meaningfully speak of clock values, and thus relational dynamics, to start with. On the other hand, this is expected to hold only approximately, for the simple reason that also the system we choose to play the role of relational clock is a quantum system and exact specification of its reading cannot be expected to hold with absolute precision.

Therefore, the above \virgolette{synchronization condition} has to be taken with care. Taking the limit $\epsilon\to 0$, in fact, would produce \cite{toappear} \emph{arbitrarily large} quantum fluctuations on the momentum of the massless scalar field. Such infinite fluctuations can not be of course included in a self-consistent framework implementing a notion of effective relational dynamics. From now on, we will therefore consider a small but \emph{finite} $\epsilon$. Formally, therefore, CPSs do not live in $\tilde{\mathcal{F}}$, though they can be thought to be \virgolette{very close} (small $\epsilon$) to such \virgolette{synchronized states}\footnote{Notice that from this perspective (coherent) perfectly \virgolette{synchronized states} living in $\tilde{\mathcal{F}}$ should not be seen as defining an appropriate effective notion of relational dynamics.}. 

Still, even with a finite, but small $\epsilon$, the relative variance of the operator $\hat\Pi$ (and similarly of the operator $\hat{H}$ to be defined below) turns out to be possibly very large \cite{toappear}. This feature should not be surprising, since $\hat{\chi}$ and $\hat{\Pi}$ are canonically conjugate. However, while one expects such large fluctuations to naturally arise in a truly pre-geometric phase of the theory, there must exist a regime in a proto-geometric phase in which they are suppressed, eventually leading to a good semi-classical description of the scalar field. For instance, this can be achieved  in the limit of very large number of GFT quanta by imposing the condition
\begin{equation}\label{eqn:smallvariance}
    \epsilon\pi_0^2\gg 1\,,
\end{equation}
\end{subequations}
on the parameters $\epsilon$ and $\pi_0$ \cite{toappear}, which we will therefore assume from now on. So, we see that the conditions \eqref{eqn:goodclocksmallvariance} are related to very different aspects of the implementation of the relational dynamics: while the good clock condition \eqref{eqn:goodclock} is important to obtain an \emph{almost} perfect \virgolette{synchronization} of the fundamental \virgolette{atoms of space} (and thus it is relevant even at the level of average values of operators), condition \eqref{eqn:smallvariance} is related only to quantum fluctuations of the conjugate clock variable. 

Second,  these states, \emph{by construction}, can not be \lq\emph{minimum uncertainty states}\rq$\,$ (MUCs) for the couple of operators $(\hat{X},\hat{\Pi})$. Indeed, it is well-known that a couple of operators $\hat{A}$ and $\hat{B}$ saturates the Heisenberg uncertainty inequality on a state $\ket{\psi}$ if and only if \cite{puri}
     \begin{equation*}
[\hat{A}+i\lambda\hat{B}]\ket{\psi}=\left[\braket{\hat{A}}_{\psi}+i\lambda\braket{B}_{\psi}\right]\ket{\psi}\,,
\end{equation*}
where $\lambda$ is a complex number. Now, it is easy to realize that the CPSs introduced above (and, more generally any coherent state of the GFT field of the form \eqref{eqn:coherentstates}) do not satisfy the above equation for the operators $\hat{X}$ and $\hat{\Pi}$, given explicitly in equations \eqref{eqn:scalarfieldoperator}, \eqref{eqn:momentumoperator}:
\begin{equation*}
\left[\hat{X}+i\lambda\hat{\Pi}\right]\ket{\sigma}\neq \left[\braket{\hat{X}}_{\sigma}+i\lambda\braket{\hat{\Pi}}_{\sigma}\right]\ket{\sigma}\,.
\end{equation*}
In particular, the state obtained at the left-hand-side of this equation does not contain the vacuum state, while the second one does. Typically, it is precisely the property of being minimum uncertainty states (MUCs) for some operators satisfying a certain algebra, which defines coherent states as being states behaving \virgolette{as classical as possible} with respect to those quantities. In this case, we are using states which are indeed coherent, but just according to the GFT operators, for which they indeed are MUCs. Still, they are not MUCs for $\hat{X}$ and $\hat{\Pi}$. It would be therefore surprising if these states turned out to fit perfectly a classical description of a massless scalar field coupled to geometry. And indeed, as we will see in Subsection \ref{subsec:volumedynamics}, this will not be the case. 

In general, we have not exploited the possibilities, offered by the second quantized formalism, to define coherent states which would minimize the uncertainty relations between $\hat{X}$ and $\hat{\Pi}$ in particular minimizing as much as possible also the variance of the collective observable $\hat{\Pi}$. We have not done so because we are not aiming, in the present context, to identify relational clocks that would also be \emph{ideal} (i.e., \virgolette{as classical as possible}), but only to define a good relational dynamics. 
Should we be interested in imposing additional and more stringent semi-classicality condition on our clock, we could for example adapt to the GFT condensate context the techniques developed in \cite{Oriti:2012us} to construct coherent states for collective variables in the LQG context.

Third, on the same line, we want to emphasize that, given the specific form of a CPS with peaking function as in \eqref{eqn:peakingfunction}, taking the limit $\epsilon\to \infty$ will not lead to a localization of the wavefunction in momentum space around $\pi_0$, as one would naively guess. In fact, the very same assumption of the factorization of the CPS wavefunction into a peaking function and a \emph{$\chi$-dependent} reduced wavefunction, implies that the wavefunction in momentum space is given by the following convolution product
\begin{equation*}
\sigma_{\text{f},\epsilon}(g_I,\pi;\chi_{0},\pi_{0})\equiv \int\diff \pi'\eta_{\epsilon}(\pi-\pi';\chi_{0},\pi_{0})\tilde{\sigma}(g_I,\pi')\,.
\end{equation*}
This shows immediately that, even if the Fourier transform $\eta_{\epsilon}(\pi;\chi_{0},\pi_{0})$ of the peaking function is peaked on $\pi_{0}$, the convolution integral is not going to be peaked on $\pi_{0}$. More precisely, when $\epsilon\to\infty$ (when $\eta_{\epsilon}(\pi;\chi_{0},\pi_{0})$ is indeed peaked) the above equation becomes
\begin{equation*}
\sigma_{\text{f},\epsilon}(g_I,\pi;\chi_{0},\pi_{0})\simeq \mathcal{N}_{\epsilon}e^{-\chi_{0}^2/(2\epsilon)}\tilde{\sigma}(g_I,\pi-\pi_{0})\,.
\end{equation*}
For instance, this implies that the expectation value of the occupation number on the factorized state in the limit $\epsilon\to\infty$ is given by (see Section \ref{subsec:validityaverageddynamics} for similar computations)
\begin{equation*}
\braket{\hat{N}}_{\sigma_{\text{f},\epsilon};\pi_{0},\chi_{0}}=\int\diff\pi\int\diff g_I\,\vert\sigma_{\text{f},\epsilon}(g_I,\pi;\chi_{0},\pi_{0})\vert^2\simeq \mathcal{N}_{\epsilon}^2e^{-\chi_{0}^2/\epsilon}\int\diff g_I\int\diff \pi\vert\tilde{\sigma}(g_I,\pi)\vert^2\,,
\end{equation*}
which does not depend at all on the variable $\pi_0$. However, as we have already mentioned, the role of $\pi_{0}$ is crucial in order to make the above states meet some semi-classicality requirements (at least in some regimes), by ensuring some control over the variance of the momentum and the Hamiltonian operator.

Lastly, we remark that, as a consequence of the above construction, the divergences that plague general $n$-point \virgolette{relational} operators in the prescription of \cite{Oriti:2016qtz}, can not be present in this framework. In fact, since we use no redefinition of second-quantized operators to define relational quantities, but rather we stick to an effective \virgolette{Schr\"odinger picture}, the commutation relations between $\hat{\varphi}$ and $\hat{\varphi}^\dagger$, which ultimately produced the ill-defined behavior of \virgolette{relational} operators as defined in \cite{Oriti:2016qtz}, are in this case always compensated by an integration. In our framework, therefore, there is no need to introduce smeared creation and annihilation operators (see equations \eqref{eqn:smearedcreation}) as proposed in \cite{Assanioussi:2020hwf} in order to tame the aforementioned divergences\footnote{Of course suitable smearing may well be needed to define rigorously the full GFT Weyl algebra of observables; simply, it is not our concern here.}. By construction, the choice of our states produces a regular and \virgolette{almost perfect localization} of the expectation values of operators, effectively reproducing the results obtained by considering expectation values of operators constructed out of creation and annihilation operators smeared with the very same peaking function appearing in \eqref{eqn:wavefunctioncps}. 

\subsection{CPSs dynamics}\label{subsec:dynamics}
Following the same procedure of \cite{Oriti:2016qtz}, we can now obtain the dynamical equations for the \emph{reduced wavefunction} $\tilde{\sigma}$ starting from the Schwinger-Dyson equation. We need then to fully specify the GFT action $S[\varphi,\bar{\varphi}]$, including a massless scalar field. If such a field is minimally coupled to gravity, one can use the symmetries of the classical action (which are assumed to be present also at the quantum level, and in the GFT amplitudes, which generate simplicial gravity path integrals including a discretized scalar field \cite{Oriti:2016qtz}) to place strict constraints on the GFT action. This, in general, can be written as
\begin{equation}
S=K+U+\bar{U}\,,
\end{equation}
where $K$ represents the kinetic term and $U$ encodes interactions. In the following, we will restrict our analysis only to the kinetic term, thus neglecting interactions. However, a brief discussion about the contributions to the model coming from simplicial interactions can be found in Subsection \ref{app:simplicialinteractions}. 

Given the aforementioned symmetry assumptions, the kinetic term takes the form  \cite{Oriti:2016qtz}
\begin{equation}\label{eqn:kineticterm}
K=\int\diff g_I\diff h_I\int \diff\chi\diff\chi'\,\bar{\varphi}(g_I,\chi)K(g_I,h_I;(\chi-\chi')^2)\varphi(h_I,\chi')\,,
\end{equation}
where the dependence of the kinetic kernel from $(\chi-\chi')^2$ is due to the fact that the scalar field propagates between neighbouring $4$-simplices along dual links, and that we are requiring the classical shift and reflection symmetries of the massless scalar field action to be preserved at the quantum level. On the other hand, the dependence of $K$ from $g_I$ and $h_I$ is determined by the specific choice of the GFT model we are interested in (see for instance \cite{Oriti:2016qtz} for an example in the case of a GFT based on the EPRL spinfoam model).
\subsubsection{Reduced wavefunction effective dynamics}
In light of the effective and approximate nature of the relational framework we want to implement (see the discussion in Subsection \ref{subsec:definingeffectivedynamics}), we content ourselves with extracting an effective mean field dynamics from the full set of Schwinger-Dyson equations, assuming that the relevant states for cosmological dynamics are CPSs, and then 
an averaged relational dynamics for interesting geometric observables from it. Thus we only impose the equation
\begin{equation}\label{eqn:simplestschwinger}
\left\langle\frac{\delta S[\hat{\varphi},\hat{\varphi}^\dagger]}{\delta\hat{\varphi}^\dagger(g_I,\chi_{0})}\right\rangle_{\sigma_{\epsilon};\chi_{0},\pi_{0}}\equiv\left\langle\sigma_{\epsilon};\chi_{0},\pi_{0}\biggl\vert\frac{\delta S[\hat{\varphi},\hat{\varphi}^\dagger]}{\delta\hat{\varphi}^\dagger(g_I,\chi_{0})}\biggr\vert\sigma_{\epsilon};\chi_{0},\pi_{0}\right\rangle=0\,,
\end{equation}
to be satisfied, where $\ket{\sigma_{\epsilon};\chi_{0},\pi_{0}}$ is the CPS with wavefunction \eqref{eqn:wavefunctioncps} and with peaking function \eqref{eqn:peakingfunction}. 
After a change of variable $\chi-\chi_{0}\to\chi$, and neglecting the contribution from GFT interactions, equation \eqref{eqn:simplestschwinger} becomes
\begin{equation*}
\int\diff h_I\diff \chi\,K(g_I,h_I;\chi^2)\eta_{\epsilon}(\chi;\pi_{0})\tilde{\sigma}(g_I,\chi+\chi_{0})=0\,.
\end{equation*}
We further assume that the kinetic kernel can be written in terms of a series expansion as
\begin{equation}
K(g_I,h_I;\chi^2)=\sum_{n=0}^\infty\frac{K^{(2n)}(g_I,h_I)}{(2n)!}\chi^{2n}\,.
\end{equation}
Let us notice that while this assumption might appear too restrictive, given the typical distributional nature of kinetic kernels in general QFTs, the situation might be different in GFT. Indeed, existing studies \cite{Li:2017uao} on the form of the action of a GFT for gravity minimally coupled to a massless scalar field suggest a form of the kinetic kernel in terms of a non-polynomial function of second derivatives with respect to the argument of the GFT field encoding the scalar field degrees of freedom (our $\chi$) acting on a $\delta$ function of the same argument. The GFT kinetic kernel so identified  admits a representation in terms of smooth functions of the arguments of the fields appearing in the action, thus it would satisfy our assumption. We believe this is an interesting point which deserve further scrutiny.  

Since, because of the function $\eta_{\epsilon}$, the integrand is peaked around $\chi=0$, we Taylor expand the \emph{reduced wavefunction} $\tilde{\sigma}$ around that point, so that the kinetic term contribution can be written as
\begin{equation*}
\sum_{n=0}^\infty\sum_{m=0}^\infty\int\diff h_I\frac{K^{(2n)}(g_I,h_I)}{(2n)!}\frac{\tilde{\sigma}^{(m)}(h_I,\chi_{0})}{m!}I_{2n+m}(\pi_{0},\epsilon)\,,
\end{equation*}
where the apex on the reduced wavefunction indicates the $m$-th derivative of the function with respect to the massless scalar field variable, and where
\begin{align*}
I_{2n+m}(\pi_{0},\epsilon)&\equiv \mathcal{N}_{\epsilon}\int\diff\chi\,\chi^{2n+m}e^{-\chi^2/(2\epsilon)+i\pi_{0}\chi}=\mathcal{N}_{\epsilon}\sqrt{2\pi\epsilon}\left(-i\right)^{2n+m}\frac{\partial^{2n+m}}{\partial\pi_{0}^{2n+m}}e^{-\pi_{0}^2\epsilon/2}\\
&=\mathcal{N}_{\epsilon}\sqrt{2\pi\epsilon}\left(i\sqrt{\frac{\epsilon}{2}}\right)^{2n+m}e^{-\pi_{0}^2\epsilon/2} H_{2n+m}\left(\sqrt{\frac{\epsilon}{2}}\pi_{0}\right)\,,
\end{align*}
where $H_{2n+m}$ are Hermite polynomials of order $2n+m$. 

We now retain only the lowest order contributions\footnote{This additional approximation should be understood in the spirit of the discussion in Subsection \ref{subsec:definingeffectivedynamics}, as one of the approximations characterizing the effective nature of the approach.}, truncating the above sum at order $\epsilon$, i.e., with the combination $2n+m\le 2$. We thus obtain
\begin{align*}
\mathcal{N}_{\epsilon}\sqrt{2\pi\epsilon}e^{-\pi_{0}^2\epsilon/2}\int\diff h_I&\,K^{(0)}(g_I,h_I)\biggl[\tilde{\sigma}(h_I,\chi_{0})\left(1-\frac{\epsilon}{4}H_{2}\left(\sqrt{\frac{\epsilon}{2}}\pi_{0}\right)\frac{K^{(2)}(g_I,h_I)}{K^{(0)}(g_I,h_I)}\right)\\&\quad+i\sqrt{\frac{\epsilon}{2}}H_{1}\left(\sqrt{\frac{\epsilon}{2}}\pi_{0}\right)\tilde{\sigma}'(h_I,\chi_{0})-\frac{\epsilon}{4}H_{2}\left(\sqrt{\frac{\epsilon}{2}}\pi_{0}\right)\sigma''(h_I,\chi_{0})\biggr].
\end{align*}

Notice that the truncation at order $2n+m=2$ might not be entirely understood as a truncation in powers of $\epsilon$. In fact, the features of the weight function $I_{2n+m}$ depend on $\epsilon$ and on $\pi_{0}$ as well, so it might well be that, in some regimes, this truncation is not allowed. However, as discussed in Appendix \ref{app:approximations}, in the case of $\pi_0\epsilon <1$, such a truncation is possible. 

 The same computation can of course be performed in the spin representation. After imposition of isotropy, one finds that the reduced wavefunction $\tilde{\sigma}_j$ obeys the following equation of motion:
\begin{equation}\label{eqn:condensateequationofmotion}
\tilde{\sigma}_{j}''(\chi_{0})-2i\tilde{\pi}_{0}\tilde{\sigma}'_j(\chi_{0})-E_{j}^2\tilde{\sigma}(\chi_{0})=0\,,
\end{equation}
where we have defined the parameters $\tilde{\pi}_0$ and $E_j^2$ as
\begin{subequations}
\begin{align}
\tilde{\pi}_{0}&=\frac{\pi_{0}}{\epsilon\pi_{0}^2-1}\,,\\
E_{j}^2&=\epsilon^{-1}\frac{2}{\epsilon\pi_{0}^2-1}+\frac{B_{j}}{A_{j}}\,,
\end{align}
\end{subequations}
where the coefficients $A_{j}$ and $B_{j}$ and $w_{j}$ are defined in the same way as in \cite{Oriti:2016qtz}. 

\paragraph*{Comments on the effective dynamics.}
Let us briefly comment on this result.

First, the dynamics we have obtained is intended to describe an evolution with respect to the quantity $\chi_{0}$, which now truly is a parameter. Different values of $\chi_{0}$ at the effective level correspond to different spacetime slices, so that the evolution has an immediate physical interpretation. This is also what happens in the effective approach of \cite{Bojowald:2010xp, Bojowald:2010qw} already mentioned in Subsection \ref{subsec:definingeffectivedynamics}, where, through a deparametrization at the quantum phase space level, one obtains a relational dynamics in which evolution is specified with respect to the expectation value of the clock. 

In fact, the above equations of motion can be obtained from the effective action for the reduced wavefunction
\begin{equation}\label{eqn:effectiveaction}
S_{\text{eff}}=\sum_{j}\int\diff\chi_{0}\biggl[\overline{\tilde{\sigma}}_{j}(\chi_{0})\tilde{\sigma}''_{j}(\chi_{0})-2i\tilde{\pi}_{0}\overline{\tilde{\sigma}}_{j}(\chi_{0})\tilde{\sigma}'_{j}(\chi_{0})-E_{j}^2\overline{\tilde{\sigma}}_{j}(\chi_{0})\tilde{\sigma}_{j}(\chi_{0})\biggr]\,,
\end{equation}
where the role of $\chi_{0}$ as evolution parameter is manifest.

Second, the introduction of the \virgolette{synchronization parameter} $\epsilon$ allowed us not to make specific requirements on the properties of the reduced wavefunction. This is an important progress with respect to the way dynamics was introduced in \cite{Oriti:2016qtz}. In fact, one of the crucial conditions used to obtain the condensate dynamics was a \virgolette{hydrodynamic} one, requiring the condensate wavefunction to be slowly varying with respect to the relational time. This is what allowed the suppression of higher order derivatives (with respect to the scalar field variable) in the Taylor expansion of the wavefunction, and, in turn, of the dynamical equations (and kinetic kernel). However, this condition may not be satisfied by condensate wavefunctions that represent a cosmological spacetime satisfying the Friedmann equation at late (clock) times; this requires an exponential behavior for the volume operator, and thus for the wavefunction as well, as the computations in \cite{Oriti:2016qtz} explicitly show.

Third, let us notice that the quantity $E_{j}^2$ will play an important role for the semi-classical limit of the theory. For the moment, let us only say that, if $\epsilon\pi_{0}^2\ge 1$, the above quantity is positive, assuming that the ratio $B_{j}/A_{j}$ is either positive as well or smaller than the first term in the expression for $E_j^2$. 

Fourth, we remark that the effective dynamics obtained from \eqref{eqn:simplestschwinger} can not be straightforwardly mapped to the effective dynamics obtained from \eqref{eqn:schwingerpast}. The reason lies precisely in the choice of the states $\ket{\sigma_\epsilon,\chi_0,\pi_{0}}$ with respect to the general coherent states $\ket{\sigma}$, used in \cite{Oriti:2016qtz} and \cite{Assanioussi:2020hwf}. In fact, the choice of the condensate wavefunction \eqref{eqn:wavefunctioncps} is highly non-trivial, and rather specific. It localizes the condensate wavefunction itself as much as it is allowed by quantum-mechanical rules, but it is still not what would correspond to an exact localization. Therefore, it will not in general lead to just a \virgolette{smeared} version of the equations of motion obtained in \cite{Oriti:2016qtz}, as in \cite{Assanioussi:2020hwf} (see equation \eqref{eqn:functionaldynamics}). For example, parameters of the peaking function enter in a non-trivial way into the parameters regulating the dynamics, i.e., $\tilde{\pi}_{0}$ and $E^2_{j}$. Of course, this should be expected, since any notion of averaged effective dynamics deeply depends on the choice of states that are expected to realize it. Nonetheless, it is remarkable that the effective dynamics obtained above is very similar to the one obtained in \cite{Oriti:2016qtz}, which is based on a quite different choice of states. This similarity will be even more manifest when analysing the relational evolution of the averaged volume (see Subsection \ref{subsec:volumedynamics}).

Lastly, let us notice that computing the equations from the Schwinger-Dyson prescription and via the definition of an effective action as 
\begin{equation*}
S_{\text{eff}}=\int\diff\chi_{0}\left\langle\sigma_{\epsilon};\chi_{0},\pi_{0}\biggl\vert\hat{S}\biggr\vert\sigma_{\epsilon};\chi_{0},\pi_{0}\right\rangle,
\end{equation*}
one does not obtain exactly the same result. Still, as described in Appendix  \ref{app:dynamicseffective}, the dynamical equations obtained in these two ways are equivalent modulo the substitution $\epsilon\to 2\epsilon$ in going from Schwinger-Dyson to the effective action prescriptions. The mathematical reason why this happens is that in the second case we have one more peaking function, while in the first case one of them is eliminated by the variation. The physical reason is that the Schwinger-Dyson equation somehow \virgolette{perfectly localizes} the equations of motion on a given slice (i.e., value of $\chi_0$), while in the second procedure, the localization happens because of the CPS itself. Nonetheless, since there is no a priori determination of the parameter $\epsilon$, it seems fair to say that the two dynamics are equivalent (and so are all the results described below). Still, there is probably more to be understood about this difference.

\subsubsection{Analysis of the effective dynamics}
As already noticed in \cite{Oriti:2016qtz}, equation \eqref{eqn:condensateequationofmotion} is best studied by splitting the reduced wavefunction into a  phase and a modulus part, i.e.\ adopting a more conventional hydrodynamic form of the condensate dynamical equations.
Defining $\tilde{\sigma}_{j}\equiv \rho_{j}\exp[i\theta_{j}]$, we find
\begin{subequations}
\begin{align}\label{eqn:phaseequation}
0&=2\theta'_{j}(\chi_{0})\rho_{j}'(\chi_{0})+\rho_{j}(\chi_{0})\theta_{j}''(\chi_{0})-2\tilde{\pi}_{0}\rho_{j}'(\chi_{0})\,,\\
0&=\rho_{j}''(\chi_{0})-\rho_{j}(\chi_{0})[\theta_{j}'(\chi_{0})]^2+2\tilde{\pi}_{0}\rho_{j}(\chi_{0})\theta'_{j}(\chi_{0})-E_{j}^2\rho_{j}(\chi_{0})\label{eqn:modulusequation}\,.
\end{align}
\end{subequations}
Multiplying both sides of equation \eqref{eqn:phaseequation} by $\rho_{j}(\chi_{0})$, we find immediately that
\begin{equation}\label{eqn:phaseqj}
\theta_{j}'(\chi_{0})=\tilde{\pi}_{0}+\frac{Q_{j}}{\rho_{j}^2(\chi_{0})}\,,
\end{equation}
which in turn can be substituted in equation \eqref{eqn:modulusequation} to get
\begin{equation}\label{eqn:fundamentalequation}
\rho''_{j}(\chi_{0})-\frac{Q_{j}^2}{\rho_{j}^3(\chi_{0})}-\mu_{j}^2\rho_{j}(\chi_{0})=0\,,
\end{equation}
where
\begin{align}
\mu_{j}^2&=E_{j}^2-\tilde{\pi}_{0}^2=\frac{\pi_{0}^2}{\epsilon\pi_{0}^2-1}\left(\frac{2}{\epsilon\pi_{0}^2}-\frac{1}{\epsilon\pi_{0}^2-1}\right)+\frac{B_{j}}{A_{j}}\,.
\end{align}
Notice that in the regime $\epsilon\pi_0^2\gg 1$, the first term is always positive, which means that $\mu_j^2$ is positive as long as either $B_{j}/A_{j}$ is positive too, or it is less (in modulus) than the first term in the above expression (which is positive). As we will see in Subsection \ref{subsec:volumedynamics}, the positivity of $\mu_j^2$ is necessary in order to obtain a Friedmann-like behavior for the volume of the universe. 

It is remarkable that our improved procedure for extracting a relational cosmological dynamics from quantum gravity has produced equations of motion with the same functional form as obtained in \cite{Oriti:2016qtz}, but with redefined parameters and conserved quantities. These parameters and conserved quantities now carry a dependence on the properties of our relational quantum clock.

An explicit solution to equation \eqref{eqn:fundamentalequation} is given by \cite{toappear}
\begin{equation}\label{eqn:solutiontoappear}
\rho_{j}^2=-\frac{\mathcal{E}_{j}}{2\mu_{j}^2}+\frac{\sqrt{\mathcal{E}_{j}^2+4\mu_{j}^2Q_{j}^2}}{2\mu_{j}^2}\cosh\left(2\mu_{j}(\chi_{0}-\chi_{0}^j)\right)\,,
\end{equation}
where $\chi_{0}^j$ are integration constants and $\mathcal{E}_j$ are conserved quantities (see below). From both this explicit expression and from equation \eqref{eqn:fundamentalequation} we can see that the behavior of the reduced condensate wavefunction does not spoil the peaking properties of the condensate wavefunction induced by $\eta_\epsilon$.

Also, again from both equations \eqref{eqn:fundamentalequation} and \eqref{eqn:solutiontoappear}, we notice that if one of the $Q_{j}$s is different from zero, the associated $\rho_{j}$ is always different from zero. The value of $Q_{j}$ might therefore be important, \emph{at least at the mean field level}, to tell if the volume of the universe ever reaches zero. Indeed, as we already know from equation \eqref{eqn:expectationvaluev} and we will see again in Subsection \ref{subsec:volumedynamics}, in an effective relational dynamics framework the volume operator is made of a sum of $\rho_{j}^2$. This is an important factor for determining whether the classical big bang singularity is replaced by a bouncing scenario in GFT condensate cosmology. However, notice that even if all the $Q_j$s are identically zero, it is still possible to have an always strictly positive volume, as long as \emph{at least} one of the $\mathcal{E}_j$ is strictly negative. We will discuss further this point below, in Subsection \ref{subsec:volumedynamics}.
\paragraph*{Symmetries and conserved quantities.}
Before moving to the study of expectation values and variances of quantum operators, it is useful to recall which quantities are conserved by the above dynamics. 

We have two conserved quantities \cite{Oriti:2016qtz}. The first one, $Q_{j}$, entering in equation \eqref{eqn:phaseqj}, is the conserved charge related to a $U(1)$ symmetry of the effective action \eqref{eqn:effectiveaction}: $\tilde{\sigma}_{j}\to\tilde{\sigma}_{j}e^{i\alpha_{j}}$, with $\alpha_{j}$ constant; this symmetry is in general only approximate, subject to the GFT interactions being negligible. It is given by
\begin{align}
Q_{j}&=-\frac{1}{2}\left[\frac{\partial\mathcal{L}_{j}}{\partial\tilde{\sigma}'_{j}(\chi_{0})}i\tilde{\sigma}_{j}(\chi_{0})+\frac{\partial\mathcal{L}_{j}}{\partial\overline{\tilde{\sigma}}'_{j}(\chi_{0})}(-i\overline{\tilde{\sigma}}_{j}(\chi_{0}))\right]\nonumber\\
&=-\frac{i}{2}\left(\overline{\tilde{\sigma}}_{j}\tilde{\sigma}'_{j}-\overline{\tilde{\sigma}}'_{j}\tilde{\sigma}_{j}-2i\tilde{\pi}_{0}\vert\tilde{\sigma}_{j}\vert^2\right)=\rho_{j}^2(\theta_{j}'-\tilde{\pi}_{0})\,.
\end{align}
The second conserved charge is obtained by multiplying equation \eqref{eqn:fundamentalequation} by $\rho'_{j}$: it is the \virgolette{bulk condensate energy}
\begin{equation}\label{eqn:bulkenergy}
\mathcal{E}_{j}=(\rho_{j}')^2+\frac{Q_{j}^2}{\rho_{j}^2}-\mu_{j}^2\rho_{j}^2\,.
\end{equation}
This, however, is not exactly the charge generating translations of the reduced wavefunction $\tilde{\sigma}_{j}$ along the effective \virgolette{time} direction $\chi_{0}$.
The latter is given by
\begin{equation}
\bar{\mathcal{E}}_{j}=-\left(\frac{\partial\mathcal{L}_{j}}{\partial\tilde{\sigma}'_{j}}\tilde{\sigma}'_{j}+\frac{\partial\mathcal{L}_{j}}{\partial\overline{\tilde{\sigma}}_{j}'}\overline{\tilde{\sigma}}'_{j}-\mathcal{L}_{j}\right)=\vert\tilde{\sigma}'_{j}\vert^2-E_{j}^2\vert\tilde{\sigma}_{j}\vert^2=\mathcal{E}_{j}+2Q_{j}\tilde{\pi}_{0}\,,
 \end{equation} 
and it represents the total \virgolette{relational energy} of the condensate. This quantity is, of course, also conserved. And we see from the above equation that such energy consists of two terms: a \virgolette{bulk condensate energy} $\mathcal{E}_{j}$, and a term given by $2\tilde{\pi}_{0}Q_{j}$, which is the result of a sort of  \virgolette{energy injection} due to the precise choice of the peaking function \eqref{eqn:peakingfunction}. 
\subsection{Simplicial interactions}\label{app:simplicialinteractions}
Even though in the main discussion below we will neglect interactions, in this subsection we briefly comment about the role of interaction terms in the above model, focusing in particular on simplicial ones. As we will see, the strength of these interactions grows very quickly as the number of particle grows, and therefore, at a certain point of the evolution of the Universe, these interactions can become important, as already emphasised in \cite{Oriti:2016qtz} and studied, in a more phenomenological approach, in \cite{deCesare:2016rsf,Pithis:2016wzf,Pithis:2016cxg}. Simplicial interactions are interpreted as five tetrahedra (i.e.\ five GFT quanta) gluing to form (the boundary of) one $4$-simplex. Crucial for this interpretation is their non-locality in the group variables, which are connected by the interaction kernel with the same combinatorial pattern of shred triangles in the gluing of the five tetrahedra \cite{Oriti:2013aqa, Krajewski:2012aw}. The requirement that the GFT Feynman amplitudes take the form of nice lattice path-integrals for gravity coupled to a discretized scalar field, on the other hand, imposes a local dependence of the interaction kernel on $\chi$. In fact, as we will see below, crucial differences appear also in the effective dynamics if we impose from the beginning locality\footnote{Here, by locality, we mean that the value of the scalar field is the same in all of the (five) interacting fields appearing in the interaction term of the Lagrangian.} of interactions with respect to the massless scalar field variable $\chi$ or if we do not make any particular assumption on them.
\paragraph*{Local interactions.}
In \cite{Oriti:2016qtz}, it has been considered a local interaction term of the form
\begin{equation}
U_{\text{loc}}=\int\diff\chi\int\left(\prod_{a=1}^5\diff g_I^a\right)\mathcal{U}_{\text{loc}}(g_I^1,\dots,g_I^5)\prod_{a=1}^5\varphi(g_I^a,\chi)\,,
\end{equation}
entering in the GFT action $S=K+U_{\text{loc}}+\bar{U}_{\text{loc}}$, with $K$ a kinetic term. The form of the above interaction term, as said, is suggested by the discrete gravity interpretation of the GFT amplitudes. The fact that the interaction kernel $\mathcal{U}_{\text{loc}}$ does not depend on $\chi$ is due to the fact that we are considering a massless free scalar field, and that we are assuming the shift symmetry of its classical action to be conserved at the quantum level as well \cite{Oriti:2016qtz}.

As in Subsection \ref{subsec:dynamics}, the mean field equations of motion can be obtained from \eqref{eqn:simplestschwinger},
\begin{equation*}
\left\langle\frac{\delta S[\hat{\varphi},\hat{\varphi}^\dagger]}{\delta \hat{\varphi}^\dagger(g_I,\chi_{0})}\right\rangle_{\sigma_{\epsilon};\chi_{0},\pi_{0}}=0\,,
\end{equation*}
and the interaction term contribution is given by
\begin{equation*}
\mathcal{N}_{\epsilon}^4\int\left(\prod_{a=1}^4\diff g_I^a\right)\bigl[\overline{\mathcal{U}}_{\text{loc}}(g_I,g_I^1,\dots,g_I^4)+\dots+\overline{\mathcal{U}}_{\text{loc}}(g_I^1,\dots,g_I^4,g_I)\bigr]\prod_{a=1}^4\tilde{\sigma}(g_I^a,\chi_{0})\,.
\end{equation*}
When we also impose isotropy of the wavefunction, the effective equation \eqref{eqn:condensateequationofmotion} becomes
\begin{equation*}
\tilde{\sigma}_{j}''(\chi_{0})-2i\tilde{\pi}_{0}\tilde{\sigma}'(\chi_{0})-E_{j}^2\tilde{\sigma}(\chi_{0})-u_j^{\text{loc}}\overline{\tilde{\sigma}}^4(\chi_0)=0\,,
\end{equation*}
where
\begin{equation*}
    u_j^{\text{loc}}=\frac{\mathcal{N}_\epsilon^3}{\sqrt{2\pi\epsilon}}e^{\pi_0^2\epsilon/2}\frac{w_j^{\text{loc}}}{A_j}\,,
\end{equation*}
where the coefficients $w_{j}^{\text{loc}}$ are defined as in \cite{Oriti:2016qtz}. 

Both the factors $\mathcal{N}_{\epsilon}^3/(\sqrt{2\pi\epsilon})$ and $e^{\pi_0^2\epsilon/2}$ are indeed large when the conditions \eqref{eqn:goodclocksmallvariance} are satisfied, so the validity of the assumption of negligible interactions with respect to the kinetic term depends crucially on the parameters of the fundamental GFT model, i.e., $A_j$ and $w_j^{\text{loc}}$ (and thus, generally speaking, on choice of the kinetic and interaction kernels). 

Let us notice that including interactions results in different coefficients governing the dynamics depending on whether one follows equation \eqref{eqn:simplestschwinger} or the \virgolette{effective action procedure} described in Appendix \ref{app:dynamicseffective} to obtain it. In fact, a quick computation shows that in this second case, the coefficient of the interaction term is given, instead, by
\begin{equation*}
    \tilde{w}_j^{\text{loc}}=\frac{\mathcal{N}_\epsilon^2}{\sqrt{5}}e^{-3\pi_0^2\epsilon/2}\frac{w_j^{\text{loc}}}{A_j}\,.
\end{equation*}
From this formula we see that the different exponential dependence might indeed allow for a regime where interactions are under better control. This is not actually surprising, since the two ways of obtaining dynamics have very different features. As we have already mentioned, the one involving the use of equation \eqref{eqn:simplestschwinger}, basically localizes the dynamics \virgolette{on a given slice} via the functional derivative and subsequently via the expectation value on the appropriate CPS. On the other hand, the effective action procedure lets the localization happen via the projection on the states themselves. Therefore, for interactions whose localization in the variable $\chi$ has been imposed beforehand, the first procedure will result in more singular contributions, while the second one will be more regular.  
\paragraph*{Non-local interactions.}
Interestingly (but given the above remark, not surprisingly), non-local interactions, even though not very well motivated from the discrete gravity (and scalar field) point of view, can be kept under control more easily. To be concrete, let us consider a model of the form
\begin{equation}
 U_{\text{non-loc}}=\int\left(\prod_{a=1}^5\diff g_I^a\diff \chi^a\right)\mathcal{U}_{\text{non-loc}}(g_I^1,\dots,g_I^5,\chi^1,\dots,\chi^5)\prod_{a=1}^5\varphi(g_I^a,\chi^a)\,,
 \end{equation}
which is the most general possibility we can consider for the case of simplicial interactions. Its contribution to the dynamical equations would be of the form
\begin{align*}
e^{-2\pi_{0}^2\epsilon}\mathcal{N}_{\epsilon}^4(2\pi\epsilon)^2\biggl[\int\left(\prod_{a=1}^4\diff h_I^a\right)\bigl[&\overline{\mathcal{U}}_{\text{non-loc}}(g_I,h_I^1,\dots,h_I^4,\chi_{0})\\
&+\dots+\overline{\mathcal{U}}_{\text{non-loc}}(h_I^1,\dots,h_I^4,g_I,\chi_{0})\bigr]\prod_{a=1}^4\tilde{\sigma}(h_I^a,\chi_{0})\biggr]\,,
\end{align*}
where again higher order contributions have been neglected. This quantity can be well behaved if the kernel $\overline{\mathcal{U}}_{\text{non-loc}}$ is a regular function of its variables. In particular, in the isotropic limit, equation \eqref{eqn:condensateequationofmotion} becomes, after the introduction of interactions,
\begin{equation*}
\tilde{\sigma}_{j}''(\chi_{0})-2i\tilde{\pi}_{0}\tilde{\sigma}'(\chi_{0})-E_{j}^2\tilde{\sigma}(\chi_{0})-u_j^{\text{non-loc}}\overline{\tilde{\sigma}}^4(\chi_0)=0\,,
\end{equation*}
where
\begin{equation*}
    u^{\text{non-loc}}_{j}=\mathcal{N}_{\epsilon}^{3}(2\pi\epsilon)^{3/2}e^{-3\pi_{0}^2\epsilon/2}\frac{w^{\text{non-loc}}_{j}}{A_{j}}\,,
\end{equation*}
where again the quantities $w^{\text{non-loc}}_{j}$ are defined as in \cite{Oriti:2016qtz}. Interactions of this form can indeed be small (even with the presence of $\mathcal{N}_{\epsilon}^4$ and regardless of the precise functional form of the interaction kernel), provided that the factor $\pi_{0}^2\epsilon$ is large enough.

\subsection{Validity of the averaged dynamics conditions}\label{subsec:validityaverageddynamics}
From our defining equations \eqref{eqn:averagedrelationaldynamics}, three operators play a crucial role in the definition of the averaged relational evolution: the internal degree of freedom chosen as relational clock $\hat{\chi}$, its conjugate momentum $\hat{\Pi}$ and the relational Hamiltonian operator $\hat{H}$. We will define this Hamiltonian operator later in this section. We now focus on these three operators, as defined in our GFT condensate cosmology context, making clear in which sense they realize an averaged relational evolution as defined in Subsection \ref{subsec:definingeffectivedynamics}. Notice, however, that while in order to meet all the \virgolette{averaged relational evolution conditions} described in Subsection \ref{subsec:definingeffectivedynamics} one would need in principle to compute \emph{all} the moments of $\hat{\Pi}$ and $\hat{H}$ on CPSs, here we will content ourselves to check only the validity of equation \eqref{eqn:piequalh}. We stress again that this condition is actually enough to guarantee an effective approximate equality between $\hat{H}$ and $\hat{\Pi}$ in the regime in which relative quantum fluctuations of these operators are negligibly small.  

Since in the following we will need to compute expectation values of these operators on CPSs, it is useful to sketch the typical computation for a generic (non-derivative) two-body operator:
\begin{align*}
\braket{\hat{O}}_{\sigma_{\epsilon};\chi_0,\pi_{0}}&=\int\diff g_I\diff h_I\int \diff \chi\,O(g_I,h_I;\chi)\vert
\sigma_{\tilde{\epsilon}}(g_I,\chi;\chi_0,\pi_0)\vert^2\\&=\!\!\int\!\!\diff g_I\diff h_I\!\!\int \!\!\diff \chi\,O(g_I,h_I;\chi)\vert
\tilde{\sigma}(g_I,\chi)\vert^2\vert\eta_{\epsilon}(\chi-\chi_0,\pi_0)\vert^2\\
&\simeq \int\diff g_I\diff h_I\,O(g_I,h_I;\chi_0)\overline{\tilde{\sigma}}(g_I,\chi_0)\tilde{\sigma}(h_I,\chi_0)\,,
\end{align*}
where in the last line we have used the lowest saddle point approximation (allowed by the form of the $\eta_{\epsilon}$ function in the scalar field variables), and resting on the \virgolette{good clock condition} $\epsilon\ll 1$. This approximation will be used to compute the expectation values on CPSs of all the relevant operators below, and its validity is discussed in Appendix \ref{app:approximations} and in much more detail in \cite{toappear}. Notice that the normalization of the peaking function has been chosen such that
\begin{equation}
\int\diff\chi\, \overline{\eta}_{\epsilon}(\chi-\chi_0,\pi_{0})\eta_{\epsilon}(\chi-\chi_0,\pi_{0})=1\,,
\end{equation}
which, in our case, implies $\mathcal{N}^2_{\epsilon}=(\pi\epsilon)^{-1/2}$. For instance, for the number operator, we have
\begin{align}
N(\chi_{0})&\equiv\braket{\hat{N}}_{\sigma;\chi_{0},\pi_{0}}=\sum_{j}\int\diff\chi \vert\sigma_{\epsilon}(g_I,\chi;\chi_{0},\pi_{0})\vert^2\simeq \sum_{j}\rho_{j}^2(\chi_{0})\,.
\end{align}
\paragraph*{Massless scalar field operator.}
As we have already explained in Subsection \ref{subsec:protogeometrictime}, the role of $\hat{\chi}$ in equation \eqref{eqn:averagedrelationaldynamics} can be taken by the operator $\hat{X}/N(\chi_0)$. Its expectation value is then given by
\begin{equation}\label{eqn:averagedchiint}
\braket{\hat{\chi}}_{\sigma_\epsilon;\chi_0,\pi_0}\equiv\frac{\braket{\hat{X}}_{\sigma;\chi_{0},\pi_{0}}}{N(\chi_{0})}\simeq \chi_{0}\,,
\end{equation}
where in the last line we have followed the same computations strategy outlined before. Let us remark, however, that in this case the lowest order saddle point approximation is not expected to hold true at any value of the parameter $\chi_0$. However, for $\pi_0$ large enough (or, equivalently, $\epsilon$ small enough), this approximation is reasonable even for \virgolette{small} values of $\chi_0$. We refer again to Appendix \ref{app:approximations} and to \cite{toappear} for a more thorough discussion. In conclusion, we see that the intensive quantity associated to the second quantized scalar field operator has an expectation value on a CPS which is indeed given by $\chi_{0}$.
\paragraph*{Momentum operator.}
The momentum operator, as defined in equation \eqref{eqn:momentumoperator}, is the conjugate variable to the massless scalar field operator, as defined by equation \eqref{eqn:scalarfieldoperator}. Indeed, their commutator gives 
\begin{equation}
    [\hat{\Pi},\hat{X}]=-i\hat{N}\,.
\end{equation}
The action of this operator on a CPS is given by
\begin{align*}
\left(\mathbb{I}+i\delta\chi\hat{\Pi}\right)\ket{\sigma_{\epsilon};\chi_{0},\pi_{0}}&=\biggl(1+\delta\chi\int\diff g_I\diff\chi\hat{\varphi}^\dagger(g_I,\chi)\partial_{\chi}\sigma_{\epsilon}(g_I,\chi;\chi_{0},\pi_{0})\biggr)\ket{\sigma_{\epsilon};\chi_{0},\pi_{0}}\\
&\simeq e^{-\Vert\sigma_{\epsilon}\Vert^2/2}\exp\left[\delta\chi\int\diff g_I\diff\chi\hat{\varphi}^\dagger(g_I,\chi)\partial_{\chi}\sigma_{\epsilon}(g_I,\chi;\chi_{0},\pi_{0})\right]\\&\quad\times\exp\left[\int\diff g_I\diff\chi\hat{\varphi}^\dagger(g_I,\chi)\sigma_{\epsilon}(g_I,\chi;\chi_{0},\pi_{0})\right]\ket{0}\,,
\end{align*}
so that, for each $\delta\chi$ small enough, we can approximately write the above quantity as
\begin{equation*}
\exp\left[\int\diff g_I\diff\chi\hat{\varphi}^\dagger(g_I,\chi)\sigma_{\epsilon}(g_I,\chi+\delta\chi;\chi_{0},\pi_{0})\right]\ket{0}\equiv \ket{\sigma_{\epsilon};\chi_{0},\delta\chi,\pi_{0}}.
\end{equation*}
From the above equation, we immediately see that the expectation value of the number operator on these new \virgolette{translated states} is given by
\begin{equation*}
\braket{\hat{N}}_{\sigma;\chi_{0},\delta\chi,\pi_{0}}=\int\diff g_I\diff\chi\, \rho_{\epsilon}^2(g_I,\chi+\delta\chi;\chi_{0},\pi_{0})=\int\diff g_I\diff\chi, \rho_{\epsilon}^2(g_I,\chi;\chi_{0},\pi_{0})=\braket{\hat{N}}_{\sigma;\chi_{0},\pi_{0}}\,.
\end{equation*}
The same result of course holds for the volume operator (and in the same fashion, for all those operators whose matrix elements do not depend on $\chi$). On the other hand, the scalar field operator satisfies
\begin{equation*}
\braket{\hat{X}}_{\sigma;\chi_{0},\delta\chi,\pi_{0}}=\int\diff g_I\diff\chi\,\chi\rho_{\epsilon}^2(g_I,\chi+\delta\chi;\chi_{0},\pi_{0})\simeq (\chi_{0}-\delta\chi)\braket{\hat{N}}_{\sigma;\chi_{0},\pi_{0}}\,,
\end{equation*}
so that the label associated to the leaf, i.e.\ $\braket{\hat{X}}_{\sigma;\chi_{0},\delta\chi,\pi_{0}}/\braket{\hat{N}}_{\sigma;\chi_{0},\pi_{0}}$, is not $\chi_{0}$, but $\chi_{0}-\delta\chi$. 

In conclusion, after the action of the momentum operator, all the physical properties of these states, except for their label (and of course all the functions of it), are not changed. This is not surprising, since the momentum operator commutes with $\hat{N}$ and $\hat{V}$. In fact, this is exactly the action we would expect from the (exponential of the) operator representing the momentum of the massless scalar field, changing the massless scalar field quantum number and leaving unchanged all the other quantum numbers associated to the other operators. 

Its expectation value on a CPS is given by
\begin{align}\label{eqn:averagepi}
\braket{\hat{\Pi}}_{\sigma;\chi_{0},\pi_{0}}&=\frac{1}{i}\int\diff\chi\sum_{j}\overline{\sigma}_{\epsilon,j}(\chi;\chi_{0},\pi_{0})\partial_{\chi}\sigma_{\epsilon,j}(\chi;\chi_{0},\pi_{0})\nonumber\\
&=\sum_{j}\int\diff\chi \rho_{j}^2(\chi)(\theta'_j(\chi)+\pi_{0})\vert\eta_{\epsilon}(\chi-\chi_{0};\pi_{0})\vert^2\nonumber\\
&= \pi_{0}\left(\frac{1}{\epsilon\pi_{0}^2-1}+1\right)N(\chi_{0})+\sum_{j}Q_{j}\,,
\end{align}
where in the last line we have used the equations of motion.
We see immediately that there are two contributions to this operator: the first one, depending on the momentum parameter $\pi_0$  assigned to each tetrahedron, is proportional to the number of spacetime atoms at the relational time $\chi_{0}$, and the second one, intensive and independent of $\chi$, which is related to the $U(1)$ charge of the effective theory. Interestingly enough, in the regime $\epsilon\pi_{0}^2\gg 1$, which, we remind, is a necessary condition to maintain quantum fluctuations small at least in some regime \cite{toappear}, the extensive contribution above reduces simply to $\pi_{0}N(\chi_{0})$. 
\paragraph*{Relational Hamiltonian.}
The operator $\hat{\Pi}$, however, does not describe in general the evolution of our CPSs with respect to the parameter $\chi_0$, which, in virtue of equation \eqref{eqn:averagedchiint} it is what enters in the derivative in the left-hand-side of equation \eqref{eqn:averageddynamics}, and thus it is the parameter describing the averaged relational dynamics. 
To characterize the relational evolution enconded in a CPS $\ket{\sigma_\epsilon;\chi_0,\pi_0}$, on the other hand, we define\footnote{Let us stress that, since the very definition of $\hat{H}$ is subject to a prior choice of states encoding the relational dynamics (in this case the CPSs), this operator should only be intended as an \emph{effective} Hamiltonian operator. Thus, it would be better denoted as $\hat{H}_{\sigma_\epsilon;\chi_0,\pi_0}$. For sake of notation, however, we will drop the subscript in the main text.} a Hermitean operator $\hat{H}$
\begin{align}
\hat{H}&\equiv -i\biggl[\int\diff g_I\int\diff\chi\,\hat{\varphi}^\dagger(g_I,\chi)\partial_{\chi}\eta_{\epsilon}(\chi-\chi_{0},\pi_{0})\tilde{\sigma}(g_I,\chi)\nonumber\\
&\quad\qquad-\int\diff g_I\int\diff\chi\,\hat{\varphi}(g_I,\chi)\partial_{\chi}\overline{\eta}_{\epsilon}(\chi-\chi_{0},\pi_{0})\overline{\tilde{\sigma}}(g_I,\chi)\biggr]-\pi_0N(\chi_0)\,,
\end{align}
constructed so that its action on $\ket{\sigma_\epsilon;\chi_0,\pi_0}$ is given by
\begin{equation}\label{eqn:defh}
\hat{H}\ket{\sigma_{\epsilon};\chi_{0},\pi_{0}}\equiv -i\left(\frac{N'(\chi_{0})}{2}+\int\diff g_I\int\diff\chi\,\hat{\varphi}^\dagger(g_I,\chi)\partial_{\chi}\eta_{\epsilon}(\chi-\chi_{0},\pi_{0})\tilde{\sigma}(g_I,\chi)\right)\ket{\sigma_{\epsilon};\chi_{0},\pi_{0}}\,.
\end{equation}
It is easy to see that, in the limit of small $\epsilon$, the state resulting from the action of $\exp[i\delta\chi\hat{H}]$ is $\ket{\sigma_{\epsilon};\chi_{0}-\delta\chi,\pi_{0}}$. This Hamiltonian operator is not directly obtained from the (averaged) quantum equations of motion, but it is certainly related to them, since its definition is based on the form of the CPSs that are supposed to approximately satisfy them. Its interpretation as effective relational evolution operator then, is only to be intended in light of the validity of equations \eqref{eqn:averageddynamics} and \eqref{eqn:piequalh}, which we now discuss.

It follows from the above definition that the expectation values of all quantum operators are now computed at relational time $\chi_{0}-\delta\chi$, in the limit of small $\epsilon$. In particular, the operator $\hat{H}$ governs a Schr\"odinger equation of the form
\begin{equation}
-i\frac{\diff}{\diff\chi_{0}}\ket{\sigma_{\epsilon};\chi_{0},\pi_{0}}=\hat{H}\ket{\sigma_\epsilon,\chi_{0},\pi_{0}}\,.
\end{equation}
Of course this implies that equation \eqref{eqn:averageddynamics} is satisfied. In order to understand whether equation \eqref{eqn:piequalh} is satisfied as well, we compute the expectation value of $\hat{H}$ on a CPS.  Defining $\hat{\bar{H}}$ the operator whose action on the CPSs is given by the second term in the round brackets in equation \eqref{eqn:defh}, we get
\begin{align*}
\braket{\hat{\bar{H}}}_{\sigma_{\epsilon};\chi_{0},\pi_{0}}&=-i\int\diff g_I\int\diff \chi\,\overline{\eta}_{\epsilon}(\chi-\chi_{0},\pi_{0})\partial_{\chi}\eta_{\epsilon}(\chi-\chi_{0},\pi_{0})\vert\tilde{\sigma}(g_I,\chi_{0})\vert^2\nonumber\\
&=\pi_{0}\int\diff g_I\int\diff\chi\vert\eta_{\epsilon}(\chi-\chi_{0},\pi_{0})\vert^2\rho^2(g_I,\chi)\\&\quad-\frac{i}{2}\int\diff g_I\int\diff\chi\,\rho^2(g_I,\chi)\partial_{\chi}\vert\eta_{\epsilon}(\chi-\chi_{0},\pi_{0})\vert^2\\
&=\pi_{0}\int\diff g_I\int\diff\chi\vert\eta_{\epsilon}(\chi-\chi_{0},\pi_{0})\vert^2\rho^2(g_I,\chi)\nonumber\\
    &\quad+\frac{i}{2}\partial_{\chi_0}\int\diff g_I\int\diff\chi\,\rho^2(g_I,\chi)\vert\eta_{\epsilon}(\chi-\chi_{0},\pi_{0})\vert^2\nonumber .
\end{align*}
Recognizing the above integrals to correspond to $N(\chi_0)$, we obtain
\begin{align*}
\braket{\hat{\bar{H}}}_{\sigma_{\epsilon};\chi_{0},\pi_{0}}&= \pi_{0}N(\chi_{0})+\frac{i}{2}N'(\chi_{0})\,.
\end{align*}
In conclusion, we have, for the operator $\hat{H}$, the relation
\begin{equation}
\braket{\hat{H}}_{\sigma_{\epsilon};\chi_{0},\pi_{0}}= \braket{\hat{\bar{H}}}_{\sigma_{\epsilon};\chi_{0},\pi_{0}}-i\frac{N'(\chi_{0})}{2}= \pi_{0}N(\chi_{0})\,.
\end{equation} 
Therefore, we see that equation \eqref{eqn:piequalh} is indeed approximately satisfied in the regime $\epsilon\pi_{0}^2\gg 1$ if we also impose
\begin{equation}\label{eqn:sumzeroq}
\sum_{j}Q_{j}=0\,,
\end{equation}
which however can be imposed without losing generality, since the $Q_{j}$s are just constants of integration. Still, it is important to notice that for large enough $N(\chi_0)$ equation \eqref{eqn:piequalh} is approximately satisfied regardless of the imposition of condition \eqref{eqn:sumzeroq}.

Thus we conclude that our CPSs indeed satisfy also the requirement \eqref{eqn:piequalh}, thus leading to a satisfying implementation of the \virgolette{averaged relational evolution conditions} of Subsection \ref{subsec:definingeffectivedynamics} (at least when fluctuations in $\hat{\Pi}$ and $\hat{H}$ are small).  

\subsection{Volume dynamics}\label{subsec:volumedynamics}
We now study the average effective relational evolution of the volume operator. Given our restriction to homogeneous and isotropic states, this effective relational evolution encodes the cosmological dynamics emergent from our fundamental quantum gravity formalism.
\paragraph*{Expectation value of the volume operator.}
Using the same techniques explained in Subsection \ref{sec:previousdynamics}, in particular equations \eqref{eqn:volumematrixelements} and \eqref{eqn:expectationvaluev}, we compute the expectation value of the volume operator on a CPS:
\begin{align}\label{eqn:expectationvaluev}
V(\chi_{0})&\equiv\braket{\hat{V}}_{\sigma;\chi_{0},\pi_{0}}=\sum_{j,\vec{m}}V_j\vert\sigma_{\{j,\vec{m}\}}(\chi;\chi_{0},\pi_{0})\vert^2=\sum_j V_j\vert\sigma_j(\chi;\chi_{0},\pi_{0})\vert^2\nonumber\\&\simeq \sum_j V_j\rho_j^2(\chi_{0})\,.
\end{align}
Once again, we have used a lowest order saddle point approximation, whose validity is discussed in Appendix \ref{app:approximations} and in more detail in \cite{toappear}. We clearly see the similarity of this equation with equation \eqref{eqn:expectationvaluevold}, leading again to the interpretation of the total volume being given by the sum over $j$ of the average number of \virgolette{isotropic atoms} with assigned spin $j$ \virgolette{at a time $\chi_0$} weighted by their individual volume contribution $V_j$.
\paragraph*{Effective relational cosmological dynamics.}
By deriving equation \eqref{eqn:expectationvaluev} and using equation \eqref{eqn:bulkenergy}, we see that 
\begin{subequations}\label{eqn:newfriedmannrelational}
\begin{align}\label{eqn:firstfriedmannrelational}
\left(\frac{V'}{3V}\right)^2&\simeq \left(\frac{2\sum_{j}V_{j}\rho_{j}\text{sgn}(\rho'_{j})\sqrt{\mathcal{E}_{j}-Q_{j}^2/\rho_{j}^2+\mu_{j}^2\rho_{j}^2}}{3\sum_{j}V_{j}\rho_{j}^2}\right)^2,\\
\frac{V''}{V}&\simeq\frac{2\sum_{j}V_{j}\left[\mathcal{E}_{j}+2\mu_{j}^2\rho_{j}^2\right]}{\sum_{j}V_{j}\rho_{j}^2}\,.
\end{align}
\end{subequations}
These are the effective cosmological equations for the GFT condensate in terms of the relational time $\chi_{0}$. Remarkably enough, they have the same functional form as the equations \eqref{eqn:oldgeneralizedfriedmann} obtained in \cite{Oriti:2016qtz}, though this time some of the coefficients in the equations depend on the CPS parameters, which are in fact part of the definition of our quantum relational clock. For instance, $\mu_j^2$ carries now a dependance on both $\epsilon$ and $\pi_0$. 
\paragraph*{Classical limit.}
We can easily check that equations \eqref{eqn:newfriedmannrelational} reproduce the expected classical limit for small energy densities.
Along the same lines as in Subsection \ref{sec:previousdynamics}, in the limit $\rho_j^2\gg\vert \mathcal{E}_j\vert/m_j^2$ and $\rho_j^4\gg Q_j^2/m_j^2$, the above equations become
\begin{subequations}
\begin{align}
\left(\frac{V'}{3V}\right)^2&\simeq \left(\frac{2\sum_{j}V_{j}\mu_{j}\rho^2_{j}\text{sgn}(\rho'_{j})}{3\sum_{j}V_{j}\rho_{j}^2}\right)^2,
\\
\frac{V''}{V}&\simeq\frac{4\sum_{j}V_{j}\left[\mu_{j}^2\rho_{j}^2\right]}{\sum_{j}V_{j}\rho_{j}^2}\,.
\end{align}
\end{subequations}
A sufficient (but not necessary) condition for the above approximate equations to coincide with the Friedmann equations (in relational time) is either that all the $\mu_{j}^2$s are equal to $3\pi \tilde{G}$, where $\tilde{G}\equiv GM^2$ is the dimensionless Newton's gravitational constant, or even just that one of the $j$s is dominating, say $j_{o}$, and its characterized by $\mu_{j_{o}}^2=3\pi \tilde{G}$ \cite{Wilson-Ewing:2018mrp, Oriti:2016qtz}.
Notice that this would amount to a \emph{definition} of the Newton's constant, from the fundamental parameters and dynamics of the quantum gravity theory. Interestingly, among the parameters conspiring to the definition of the Newton's constant, we find both $\pi_0$ and $\epsilon$, which are directly related to the \virgolette{bona fide slice properties} of our CPSs, and to the quantum properties of our relational clock. In this sense, we find an interesting hint of a connection between the relational dynamics, and the choice of quantum clock defining it, and the emergent classical gravitational physics. 
This connection, and the dependence of the effective gravitational coupling from the properties of the chosen quantum clock, are certainly worth exploring further.

\

\paragraph*{Bounce.}
Analogously to the framework of \cite{Oriti:2016qtz}, also in our improved relational cosmological dynamics we have that, if at least one of the $Q_{j}$s is not zero, or at least one of the $\mathcal{E}_j$ is strictly negative, then the expectation value of the volume operator never vanishes. This would lead to a bouncing scenario replacing the cosmological big bang singularity, in the very early universe. 

However, there is a key difference with respect to \cite{Oriti:2016qtz}. In that case, the sum of the $Q_j$s was equal to the expectation value of the \virgolette{relational massless scalar field momentum}. The latter could not vanish, for physical reasons, since it would make the whole relational setting unjustified (with no matter energy density, one would expect a flat or constantly curved spacetime). 

In this case there seems to be no physical obstruction to requiring that sum to be zero. In fact, it is reasonable to actually require the condition \eqref{eqn:sumzeroq}, since in this framework it has to be imposed in order to have fully coherent relational dynamics even at \virgolette{early times} (i.e.\ not arbitrarily large $N(\chi_0)$)\footnote{In particular, notice that, in the specific case of a single-spin scenario, the constraint \eqref{eqn:sumzeroq} implies that the single remaining $Q_{j_{o}}$ has to vanish.}. As a consequence, there might be an interplay between the requirement of having a bounce at early times and the condition that the momentum of the scalar field used as a clock behaves as a good relational Hamiltonian. The dependence of the resolution of the initial singularity on the properties of the clock used to define evolution has been also highlighted in \cite{Gielen:2020abd}.

As a conclusion, while in \cite{Oriti:2016qtz} the bounce appeared as a fully general result of the volume dynamics, in this improved relational framework the presence of a bounce depends on the integration constants $\mathcal{E}_j$ and $Q_j$, meaning that in this context there is no necessary reason to select a bouncing solution, although it remains rather generic.

In addition, we remark that such a bounce, were it to be present in the chosen solution, would be in any case only an average result. That is, it would be a feature of the dynamics of the \emph{mean value} of the volume operator in the chosen state. In order to give a more solid ground for its physical interpretation, one has to check for the behaviour of quantum fluctuations in the same regime of the effective dynamics. Leaving a detailed analysis for \cite{toappear}, one can already expect that the dynamics of mean values is reliable only in the regime in which $N(\chi_0)\gg 1$  (see equations \eqref{eqn:semi-classicality}) i.e., until the number of GFT \virgolette{atoms of space} is large. It is not obvious that this would be the case in the very early universe\footnote{Notice that it may be possible that in this very same regime also quantum fluctuations of $\hat{\Pi}$ and $\hat{H}$ are important. If that is the case, one should check the equality of all of their moments on a given CPS in order to establish the validity of a consistent relational dynamics framework.}. Finally, all the results obtained so far heavily rely on a lowest order saddle point approximation (i.e., almost \virgolette{perfect peaking} of the condensate wavefunction). When these two conditions (small fluctuations of relevant operators and almost perfect peaking) are not satisfied, we might lose the ability to interpret $\chi_0$ as a relational parameter (possibly because fluctuations on the massless scalar field operator are large or because the expectation value of the massless scalar field is not $\chi_0$ or both) and the expectation value of $\hat{V}$ might not be able to capture the relevant features of the volume anymore. 

A careful analysis of these issues has be performed in \cite{toappear}, with a particular focus to the bounce and the classical regime discussed above. 

\

\paragraph*{Single-spin scenario.} 
The special case in which all the $\rho_j$s are identically zero except for a non-zero $\rho_{j_o}$ is interesting for three main reasons: first, it was shown to reproduce the effective dynamics of Loop Quantum Cosmology (up to a term that could be fixed to zero as a choice on the relevant class of solution); second, the dominance of single-spin configurations has been shown to arise dynamically in several analyses of the GFT condensate dynamics \cite{Gielen:2016uft,Pithis:2016wzf,Pithis:2016cxg}; third, it is obviously a technical simplification allowing to push much further the analysis of the emergent cosmological dynamics, in particular when including the effect of GFT interactions. 

This case can immediately be obtained from equations \eqref{eqn:newfriedmannrelational}. Similarly to equations \eqref{eqn:oldsinglespin}, we fin
\begin{subequations}\label{eqn:newsinglespin}
\begin{align}
    \left[\frac{ V'}{3V}\right]^2&=\frac{4\pi \tilde{G}}{3}-\frac{4V_{j_o}^2Q_{j_o}^2}{9V^2}+\frac{4 V_{j_o}\mathcal{E}_{j_o}}{9V}\,,\\
    \frac{V''}{V}&=12\pi \tilde{G}+\frac{2V_{j_o}\mathcal{E}_{j_o}}{V}\,.
\end{align}
\end{subequations}
The first of these two equations can be recast as \begin{equation}\label{eqn:singlespinlqc}
    \left[\frac{ V'}{3V}\right]^2=\frac{4\pi \tilde{G}}{3}\left(1-\frac{\rho}{\rho_c}\right)-\frac{\pi_0^2}{2V_{j_o}^2}+\frac{1}{V}\left(\frac{4V_{j_o}\mathcal{E}_{j_o}}{9}-\pi_0Q_{j_o}\right),
\end{equation}
where we have defined
\begin{equation}\label{eqn:rhoc}
    \rho\equiv \frac{\braket{\hat{\Pi}}_{\sigma_\epsilon;\chi_0,\pi_0}^2}{2V^2}\,,\qquad\rho_c\equiv \frac{3\pi\tilde{G}}{2V_{j_o}^2}\,.
\end{equation}
Equation \eqref{eqn:singlespinlqc} resembles the effective Friedmann dynamics of LQC \cite{Taveras:2008ke}, with two additional contributions: a constant one, and one scaling as $V^{-1}$. 

They lead to a further modification of the Friedmann dynamics and may possibly have an interpretation in terms of effective matter or geometry contributions. 

Notice, however, that if we impose the condition $\sum_j Q_j=0$, which in this case translates into $Q_{j_o}=0$, and which we have seen is required for a fully coherent relational interpretation of the cosmological dynamics, the first equation becomes 
\begin{equation*}
    \left[\frac{V'}{3V}\right]^2=\frac{4\pi \tilde{G}}{3}+\frac{4 V_{j_o}\mathcal{E}_{j_o}}{9V}\,,
\end{equation*}
which is different from equation \eqref{eqn:oldisinglespin1}, as well as from the effective LQC dynamics. The reason, indeed, lies in the different role of the constants $Q_j$ with respect to \cite{Oriti:2016qtz}, due to the fact that equation \eqref{eqn:momentumq} for the scalar field momentum is, in this framework, substituted by equation \eqref{eqn:averagepi}. When the condition is imposed, thus, the bounce implied by the LQC dynamics disappears. However, it might still be possible to have a bouncing solution, when $\mathcal{E}_{j_o}<0$, though it would be implemented via a very different physical mechanism. 

Also on this point, a deeper analysis of the effective cosmological dynamics, and of the physical meaning of the various conserved charges associated to it, is needed.  

\

\paragraph*{On the Hamiltonian and the momentum.}
Even though the averaged relational dynamics yields the correct classical limit for the relational evolution of the volume operator,  it is interesting to check if a self-consistent classical description of the effective dynamical system represented by our cosmological observables can be constructed, in the late times regime. How to construct such a description from the full quantum theory is, however, not entirely clear (see Appendix \ref{app:classical} for a review of the dynamics and the Hamiltonian analysis of a flat FRW spacetime in the harmonic gauge where the massless scalar field is used as a clock). In fact, notice that $\braket{\hat{H}}_{\sigma_\epsilon,\chi_{0},\pi_{0}}$ retains a $\chi_{0}$-dependence from the factor $N(\chi_{0})$. This has twofold consequences.

First, it creates a tension if one wants to apply equation \eqref{eqn:averageddynamics} to the expectation value of $\hat{H}$ itself. In fact, the right-hand-side would give precisely zero, while the left-hand-side is non-zero because $N(\chi_{0})$ depends on $\chi_{0}$. However, let us recall that, in standard Quantum Mechanics, the Ehrenfest theorem (to which equation \eqref{eqn:averageddynamics} is inspired) includes a term of the form $\braket{\partial\hat{O}/\partial t}_{\Psi}$ which becomes relevant when the operator $\hat{O}$ depends explicitly on time. Such a term was not included in equation \eqref{eqn:averageddynamics}, since that equation is expected to hold for geometric observables, which in a clock neutral picture should by definition not depend on clock variables. On the other hand, in the spirit of obtaining an evolution equation for the (expectation value of) the Hamiltonian operator, such a term might end up to be needed. Even though we obviously can not directly include a term of the form $\braket{\partial\hat{O}/\partial t}_{\Psi}$ in equation \eqref{eqn:averageddynamics} because we do not have a notion of (effective) relational time before taking the expectation value on an appropriate class of states, by construction we are choosing the massless scalar field as a relational clock. So, one could replace the term $\braket{\partial\hat{O}/\partial t}_{\Psi}$ with $\braket{[\hat{O},\Pi]}_{\Psi}$ in our clock-netrual framework, which indeed realizes a \virgolette{derivative} of the operator $\hat{O}$ with respect to $\hat{\chi}$ once the former is expanded in a formal power series in the latter. With such a modification, equation \eqref{eqn:averageddynamics} would read 
\begin{equation}\label{eqn:modifiedehrenfest}
i\frac{\diff}{\diff\braket{\hat{\chi}}_{\Psi}}\braket{\hat{O}_{\alpha}}_{\Psi}=\braket{[\hat{H},\hat{O}_{\alpha}]}_{\Psi}+\braket{[\hat{O}_{a},\hat{\Pi}]}_{\Psi}\,,
\end{equation}
where now we are including $\hat{H}$ among the set of operators (labelled by $\alpha$, and formerly only including geometric operators) for which the above equation is expected to hold. While for any geometric operator $\hat{O}_a$, the above equation reduces to equation \eqref{eqn:averageddynamics} in virtue of $[\hat{O}_a,\Pi]=0$, the dynamical equation for $\hat{H}$ would now read
\begin{equation*}
i\frac{\diff}{\diff\braket{\hat{\chi}}_{\Psi}}\braket{\hat{H}}_{\Psi}=\braket{[\hat{H},\hat{\Pi}]}_{\Psi}=i\frac{\diff}{\diff\braket{\hat{\chi}}_{\Psi}}\braket{\hat{\Pi}}_{\Psi}\,,
\end{equation*}
the last identity following from the assumed identification of all the moments of $\hat{\Pi}$ and $\hat{H}$ on $\ket{\Psi}$. Notice, in particular, that this implies that equation \eqref{eqn:modifiedehrenfest} applies to $\hat{\Pi}$ as well. So, this realizes a self consistent extension of the framework proposed in equation \eqref{eqn:averageddynamics}.

Second, the non-constancy with respect to $\chi_0$ of the expectation value of the Hamiltonian implies that the dynamical trajectories generated by $\hat{H}$ through an averaged quantum evolution, and which effectively produce a classical Friedmann dynamics for the volume operator at late times, are not the same that are generated by the average Hamiltonian itself\footnote{Interestingly, similar problems appear also when comparing the effective dynamics in LQC with the quantum one obtained (with currently available tools) from the full LQG \cite{Dapor:2019mil}.}. Phrased differently, the effective Hamiltonian generating the classical Friedmann dynamics at late times is not the expectation value of the quantum Hamiltonian\footnote{\label{footnote:momentum}Of course, by construction, the same kind of issue appears for the momentum of the massless scalar field. However, in a truly relational framework, clock variables should not be accessible. We thus prefer formulating the problem only in terms of $\hat{H}$.} $\hat{H}$. Such a difference between the expectation value of the quantum Hamiltonian and the classical one is encoded in the $N(\chi_0)$ dependency of the former which, being $N(\chi_0)$ related to the miscroscopic nature of spacetime and geometry, has no straightforward classical counterpart in General Relativity. From a simplicial perspective, one can think of $N$ as the number of sites of a lattice covering a given homogeneous patch \cite{Bojowald:2008ik}, i.e., $N=V_0/\ell_0^3$, where $\ell_0$ is a coordinate length and $V_0$ is the coordinate volume of the homogeneous patch under consideration. From this point of view, having a variable $N$ means performing a \virgolette{lattice refinement}\footnote{Comparing again the situation to the LQC case, it has been  shown that the lattice refinement has non-trivial consequences on the matter sector as well \cite{Nelson:2007um}.} (or, equivalently, a \virgolette{running} of $V_0$ \cite{Bojowald:2020wuc}). 

Even though it is possible that this mismatch between the classical and the effective Hamiltonian (and, of course, between the classical and the effective massless scalar field momentum, see footnote \ref{footnote:momentum}), is unavoidable and deeply connected to the granularity of spacetime in the GFT approach to quantum gravity, let us recall that the implementation of the effective relational dynamics defined above always relies on a choice of states. So, a different choice of states might indeed lead to a dynamics which can be better framed into a purely classical one. The possible inconsistency to which the CPSs lead to, i.e., the non-constancy in $\chi_0$ of the expectation values of $\hat{H}$ and $\hat{\Pi}$ on such CPSs is related to the introduction of the scale $\pi_0$. In turns, $\pi_0$ was introduced in order to allow fluctuations on the momentum operator to stay small, at least in some regimes. Indeed, one can see \cite{toappear} that such fluctuations take a contribution (among others) of the form $N/\epsilon$. By taking $\pi_0\to 0$ 
relative fluctuations on the momentum operator grow large for large $N$, which was instead expected to characterize a classical regime. This behavior of relative fluctuations is actually expected and it is a clear feature of our choice of states: since we are localizing each wavefunction of every single GFT atom around $\chi_0$ with a (good) precision $\epsilon$, fluctuations in the variable conjugated to $\hat{\chi}/N$ (i.e., $\hat{\Pi}$) are expected to grow as $1/\epsilon\times N$, and this is the very same behavior of relative fluctuations as well, if the expectation value of $\hat{\Pi}$ on a CPS is not extensive (if it was the case, it would depend on $N$ and thus on $\chi_0$). The way out of this problem, thus, may be to construct more realistic condensate states encoding non-trivial correlations among the fundamental GFT quanta (thus possibly allowing for a suppression of quantum fluctuations on the momentum operator) while still implementing a notion of $\chi$-localization of macroscopic observables. 

\section{Conclusions}
In this paper we have offered, first of all, a general perspective on the problem of time in quantum gravity and, more specifically, on the relational
strategy to solve it. We have discussed the different ways in which to implement this strategy, emphasizing the distinction between a classical implementation and one taking place at the quantum level, which we argued should be preferred. We have also pointed out the fundamentally new dimension that the problem takes in a quantum gravity context in which spacetime and geometry are understood as emergent notions from a different type of pre-geometric entities, and argued that the relational dynamics should be seen, in such context, as emergent as well, thus obtained only at an effective, approximate level. 

Then, we have realized concretely the general relational strategy we have advocated in the context of the tensorial group field theory formalism for quantum gravity, leading to the extraction of an effective relational cosmological dynamics from quantum geometric models, in which the universe is described as a quantum many-body system of simplicial building blocks, and a continuum cosmological dynamics with the correct classical limit can be extracted using the effective relational strategy.

We have then analysed in some detail the emergent cosmological dynamics, highlighting: a) the improvements over previous work, at both technical and conceptual level; b) the modifications that our effective relational strategy implies for the emergent cosmological dynamics, in particular c) the contribution of the quantum properties of the relational clock to it; d) the delicate interplay between the conditions ensuring a bona fide relational dynamics throughout the cosmological evolution and the existence of a quantum bounce in the early universe replacing the classical big bang singularity. 

\

Many physical and conceptual issues, concerning cosmology seen from a fundamental quantum gravity perspective, can now be tackled on a more solid basis, starting from the issue of quantum fluctuations of relevant geometric observables along the cosmological dynamics and of the limits of validity of the hydrodynamic approximation within which this dynamics has been extracted in our quantum gravity context \cite{toappear}.

\section*{Acknowledgements}
The authors would like to thank an anonymous referee for useful and insightful comments; Steffen Gielen, Axel Polaczek, Isha Kotecha and Giancarlo Cella for discussions on the extraction of relational dynamics and on GFT condensate cosmology. Financial support from the Deutsche Forschunggemeinschaft (DFG) is gratefully acknowledged. LM thanks the University of Pisa and the INFN (section of Pisa) for financial support, and the Ludwig-Maximilians-Universität Munich for the hospitality. 

\appendix

\section{Approximations and scales}\label{app:approximations}
It is important to analyze for which values of the parameters introduced above the approximations performed give rise to a self-consistent framework. In order to make the analysis simpler, we will consider all the \virgolette{purely geometric quantities} i.e., all the $K^{(2n)}$s and the derived objects $B_{j}$ and $A_{j}$ to be of order $1$. Moreover, we will neglect interactions and we will stick to the large $\rho_{j}$s case, which is of crucial important for the semi-classical limit. While in this appendix we will provide only a simplified discussion, we refer to \cite{toappear} (where, for instance, the assumption of large $\rho_j$s is dropped) for more details.

The first approximation we did was to obtain the dynamic equation \eqref{eqn:condensateequationofmotion}. The crucial object weighting differently derivatives of different order of the condensate wavefunction is the integral $I_{2n+m}$, which is basically given by
\begin{equation}
I_{2n+m}(\epsilon,\pi_{0})\propto \left(\frac{\epsilon}{2}\right)^{(2n+m)/2}H_{2n+m}\left(\frac{\sqrt{\epsilon}}{2}\pi_{0}\right)\,.
\end{equation}
As we have already noticed, a condition which allow us to keep variances of quantum operators small in the limit of large number of particles, is the condition
\begin{equation}
\epsilon\pi_{0}^2\gg 1\,.
\end{equation}
If this condition is satisfied, we can approximate the Hermite polynomial just with its higher order term, which is going to be proportional to $(\epsilon\pi_{0}^2/2)^{(2n+m)/2}$. Thus, the factor suppressing higher order derivatives can now be approximated as
$I_{2n+m}\sim (\epsilon\pi_{0})^{2n+m}$. So we see that the condition 
\begin{equation}
\pi_{0}\epsilon< 1
\end{equation}
allows us to perform the truncation as discussed above. 

The second important approximation we made was to use a saddle point approximation to compute the expectation value and the variances of the operators. For example, for the expectation value of the number operator, we have neglected a term of the form $[\rho_{j}^2]''(\chi_{0})(\epsilon/4)$ with respect to the term $\rho_{j}^2(\chi_{0})$. The neglected quantity, on shell, is
\begin{equation*}
\epsilon\left(\mathcal{E}_{j}/(2\rho_{j}^{2})+\mu_{j}^2\right)\,.
\end{equation*}
So we see that we have two contributions, the first one which is obviously negligible for large enough $\rho_{j}$s, and a second one which is actually negligible only when
\begin{equation}\label{eqn:saddlepointcondition}
\mu_{j}^2\epsilon=\frac{\epsilon\pi_{0}^2}{\epsilon\pi_{0}^2-1}\left(\frac{2}{\epsilon\pi_{0}^2}-\frac{1}{\epsilon \pi_{0}^2-1}\right)+\epsilon\frac{B_{j}}{A_{j}}\ll 1\,.
\end{equation}
We see immediately that under our assumptions $\epsilon\pi_{0}^2\gg 1$ and $B_{j}/A_{j}\ll \epsilon^{-1}$, this second term is always negligible. The same arguments of course hold for the expectation value of the volume operator and for the effective hamiltonian $\hat{H}$ and the momentum operator as well (in fact, the expectation value of these two operators actually reduce to just a computation of the expectation value of the number operator).

About the expectation value of the scalar field operator, the situation is very similar, but this time the term we neglected is actually of the form
\begin{equation*}
\frac{\epsilon}{4\chi_{0}}\frac{(\chi\rho_{j}^2)''(\chi_{0})}{\rho_{j}^2(\chi_{0})}=\frac{\epsilon}{4\chi_{0}}\frac{[\chi_{0}(\rho_{j}^2)''(\chi_{0})+2(\rho_{j}^2)'(\chi_{0})]}{\rho_{j}^2(\chi_{0})}\,.
\end{equation*}
For the first term the same arguments as before hold, while for the second one, the main contribution for large $\rho_{j}$s is given in modulus by $\epsilon\vert\mu_{j}\vert/\vert\chi_{0}\vert$, from which we deduce that, in this limit, it is necessary that 
\begin{equation}
\vert\chi_{0}\vert\gg\epsilon\vert\mu_j\vert\,,
\end{equation}
in order to make the approximation meaningful. In particular, since $\epsilon\mu_j\sim \pi_0^{-1}$ when $\epsilon\pi_0^2\gg 1$, this implies that $\epsilon\vert\mu_j\vert\ll \sqrt{\epsilon}$. Thus the above condition is satisfied as long as $\vert\chi_0\vert \gg\sqrt{\epsilon}$. Notice, that $\sqrt{\epsilon}$ can be seen as the statistical relative variance of a Gaussian distributed variable with mean $\chi_{0}$ and width of order $\epsilon$. The requirement $\vert\chi_0\vert\gg\sqrt{\epsilon}$ seems therefore a very natural one in our \virgolette{peaked framework}. 
\section{Dynamics from an effective action}\label{app:dynamicseffective}
Here we want to show how a different procedure to define dynamics actually leads, in the limit where interactions are neglected to the same dynamical equations as in \eqref{eqn:condensateequationofmotion}, but with different parameters.

The way we want to obtain the dynamics, here, consists in writing an effective action of the form
\begin{equation*}
 S_{\text{eff}}=\int\diff \chi_{0}\braket{\hat{S}}_{\sigma_{\epsilon};\chi_{0},\pi_{0}}\,.
 \end{equation*}
We will restrict, for simplicity, just to the case of negligible interactions. First, we compute 
\begin{align*}
\braket{\hat{K}}_{\sigma_{\epsilon};\chi_{0},\pi_{0}}&=\int\diff g_I\diff h_I\int\diff\chi\diff\chi'\,\overline{\tilde{\sigma}}(g_I,\chi)\overline{\eta}_{\epsilon}(\chi-\chi_{0},\pi_{0})\\
&\quad\times K(g_I,h_I;(\chi-\chi')^2)\tilde{\sigma}(h_I,\chi')\eta_{\epsilon}(\chi'-\chi_{0},\pi_{0})\\
&=\int\diff g_I\diff h_I\int\diff\chi\diff u\,\overline{\tilde{\sigma}}(g_I,\chi)\overline{\eta}_{\epsilon}(\chi-\chi_{0},\pi_{0})\\
&\quad\times K(g_I,h_I;u^2)\tilde{\sigma}(h_I,\chi+u)\eta_{\epsilon}(\chi-\chi_{0}+u,\pi_{0})\,.
\end{align*}
Next, we notice that, by changing variables and defining $x\equiv \chi-\chi_{0}+u/2$, we have that the product of peaking functions is given by
\begin{equation*}
\overline{\eta}_{\epsilon}(x-u/2,\pi_{0})\eta_{\epsilon}(x+u/2,\pi_{0})=\mathcal{N}_{\epsilon}^2e^{-u^2/4\epsilon}e^{-x^2/\epsilon}e^{i\pi_{0}u}\,,
\end{equation*}
so that the kinetic term becomes
\begin{align*}
\braket{\hat{K}}_{\sigma_{\epsilon};\chi_{0},\pi_{0}}&=\mathcal{N}_{\epsilon}^2\!\!\int\!\diff g_I\diff h_I\!\!\int\! \diff u\,K(g_I,h_I;u^2)e^{-u^2/4\epsilon+i\pi_{0} u}\\
&\quad\times\int\diff x\,\overline{\tilde{\sigma}}\left(g_I,x+\chi_{0}-\frac{u}{2}\right)\tilde{\sigma}\left(h_I,x+\chi_0+\frac{u}{2}\right)e^{-x^2/\epsilon}\,.
\end{align*}
The factor in the second line can be expanded in a Taylor series around $x=0$, given the exponential peaking function around that point. The lowest order in $\epsilon$ containing derivative terms is given by
\begin{equation*}
 \frac{\epsilon}{2}\biggl[\frac{1}{2}\left[\overline{\tilde{\sigma}^{(2)}}(g_{I},\chi_{0,-})\tilde{\sigma}(h_{I},\chi_{0,+})+\overline{\tilde{\sigma}}(g_{I},\chi_{0,-})\tilde{\sigma}^{(2)}(h_{I},\chi_{0,+})\right]+\tilde{\sigma}^{(1)}(g_I,\chi_{0,-})\tilde{\sigma}^{(1)}(h_{I},\chi_{0,+})\biggr],
 \end{equation*} 
where $\chi_{0,\pm}\equiv \chi_{0}\pm u/2$. Now, these quantities are already of higher order, so for the $u$ integration we can just use the lowest order and evaluate all the above quantities in $u=0$. It is easy to see that this quantity, which now depends entirely on $\chi_{0}$, when it is integrated on $\chi_{0}$  (and when isotropy is assumed) to obtain the effective action, gives exactly zero by integration by parts (it is a total derivative). Since higher derivative contributions will be of order higher than $\epsilon$, which is where we want to stop our perturbative expansion, we are going to neglect them and just consider the non-derivative term, which is given by 
\begin{equation*}
\braket{K}_{\sigma_{\epsilon};\chi_0,\pi_{0}}=\sqrt{\pi\epsilon}\mathcal{N}_\epsilon^2\int\!\diff g_I\diff h_I\!\int\! \diff u\,K(g_I,h_I;u^2)e^{-u^2/4\epsilon} e^{-i\pi_0 u}\overline{\tilde{\sigma}}(g_I,\chi_{0}-u/2)\tilde{\sigma}(h_I,\chi_0+u/2)\,.
\end{equation*}
We next assume that 
\begin{equation*}
K(g_I,h_I;u^2)\equiv \sum_{n=0}^\infty \frac{K^{(2n)}(g_I,h_I)}{(2n)!}u^{2n}\,.
\end{equation*}
Now, we use again that the exponential peaks on $u=0$ and we expand the reduced wave-functions in Taylor series around $u=0$, so that we find
\begin{align*}
\braket{K}_{\sigma_{\epsilon};\chi_0,\pi_{0}}&=\sqrt{\pi\epsilon}\mathcal{N}_{\epsilon}^2\int\diff g_I\diff h_I\int \diff u\,e^{-u^2/4\epsilon}e^{i\pi_0 u}\sum_{n=0}^\infty \frac{K^{(2n)}(g_I,h_I)}{(2n)!}u^{2n}\sum_{m=0}^\infty\frac{(-1)^m}{m!}\frac{u^m}{2^m}\\
&\quad\quad\times\left[\frac{\partial^m}{\partial\chi^m}\overline{\tilde{\sigma}}(g_I,\chi)\right]_{\chi_{0}}\sum_{l=0}^\infty\frac{1}{l!}\frac{u^l}{2^l}\left[\frac{\partial^l}{\partial\chi^l}\tilde{\sigma}(g_I,\chi)\right]_{\chi_{0}}\\
&\quad=\sqrt{\pi\epsilon}\mathcal{N}_{\epsilon}^2\int\diff g_I\diff h_I\sum_{n,m,l}\frac{K^{(2n)}(g_I,h_I)}{(2n)!}\frac{(-1)^m}{m!l!}\frac{1}{2^{m+l}}\\
&\quad\quad\times\left[\frac{\partial^m}{\partial\chi^m}\overline{\tilde{\sigma}}(g_I,\chi)\right]_{\chi_{0}}\left[\frac{\partial^l}{\partial\chi^l}\tilde{\sigma}(g_I,\chi)\right]_{\chi_{0}}\!\!J_{2n+l+m}(\epsilon,\pi_{0})\,,
\end{align*}
where
\begin{align*}
I_{2n+l+m}(\tilde{\pi}_0,\tilde{\epsilon})&=\int\diff u\,e^{-u^2/4\epsilon}e^{i\pi_0 u} u^{2n+m+l}=(-i)^{2n+m+l}\frac{\partial^{2n+m+l}}{\partial \pi_0^{2n+m+l}}\int\diff u\,e^{-u^2/4\epsilon}e^{i\pi_0 u}\\
&=(-i)^{2n+m+l}2\sqrt{\pi\epsilon}\frac{\partial^{2n+m+l}}{\partial \pi_0^{2n+m+l}}e^{-\pi_0^2\epsilon}=(-i)^{2n+m+l}2\sqrt{\pi\epsilon}\epsilon^{(2n+m+l)/2}\frac{\partial^{2n+m+l}}{\partial x^{2n+m+l}}e^{-x^2}\\
&=(i)^{2n+m+l}2\sqrt{\pi\epsilon}\epsilon^{(2n+m+l)/2}H_{2n+m+l}(\pi_0\sqrt{\epsilon})e^{-\pi_0^2\epsilon}\,.
\end{align*}
Putting everything together, we are left with
\begin{align*}
\braket{K}_{\sigma_{\epsilon};\chi_0,\pi_{0}}&=2\pi\epsilon\mathcal{N}_{\epsilon}^2\!\!\int\!\!\diff g_I\diff h_I\sum_{n,m,l}\frac{K^{(2n)}(g_I,h_I)}{(2n)!}\frac{(-1)^m}{m!l!}\frac{i^{2n+m+l}}{2^{m+l}}\epsilon^{(2n+m+l)/2}H_{2n+m+l}(\pi_0\sqrt{\epsilon})\\
&\quad\times e^{-\pi_0^2\epsilon}\overline{\tilde{\sigma}}^{(m)}(g_I,\chi_{0})\tilde{\sigma}^{(l)}(h_I,\chi_{0})\,.
\end{align*}
Consistently with what we did before, we keep only terms with $2n+m+l=2$. We have:
\begin{subequations}
\begin{itemize}
\item \emph{Zeroth order}: this means $n=m=l=0$. Thus we have immediately that
\begin{equation*}
\braket{K}_{\sigma_{\epsilon};\chi_0,\pi_{0}}^{(0,0,0)}=2\pi\epsilon\mathcal{N}_{\epsilon}^2K^{(0)}e^{-\pi_0^2\epsilon}\int\diff g_I\diff h_I\overline{\tilde{\sigma}}(g_I,\chi_{0})\tilde{\sigma}(h_I,\chi_{0})\,.
\end{equation*}
\item \emph{First order}: this means $n=0$ and $m=0$, $l=1$ or $m=1$, $l=0$. The two contributions are (recall that $H_{1}(x)=2x$)
\begin{align*}
\braket{K}_{\sigma_{\epsilon};\chi_0,\pi_{0}}^{(0,1,0)}&=-2i\pi\epsilon\mathcal{N}_{\epsilon}^2K^{(0)}e^{-\pi_0^2\epsilon}\sqrt{\epsilon}\pi_0\sqrt{\epsilon}\int\diff g_I\diff h_I\overline{\tilde{\sigma}}^{(1)}(g_I,\chi_0)\tilde{\sigma}(h_I,\chi_0)\,,\\
\braket{K}_{\sigma_{\epsilon};\chi_0,\pi_{0}}^{(0,1,0)}&=2i\pi\epsilon\mathcal{N}_{\epsilon}^2K^{(0)}e^{-\pi_0^2\epsilon}\sqrt{\epsilon}\pi_0\sqrt{\epsilon}\int\diff g_I\diff h_I\overline{\tilde{\sigma}}(g_I,\chi_0)\tilde{\sigma}^{(1)}(h_I,\chi_0)\,.
\end{align*}
\item \emph{Second order}: this means $n=1$ and $m=l=0$ or $n=0$ and $m+l=2$, i.e., $m=2,l=0$, $m=1, l=1$, $m=0$, $l=2$. We list all these four possibilities (recall that $H_{2}(x)=4x^2-2$):
\begin{align*}
\braket{K}_{\sigma_{\epsilon};\chi_0,\pi_{0}}^{(1,0,0)}&=-2\pi\epsilon\mathcal{N}_{\epsilon}^2\frac{K^{(2)}}{2}(4\epsilon\pi_0^2-2)e^{-\pi_0^2\epsilon}\epsilon\times\int\diff g_I\diff h_I\overline{\tilde{\sigma}}(g_I,\chi_0)\tilde{\sigma}(h_I,\chi_0)\,,\\
\braket{K}_{\sigma_{\epsilon};\chi_0,\pi_{0}}^{(0,2,0)}&=-2\pi\epsilon\mathcal{N}_{\epsilon}^2\frac{K^{(0)}}{8}e^{-\pi_0^2\epsilon}(4\epsilon\pi_0^2-2)\epsilon\int\diff g_I\diff h_I\overline{\tilde{\sigma}}^{(2)}(g_I,\chi_0)\tilde{\sigma}(h_I,\chi_0)\,,\\
\braket{K}_{\sigma_{\epsilon};\chi_0,\pi_{0}}^{(0,0,2)}&=-2\pi\epsilon\mathcal{N}_{\epsilon}^2\frac{K^{(0)}}{8}e^{-\pi_0^2\epsilon}(4\epsilon\pi_0^2-2)\epsilon\int\diff g_I\diff h_I\overline{\tilde{\sigma}}(g_I,\chi_0)\tilde{\sigma}^{(2)}(h_I,\chi_0)\,,\\
\braket{K}_{\sigma_{\epsilon};\chi_0,\pi_{0}}^{(0,1,1)}&=2\pi\epsilon\mathcal{N}_{\epsilon}^2\frac{K^{(0)}}{4}e^{-\pi_0^2\epsilon}(4\epsilon\pi_0^2-2)\epsilon\int\diff g_I\diff h_I\overline{\tilde{\sigma}}^{(1)}(g_I,\chi_0)\tilde{\sigma}^{(1)}(h_I,\chi_0)\,.
\end{align*}
\end{itemize}
\end{subequations}
Now we write the action as the integral of $\braket{K}_{\sigma_{\epsilon};\chi_0,\pi_{0}}$ over $\tilde{\chi}_{0}$, and we find
\begin{align*}
S_{\text{eff}}&\simeq 2\pi\epsilon\mathcal{N}^2_{\epsilon}e^{-\pi_0^2\epsilon}\int\diff g_I\diff h_I\int\diff\chi_{0}\biggl[\overline{\tilde{\sigma}}(g_I,\chi_{0})\tilde{\sigma}(h_I,\chi_{0})\left(K^{(0)}-K^{(2)}(2\epsilon\pi_0^2-1)\epsilon\right)\nonumber\\
&\quad\quad\quad+2iK^{(0)}\epsilon\pi_0\overline{\tilde{\sigma}}(g_I,\chi_{0})\tilde{\sigma}'(g_I,\chi_{0})-K^{(0)}(2\epsilon\pi_0^2-1)\epsilon\overline{\tilde{\sigma}}(g_I,\chi_{0})\tilde{\sigma}''(g_I,\chi_{0})\biggr]\,,
\end{align*}
where a prime denotes a derivative with respect to $\chi_{0}$ and we have integrated by parts some terms. From this we find the following equations of motion for the reduced wave-function:
\begin{equation}\label{eqn:condensateequationfromeffective}
\tilde{\sigma}''_{j}(\chi_{0})-2i\tilde{\pi}_{0}\tilde{\sigma}'_{j}(\chi_{0})-E^2_{j}\tilde{\sigma}_{j}(\chi_{0})=0\,,
\end{equation}
where
\begin{equation}
\tilde{\pi}_{0}\equiv\frac{\pi_{0}}{2\pi_{0}^2\epsilon-1}\,,\qquad E_j^2\equiv \frac{1}{\epsilon}\frac{1}{2\epsilon\pi_{0}^2-1}-\frac{B_{j}}{A_{j}}\,.
\end{equation}
We see that equations \eqref{eqn:condensateequationfromeffective} are identical to equations \eqref{eqn:condensateequationofmotion} with the replacement $\epsilon\to 2 \epsilon$.

\section{Classical relational dynamics of flat FRW spacetime}\label{app:classical}
In order to compare our effective relational results with the classical ones, in this appendix we review the classical relational setting. We will start from the total Hamiltonian of a flat FRW spacetime with a minimally coupled massless scalar field. This is given by \cite{Bojowald:2008zzb}
\begin{equation*}
S=\frac{3}{8\pi G}\int\diff t\,N\left(-\frac{aV_{0}\dot{a}^2}{N^2}+\frac{V}{N}\frac{\dot{\chi}^2}{2N}\right)=-\frac{3}{8\pi G}\int\diff t NV\left(\frac{H^2}{N^2}-\frac{4\pi G}{3}\frac{\dot{\chi^2}}{N^2}\right)\,,
\end{equation*}
where $\chi$ is the massless scalar field, a dot denotes a derivative with respect to $t$ and $V_{0}$ is the fiducial coordinate volume (so that $V\equiv V_{0}a^3$). The constraint obtained from an Hamiltonian analysis of the above action is \begin{equation}
\mathcal{C}=-\frac{3}{8\pi G}NVH^2+\frac{N\pi_{\chi}^2}{2V}=0\,.
\end{equation}
Together with the Poisson brackets $\{H,V\}=4\pi G$ and $\{\chi,\pi_{\chi}\}=1$, the above constraint implies that the equation of motion for the massless scalar field and the volume are
\begin{equation*}
\dot{\chi}=\{\chi,\mathcal{C}\}=N\pi_\chi/V\,,\qquad
\dot{V}=\{V,\mathcal{C}\}=3NVH\,,
\end{equation*}
and inserting the former into the latter, together with the constraint equation, we obtain
\begin{equation}
\left(\frac{1}{3V}\frac{\diff V}{\diff\chi}\right)^2\equiv\left(\frac{V'}{3V}\right)^2=\frac{4\pi G}{3}\,.
\end{equation}
By deriving this equation with respect to $\chi$, we find
\begin{equation}
V''/V=\left(V'/V\right)^2=12\pi G\,.
\end{equation}
These are the relational equations for a spatially flat FRW spacetime.
\paragraph*{Gauge fixing.}
Let us now perform a gauge fixing, choosing $\chi$ as our time, i.e.,  choosing $N=V\dot{\chi}/\pi_{\chi}$. In this way, we obtain
\begin{equation}
S=-\frac{3}{8\pi G}\int\diff t\dot{\chi} \frac{V^2}{\pi_{\chi}}\left(\frac{H^2\pi_{\chi}^2}{V^2\dot{\chi}^2}-\frac{4\pi G}{3}\frac{\dot{\chi^2}\pi_{\chi}^2}{V^2\dot{\chi}^2}\right)=-\frac{3\pi_{\chi}}{8\pi G}\int\diff\chi\left(\mathcal{H}^2-\frac{4\pi G}{3}\right)\,.
\end{equation}
The equations of motion generated by this action are easily obtained by writing $\mathcal{H}=V'/(3V)$:
 \begin{equation*}
  V''/V=(V')^2/V^2\,,
  \end{equation*}
which is the second Friedmann equation, and which gives indeed the correct dynamics. The Hamiltonian obtained from the above Lagrangian, therefore, neglecting irrelevant constants, can be written immediately as
\begin{equation}\label{eqn:classicalrelationalhamiltonian}
H_{\text{rel}}=-\frac{3\pi_{\chi}}{8\pi G}\mathcal{H}^2\,.
\end{equation}

\nocite{*}

\bibliographystyle{jhep}
\bibliography{paper_effective}

\end{document}